\documentclass{aa}  

\usepackage{natbib,twoopt}
\usepackage{graphicx,xcolor,textpos}
\usepackage{rotating}
\usepackage{longtable,ltcaption}
\usepackage{txfonts}
\usepackage{hyperref}
\usepackage{lscape} 
\usepackage[utf8]{inputenc}
\DeclareUnicodeCharacter{2212}{-}
\usepackage{gensymb}
\DeclareUnicodeCharacter{00A0}{~}
\bibpunct{(}{)}{;}{a}{}{,}

\begin{document} 
\title{Barium \& related stars and their white-dwarf companions 
\thanks{Based on observations carried out with the Flemish Mercator Telescope at the Spanish Observatorio del Roque de los Muchachos (ORM, La Palma, Spain), the Swiss 1m-telescope at Haute Provence Observatory (OHP, France) and the 1.54-m Danish telescope and the Swiss 1.2m telescope at the European Southern Observatory (ESO, La Silla, Chile).},
\thanks{Based on observations obtained with the HERMES spectrograph, which is supported by the Fund for Scientific Research of Flanders (FWO), Belgium, the Research Council of KU Leuven, Belgium, the Fonds National de la Recherche Scientifique (F.R.S.-FNRS), Belgium, the Royal Observatory of Belgium, the Observatoire de Gen\`{e}ve, Switzerland and the Th\"{u}ringer Landessternwarte Tautenburg, Germany.}\\
II. Main-sequence and subgiant stars
}

\author{A. Escorza\inst{1,2} \and
D. Karinkuzhi\inst{2,3} \and
A. Jorissen\inst{2} \and
L. Siess\inst{2} \and
H. Van Winckel\inst{1} \and
D. Pourbaix\inst{2} \and
C. Johnston\inst{1} \and
B. Miszalski\inst{4,5} \and
G-M. Oomen\inst{1,6} \and
M. Abdul-Masih\inst{1}\and
H.M.J. Boffin\inst{7} \and
P. North\inst{8} \and 
R. Manick\inst{1,4} \and
S. Shetye\inst{2,1} \and
J. Miko{\l}ajewska\inst{9}
}

\offprints{A. Escorza, \\ \email{ana.escorza@kuleuven.be}}

\institute{
    Institute of Astronomy, KU Leuven, Celestijnenlaan 200D, B-3001 Leuven, Belgium \\
    \email ana.escorza@kuleuven.be \and
    Institut d'Astronomie et d'Astrophysique, Universit\'{e} Libre de Bruxelles, Boulevard du Triomphe, B-1050 Bruxelles, Belgium \and
    Department of physics, Jnana Bharathi Campus, Bangalore University, Bangalore 560056, India \and
    South African Astronomical Observatory, PO Box 9, Observatory 7935, South Africa \and
	Southern African Large Telescope Foundation, PO Box 9, Observatory 7935, South Africa \and
	Department of Astrophysics/IMAPP, Radboud University, P.O. Box 9010, 6500 GL Nijmegen, The Netherlands \and
  	ESO, Karl-Schwarzschild-str. 2, 85748 Garching bei M\"{u}nchen, Germany \and
    Institut de Physique, Laboratoire d’astrophysique, \'{E}cole Polytechnique F\'{e}d\'{e}rale de Lausanne (EPFL), Observatoire, 1290 Versoix, Switzerland \and
    N. Copernicus Astronomical Center, Polish Academy of Sciences, Bartycka 18, 00-716 Warsaw, Poland
}

\date{Accepted}

\abstract{Barium (Ba) dwarfs and CH subgiants are the less-evolved analogues of Ba and CH giants. They are F- to G-type main-sequence stars polluted with heavy elements by a binary companion when the latter was on the Asymptotic Giant Branch (AGB). This companion is now a white dwarf that in most cases cannot be directly detected. We present a large systematic study of 60 objects classified as Ba dwarfs or CH subgiants. Combining radial-velocity measurements from HERMES and SALT high-resolution spectra with radial-velocity data from CORAVEL and CORALIE, we determine the orbital parameters of 27 systems. We also derive their masses by comparing their location in the Hertzsprung-Russell diagram with evolutionary models. We confirm that Ba dwarfs and CH subgiants are not at different evolutionary stages and have similar metallicities, despite their different names. Additionally, Ba giants appear significantly more massive than their main-sequence analogues. This is likely due to observational biases against the detection of hotter main-sequence post-mass-transfer objects. Combining our spectroscopic orbits with the Hipparcos astrometric data, we derive the orbital inclinations and the mass of the WD companion for four systems. Since this cannot be done for all systems in our sample yet (but should be with upcoming Gaia data releases), we also analyse the mass-function distribution of our binaries. We can model this distribution with very narrow mass distributions for the two components and random orbital orientation on the sky. Finally, based on BINSTAR evolutionary models, we suggest that the orbital evolution of low-mass Ba systems can be affected by a second phase of interaction along the Red Giant Branch of the Ba star, impacting on the eccentricities and periods of the giants.\\}

\keywords{stars: late-type - stars: chemically peculiar - stars: binaries: spectroscopic - techniques: spectroscopic}

\titlerunning{Orbits of Ba dwarfs}
\authorrunning{Escorza et al.}

\maketitle

\section{Introduction}\label{sec:intro}

Barium (Ba) stars or \ion{Ba}{II} stars, as they were originally named, are G- and K-type giants with peculiarly strong absorption lines of slow-neutron-capture (s)-process elements in their spectra in combination with enhanced CH, CN and C$_{\rm 2}$ molecular bands. They were first identified as chemically peculiar by \cite{BidelmanKeenan51}, who discussed their distinctive spectroscopic characteristics and particularly stressed the extraordinary strength of the resonance line of ionised barium at 4554\,\text{\AA}. The resulting overabundance of barium and other s-process elements on the surface of red giant stars could not be explained from an evolutionary point of view because the s-process of nucleosynthesis takes place in the interiors of Asymptotic Giant Branch (AGB) stars. Ba giants occupy the first giant branch (RGB) or the Red Clump (\citealt{Escorza17}; and references therein) and are hence not advanced enough in their evolution to produce and dredge up s-process elements.

However, Ba stars are now understood to originate from a binary evolution channel. According to this formation scenario, the carbon and the s-process elements were transferred to the current primary from a more evolved companion when the latter was in its AGB phase. This implies that the companions of Ba stars are white dwarfs (WD). The presence of these WD companions has since been indirectly and directly supported. For example, \cite{Webbink86}, \cite{McClureWoodsworth90}, \cite{Jorissen98}, \cite{Merle16}, and \cite{Jorissen19} found that the mass-function distribution of Ba giants is consistent with a narrow distribution of companion masses peaking at 0.6 \textit{M}$_{\sun}$. Additionally, \cite{Bohm-Vitense84, Bohm-Vitense00}, and \cite{Gray11}, among others, detected UV excess flux from some Ba star systems which can be attributable to a WD companion.

The orbital properties of Ba giants (gBa) have been intensively and systematically studied (e.g \citealt{McClure84}; \citealt{McClureWoodsworth90}; \citealt{Udry98}; \citealt{Jorissen98, Jorissen16, Jorissen19}) since they are a prototypical family of post-mass-transfer low- or intermediate-mass binary systems. A remaining long-standing problem concerning these objects is that binary evolutionary models cannot account for their observed orbital properties. Distributions of observed periods and eccentricities, and abundances of s-process elements are not well reproduced by these models (e.g., \citealt{Pols03}, \citealt{BonacicMarinovic08}, \citealt{Izzard10}). This is a common problem among post-interaction binary systems, such as post-AGB stars (e.g., \citealt{Oomen18}), CH stars (e.g., \citealt{McClureWoodsworth90}), carbon-enhanced metal-poor stars (e.g., \citealt{Izzard10}, \citealt{Jorissen16}, \citealt{Abate18}), blue stragglers (e.g., \citealt{MathieuGeller15}), symbiotic stars (e.g., \citealt{Mikoajewska12}), and subdwarf B-type binaries (e.g., \citealt{Vos17}).

CH stars are closely related to Ba stars and are seen as their Population II analogues. They are also evolved stars with similar enhancement of s-process elements and strong CH molecular bands. However, they have overall weaker metal lines because they belong to an older and more metal-deficient population. They were identified as chemically peculiar for the first time by \cite{Keenan42}.

At lower luminosities, dwarf barium (dBa) stars have been much less intensively studied. They are thought to be the less evolved analogues of barium giants as they are F- or G-type main sequence stars with the same enhancement of carbon and heavy elements (\citealt{NorthDuquennoy91}, \citealt{NorthLanz91}, \citealt{JorissenBoffin92}, \citealt{North94}, \citealt{North00}). Similarly, subgiant CH stars (sgCH), which were first identified by \cite{Bond74}, are the low-metallicity counterparts of barium dwarfs and historically thought to be slightly more evolved. However, the distinction between these two subclasses is not very straightforward. We recently suggested that there is no separation between subgiant CH stars and dwarf Ba stars in the Hertzsprung–Russell diagram (HRD), contrary to what is implied by their designations \citep{Escorza17}. 

In this paper, we focus on the stellar and binary properties of Ba dwarfs and CH subgiants. We combine new radial-velocity measurements obtained from high-resolution spectra with archival radial-velocity data and determine the orbital parameters of a sample of these objects. In this way, we significantly increase the number of known orbits of main sequence and subgiant Ba and CH stars, and we can compare these populations with their evolved analogues. We describe the stellar sample and the data sets in Sect. \ref{sec:data} and our method to determine the best orbital solution in Sect. \ref{sec:orbits}. The results are presented in Sect. \ref{sec:results}, where we also describe some individual targets that deserve special attention. Additionally, in Sect. \ref{sec:hrd}, we use stellar parameters derived from high-resolution spectra and the distances derived by \cite{Bailer-Jones18} from \textit{Gaia} DR2 parallaxes \citep{Lindegren18} to plot our targets on an HRD and derive their masses by comparing their location with STAREVOL evolutionary models (\citealt{Siess00}; \citealt{SiessArnould08}). Finally in Sect. \ref{sec:disc}, we discuss the stellar and orbital properties of our main sequence stars and compare them with a well-studied sample of Ba giants. We also analyse the derived binary mass-functions to obtain information about the WD companion. For four of our objects, we derive absolute WD masses using orbital inclinations obtained by reprocessing Hipparcos astrometric data. Finally, we use BINSTAR binary evolution models \citep{BINSTAR13} to explore the evolutionary link between our main sequence and subgiants stars and the better-known Ba and CH giants.


\section{Sample overview and data description}\label{sec:data}

In 1984, a monitoring campaign of barium and related stars was initiated with the CORAVEL spectrometers \citep{Baranne79}. Some results of this programme were presented in  \cite{Jorissen88}, \cite{Jorissen98} and \cite{Udry98}. Due to the long periods of some of these binaries, the CORAVEL monitoring was not long enough to derive the orbital parameters of all the objects in the sample. Several years after the CORAVEL programme was interrupted, the monitoring of these families of binaries was resumed with the HERMES spectrograph \citep{Raskin11}. Combining the older CORAVEL radial-velocity measurements with the more recent HERMES data, the total time coverage amounts to more than 30 years, and a full orbital cycle can now be covered for some long-period binaries for the first time. The combination of these two data sets has proven successful in past studies (e.g., \citealt{Jorissen16}; \citealt{Oomen18}). Additionally, thanks to the higher accuracy of the HERMES radial velocities, systems with much lower-amplitude radial-velocity variations could also be revealed. The HERMES spectrograph is limited to observing objects with declination $\gtrsim$\,-\,30$^\circ$ thus, the southern objects in our sample were observed with the 11-m Southern African Large Telescope (SALT; \citealt{Buckley06}; \citealt{ODonoghue06}).


\subsection{Radial-velocity measurements with CORAVEL and CORALIE}\label{ssec:CORAVEL}

\begin{figure}[t]
\centering
\includegraphics[width=0.49\textwidth]{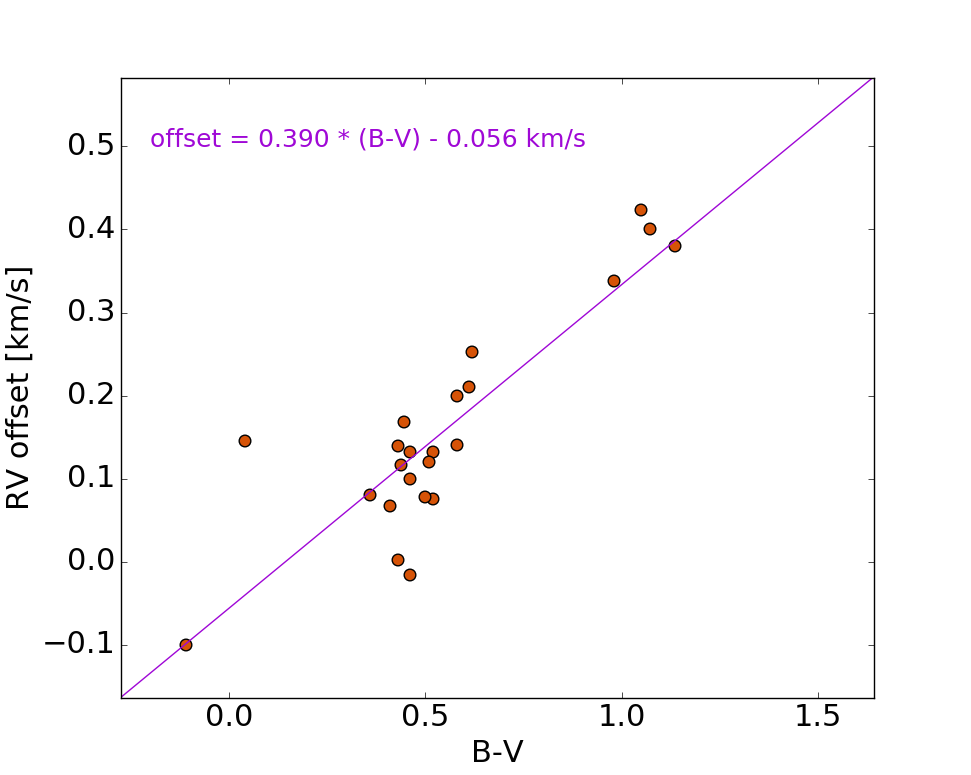}
\caption{\label{offset} Relation between the $B-V$ index of the stars and the radial-velocity offset applied on the old CORAVEL data to make it compatible with datasets tied to the new ELODIE system.}
\end{figure}

The monitoring campaign of Ba stars started with data from the two CORAVEL spectrometers: one installed on the 1-m Swiss telescope at the Haute-Provence Observatory for the northern sky and the other installed on the 1.54-m Danish telescope at ESO - La Silla (Chile) for the southern sky \citep{Baranne79}. The CORAVEL radial velocities were obtained by cross-correlation and Gaussian-fitting techniques, on the data obtained via the instrument, which has a hardware spectral mask based on a spectrum of Arcturus (K1.5III; see \citealt{Baranne79}, \citealt{Duquennoy91} or \citealt{Udry98} for more details).

The radial-velocity standard stars used to fix the zero-point of the HERMES spectrograph are tied to the ELODIE velocity system \citep{Udry1999}. However, some of our CORAVEL radial velocities are from before 1999 and are tied to an old velocity system, previous to the ELODIE system. Hence, a zero-point offset needs to be applied to these old measurements in order to convert them to the ELODIE system and combine them with HERMES data. Fortunately, we could access reprocessed pre-1999 CORAVEL data for some objects and use the difference between the reprocessed and the uncorrected data to define an offset for the rest of the stars. The zero-point offset depends on the B-V index \citep{Udry1999} and Fig.~\ref{offset} shows the relation that we obtained and applied to our other targets. Note that the scatter is not larger than the average CORAVEL errorbar ($\sim$~0.3~$\mathrm{km\,s}^{-1}$).

Additionally, a few objects have also been observed with the CORALIE spectrograph, installed on the Swiss 1.2-metre Leonhard Euler Telescope at ESO - La Silla (Chile) after CORAVEL was decommissioned in 1998. For one target, HD\,48565, we also have four radial-velocity measurements from ELODIE, an echelle spectrograph installed at the Observatoire de Haute-Provence 1.93m reflector in south-eastern France.


\subsection{Radial-velocity monitoring with the HERMES spectrograph}\label{ssec:HERMES}

The targets with declinations higher than -30$\degree$ have been observed with HERMES, the state-of-the-art fiber echelle spectrograph mounted on the 1.2-m Flemish Mercator telescope at the Spanish Observatorio del Roque de los Muchachos in La Palma. These targets are part of the long-term monitoring program of binaries \citep{Gorlova13} carried out with this instrument thanks to the agreement reached by all consortium partners (KU Leuven, Universit\'e libre de Bruxelles, Royal Observatory of Belgium, Observatoire de Gen\`{e}ve and Th\"{u}ringer Landessternwarte Tautenburg). As shown by many studies, scientific advancement in the study of evolved wide binaries is only made possible by regular long time-base monitoring, such as that guaranteed by HERMES@Mercator (e.g., \citealt{VanWinckel10}, \citealt{Vos15}, \citealt{Manick17}, \citealt{Oomen18}).

The instrument, which is fully described in \cite{Raskin11}, covers a spectral range from 380 nm to 900 nm and has a spectral resolution of about 85\,000 for the high-resolution science fibre. The temperature-controlled environment ensures a good long-term stability of the radial velocities. The radial velocities are obtained by performing a Gaussian fit on the cross-correlation function (CCF) obtained with a spectral mask matching the spectral type of the target stars. In our case, depending on the spectral type reported in the literature, we used the F0 or G2 masks in orders 55-74 (478\,nm\,-\,653\,nm), as these give the best compromise between absence of telluric lines and maximum signal-to-noise ratio for F-G-K type stars.

Crucial to obtain accurate radial velocities is the wavelength calibration of HERMES. Two lamps (ThAr and Ne) are used to obtain several wavelength reference exposures over the night to correct for the possible pressure drifts (see Fig. 9 of \citealt{Raskin11}). Additionally, the long-term accuracy may be estimated from the stability of RV standard stars (from the list of \citealt{Udry1999}) monitored along with the science targets. The distribution of the radial-velocity standards gives a 1\,$\sigma$ dispersion of 63\,m\,s$^{-1}$ (see Fig. 10 of \citealt{Raskin14}), which we adopt as the typical uncertainty on the radial velocities over the long term.


\subsection{Radial velocities with SALT HRS}\label{ssec:SALT}

We complemented our data sets with spectra obtained with the SALT High Resolution Spectrograph (HRS; \citealt{Bramall10, Bramall12}; \citealt{Crause14}). The targets were included as poor-weather targets in the SALT programme 2017-1-MLT-010 (PI: Miszalski; see e.g. \citealt{Miszalski18_1} and \citealt{Miszalski18_2}). The HRS is a dual-beam, fibre-fed echelle spectrograph enclosed in a vacuum tank located in an insulated, temperature controlled room of the 11-m Southern African Large Telescope (\citealt{Buckley06}; \citealt{ODonoghue06}). The medium resolution (MR) mode of HRS was used and the spectra covered a wavelength range from 370 nm to 890 nm with resolving powers R = $\lambda / \Delta\lambda$ of 43\,000 and 40\,000 for the blue and red arms, respectively. Regular bias, ThAr arc and quartz lamp flat-field calibrations are taken. The basic data products \citep{Crawford10} were reduced with the MIDAS pipeline developed by \cite{Kniazev16} which is based on the ECHELLE \citep{Ballester92} and FEROS \citep{Stahl99} packages. Heliocentric corrections were applied to the data using VELSET of the RVSAO package \citep{KurtzMink98}. We used the same spectral masks that we used to derive the radial velocities for the HERMES spectra by means of a Gaussian fit to the CCF. 

To make sure we could combine our datasets meaningfully, in 2018 we observed a radial-velocity standard star with both SALT HRS and HERMES. We chose HD\,156365 because its declination $\delta$\,$\sim$\,-\,24$^\circ$ makes it observable with both instruments. In the catalogue of radial-velocity standard stars for \textit{Gaia} \citep{Soubiran13}, the radial-velocity measurements of HD\,156365 range from -13.06 to -13.11 $\mathrm{km\,s}^{-1}$. We obtained RV$_{\mathrm{SALT\,HRS}}=-13.30\pm0.17\mathrm{km\,s}^{-1}$ and RV$_{\mathrm{HER18}}=-12.98\pm0.07\mathrm{km\,s}^{-1}$. In the data archive of HERMES, there was another spectrum of this star from 2013 for which we obtained RV$_{\mathrm{HER13}}=-13.08\pm0.07\mathrm{km\,s}^{-1}$. We conclude that this limited experiment shows a slight offset of about 260 $\mathrm{m\,s}^{-1}$ between SALT HRS and HERMES. As, however, the amount of HERMES and CORAVEL observations is so large, we decided not to apply any systematic offset to the SALT-HRS data.


\subsection{Stellar sample}\label{ssec:sample}

Our sample includes 60 stars that have been identified as Ba dwarfs or CH subgiants in past studies or as suspected candidates. An important fraction of the present sample was already included in \cite{North00}, where the authors used radial velocities from CORAVEL and from the high-resolution CES (Coud\'{e} Echelle Spectrometer, installed at the ESO CAT 1.4-m telescope) spectra. We complement their data set with HERMES, CORALIE and SALT data. Several new targets were also added to the sample. The following objects were monitored:
\begin{itemize}
\item 13 confirmed dBa stars from \cite{North94}: BD+18$^\circ$5215, HD\,15306, HD\,48565, HD\,76225, HD\,92545, HD\,106191, HD\,107574, HD\,113402, HD\,147609, HD\,188985, HD\,202400, HD\,221531, and HD\,222349;\\
\item 7 targets from \cite{Houk&Cowley75}. Five of them (HD\,18853, HD\,24864, HD\,26455, HD\,31732, and HD\,69578) were reported to have "strong Sr 4077", as confirmed in \cite{North00}. HD\,9529 presents a small overabundance of yttrium, and BD-18$^\circ$255 was afterwards classified as a CH subgiant by \cite{LuckBond91} and again as a dBa star by \cite{North95};\\
\item 2 dBa stars from the catalogue of \cite{Lu83}: HD\,50264 and HD\,87080. They have been afterwards classified as CH subgiants by \cite{PereiraJunqueira03};\\
\item 3 gBa stars from the catalogue of \cite{Lu83} which seem to be either main sequence or subgiant stars in our HRD (see Sect. \ref{HRD}): HD\,22589, HD\,120620, and HD\,216219. Their orbital elements were already determined by \cite{Udry98} and \cite{Udry98II}. However, since we are now able to add new radial velocity data points, we refit their orbits and include them as barium dwarfs in our sample;\\
\item 1 dBa star from \cite{Tomkin89}: HD\,2454;\\
\item 1 dBa star from \cite{Houk78}: HD\,109490;\\
\item 2 dBa stars identify by \cite{Gray11}: HD\,34654 and HD\,114520;\\
\item 9 mild Ba dwarfs from \cite{Edvardsson93} with marginal overabundance of s-process elements with respect to iron, typically of the order of [s/Fe] $\approx$ 0.2: HD\,6434, HD\,13555, HD\,35296, HD\,60532, HD\,82328, HD\,95241, HD\,98991, HD\,124850, HD\,220117;\\
\item 10 objects listed as suspected Ba dwarfs in Table 2 of \cite{Lu91}: HD\,101581, HD\,103840, HD\,104342, HD\,105671, HD\,117288, HD\,146800, HD\,170149, HD\,177996, HD\,205156, HD\,219899. One of them, HD\,177996, has been reported as an active pre-main-sequence star by \cite{Henry96} and as a double-lined spectroscopic binary (SB2) by \cite{Soderblom98}. We confirm the latter classification from the SALT-HRS data;\\
\item 11 CH subgiants from \cite{LuckBond91}: BD-10$^\circ$4311, BD-11$^\circ$3853, CD-62$^\circ$1346, HD\,89948, HD\,123585, HD\,127392, HD\,141804, HD\,150862, HD\,182274, HD\,207585, and HD\,224621;\\
\item One CH subgiant from \cite{McClureWoodsworth90}: HD\,130255.
\end{itemize}

To our knowledge, with a sample of 60 objects, this is the largest systematic radial-velocity survey of dwarf and subgiant Ba and CH stars. Table \ref{info} collects information about the stars and about the available observations.


\section{Orbital analysis}\label{sec:orbits}

The orbital parameters, i.e., the period ($P$), the eccentricity ($e$), the time of periastron passage ($T\rm_0$), the longitude of periastron ($\omega$), the velocity semi-amplitude ($K_1$) and the systemic velocity of the system ($\gamma$), were determined for each target by fitting a Keplerian orbit to its radial-velocity curve (see Table \ref{table_dBa}). We obtained an initial model by applying a minimisation method using the dominant frequency found in the periodogram of the radial-velocity data. We then computed a second model by means of an iterative non-linear least-square minimisation procedure, which uses as convergence criterion the comparison of consecutive variance reductions. Additionally, an extra model was fitted to the data for which we imposed a circular orbit. For each target, we applied the Lucy\,\&\,Sweeney test \citep{Lucy&Sweeney71} to avoid the determination of spurious eccentricities for truly circular orbits. This test compares the sum of the squared residuals of the best-fitting eccentric orbit (indicated with the subscript \textit{ecc}) with those of a circular fit (indicated with the subscript \textit{circ}) through the determination of the parameter $p$ as indicated in Eq. \ref{eq:L&Sp}:

\begin{equation}\label{eq:L&Sp}
p = \left( \frac{\Sigma (o-c)^2_{\text{ecc}}}{\Sigma (o-c)^2_{\text{circ}}} \right)^{(n-m)/2}
\end{equation}

\noindent where $n$ is the  total  number of  observations, and $m$\,=\,6 is the number of free parameters in an eccentric fit. \cite{Lucy&Sweeney71} argued that only when $p$\,<\,0.05 can the orbit be considered significantly eccentric. 

Once the mentioned parameters were determined from the radial-velocity fitting, we could also compute the mass function, $f(m)$, and the apparent orbital separation, $a_1\sin{i}$, which depend on $P$, $e$ and $K_1$. The results are also included in Table \ref{table_dBa}. 

We used a Monte Carlo method to determine the uncertainties of the orbital parameters. We generated 1000 RV curves for each fitted binary by randomly sampling new RV data normally distributed around the best-fitting model and within the standard deviation of the residuals of the fit. The 1$\sigma$-uncertainties were chosen as the standard deviation of each parameter after fitting these 1000 RV curves. Note that circular orbits have a one-sided uncertainty and do not have an argument of periastron. The time of periastron passage is replaced by the epoch of maximum velocity.

Finally, due to the better quality of the HERMES data (see the error bars of the two data sets in Fig. \ref{BD-18_255} as an example), whenever an orbit could be fully covered with HERMES radial velocities, we decided to determine the orbital parameters using only HERMES data. Then data points from other instruments were overplotted to confirm the fit. This was the case for BD-18$^\circ$255, HD\,107574, HD\,147609 and HD\,221531 and it was tested that the orbital parameters remain the same within errors when the other datasets were included. This can also be seen as a confirmation of the offset applied to CORAVEL data and of the potential of including SALT HRS to determine the orbital parameters of the long and uncovered southern Ba star systems. 

\section{Results}\label{sec:results}
Our analysis confirmed binarity for 40 out of our 60 objects and we constrained 27 spectroscopic orbits. Table \ref{table_dBa} lists the orbital elements derived in this work. Figures with the best-fitting models and the residuals of the fit are included in Appendix \ref{Sect:AppendixdBa}.

\setlength{\tabcolsep}{3pt}
\renewcommand{\arraystretch}{1.3}
\begin{sidewaystable*}
\caption{Orbital parameters (i.e., period, $P$; eccentricity, $e$; time of periastron passage, $T\rm_0$; longitude of periastron, $\omega$; velocity semi-amplitude, $K_1$; and the systemic velocity of the system, $\gamma$) derived in this study. The mass function $f(m)$ and the apparent orbital separation $a_1\sin{i}$, which depend on $P$, $e$ and $K_1$ are also included. The last column shows the standard deviation of the residuals of the fit. In the upper part of the table, we list the 25 dwarf and subgiant Ba and CH stars that are in SB1 systems. In the bottom part, we include the orbital parameters of the inner and outer orbits of the confirmed triple system, HD\,48565 (see Sect. \ref{ssec:HD48565}), and the parameters of the constrained Ba SB2 system, HD\,114520 (see Sect. \ref{ssec:HD114520}).}\label{table_dBa}
\centering
\begin{tabular}{lcccccccccc}
\hline\hline
\rule[0mm]{0mm}{4mm}
Name & Type & Period [d] & e & T$_0$ [HJD] & $\omega$ [$^{\circ}$] & $K_{\rm{1}}$ [$\mathrm{km\,s}^{-1}$] & $\gamma$ [$\mathrm{km\,s}^{-1}$] & f(m) [M$_{\sun}$] & $a_1\sin{i}$ [AU] & $\sigma$(O-C) [$\mathrm{km\,s}^{-1}$]\\
\hline
BD-10$^\circ$4311 & sgCH & 4865 $\pm$ 20 & 0.075 $\pm$ 0.019 & 2451183 $\pm$ 2500 & 157 $\pm$ 17 & 5.29 $\pm$ 0.11 & 50.95 $\pm$ 0.08 & 0.074 $\pm$ 0.005 & 2.36 $\pm$ 0.05 & 0.48\\
BD-18$^\circ$255 & dBa & 1585 $\pm$ 6 & 0.075 $\pm$ 0.011 & 2458136 $\pm$ 730 & 124 $\pm$ 8 & 7.25 $\pm$ 0.07 & 34.65 $\pm$ 0.05 & 0.0620 $\pm$ 0.0017 & 1.053 $\pm$ 0.010 & 0.28\\
BD+18$^\circ$5215 & dBa & 1516 $\pm$ 11 & 0.13 $\pm$ 0.06 & 2447317 $\pm$ 800 & 72 $\pm$ 28 & 3.09 $\pm$ 0.16 & -30.19 $\pm$ 0.10 & 0.0045 $\pm$ 0.0007 & 0.43 $\pm$ 0.02 & 0.53\\
HD\,15306 & dBa & 8383 $\pm$ 48 & 0.20 $\pm$ 0.03 & 2448335 $\pm$ 2400 & 140 $\pm$ 16 & 3.67 $\pm$ 0.14 & 46.82 $\pm$ 0.14 & 0.040 $\pm$ 0.005 & 2.77 $\pm$ 0.11 & 0.46\\
HD\,22589 & gBa & 5573 $\pm$ 33 & 0.232 $\pm$ 0.016 & 2448567 $\pm$ 2800 & 333 $\pm$ 4 & 1.90 $\pm$ 0.03 & -27.51 $\pm$ 0.03 & 0.00367 $\pm$ 0.00018 & 0.949 $\pm$ 0.017 & 0.13\\
HD\,24864 & dBa & 430 $\pm$ 3 & 0.0 + 0.05 & 2449475 $\pm$ 600 & - $\pm$ - & 7.6 $\pm$ 0.8 & 13.1 $\pm$ 0.3 & 0.019 $\pm$ 0.005 & 0.30 $\pm$ 0.03 & 0.98\\
HD\,34654 & dBa & 975.6 $\pm$ 0.6 & 0.110 $\pm$ 0.007 & 2457161 $\pm$ 130 & 326 $\pm$ 2 & 8.90 $\pm$ 0.04 & -6.89 $\pm$ 0.02 & 0.0699 $\pm$ 0.0011 & 0.793 $\pm$ 0.004 & 0.05\\
HD\,50264 & dBa & 909.9 $\pm$ 1.2 & 0.091 $\pm$ 0.03 & 2449541 $\pm$ 60 & 230 $\pm$ 15 & 9.52 $\pm$ 0.14 & 63.27 $\pm$ 0.19 & 0.080 $\pm$ 0.003 & 0.793 $\pm$ 0.012 & 0.35\\
HD\,76225 & dBa & 2410 $\pm$ 2 & 0.098 $\pm$ 0.005 & 2451159 $\pm$ 400 & 267 $\pm$ 3 & 6.11 $\pm$ 0.04 & 30.34 $\pm$ 0.02 & 0.0561 $\pm$ 0.0010 & 1.347 $\pm$ 0.008 & 0.09\\
HD\,87080 & sgCH & 274.3 $\pm$ 1.9 & 0.161 $\pm$ 0.007 & 2448921 $\pm$ 20 & 128 $\pm$ 2 & 12.17 $\pm$ 0.07 & -2.94 $\pm$ 0.05 & 0.0492 $\pm$ 0.0008 & 0.3029 $\pm$ 0.0017 & 0.14\\
HD\,89948 & sgCH & 673 $\pm$ 3 & 0.12 $\pm$ 0.03 & 2448225 $\pm$ 300 & 135 $\pm$ 7 & 9.9 $\pm$ 0.2 & 14.91 $\pm$ 0.14 & 0.065 $\pm$ 0.005 & 0.606 $\pm$ 0.016 & 0.32\\
HD\,95241 & dBa & 5244 $\pm$ 56 & 0.82 $\pm$ 0.02 & 2450499 $\pm$ 2000 & 112 $\pm$ 3 & 4.5 $\pm$ 0.2 & -7.98 $\pm$ 0.05 & 0.0091 $\pm$ 0.0013 & 1.23 $\pm$ 0.05 & 0.2\\
HD\,98991 & dBa & 2836 $\pm$ 99 & 0.317 $\pm$ 0.008 & 2455888 $\pm$ 1400 & 25 $\pm$ 20 & 4.70 $\pm$ 0.08 & 14.06 $\pm$ 0.04 & 0.0259 $\pm$ 0.0007 & 1.16 $\pm$ 0.03 & 0.11\\
HD\,106191 & dBa & 1314 $\pm$ 13 & 0.15 $\pm$ 0.10 & 2446527 $\pm$ 1300 & 120 $\pm$ 34 & 7.1 $\pm$ 0.5 & -3.3 $\pm$ 0.2 & 0.048 $\pm$ 0.008 & 0.85 $\pm$ 0.07 & 0.58\\
HD\,107574 & dBa & 1384 $\pm$ 2 & 0.084 $\pm$ 0.004 & 2457910 $\pm$ 160 & 218 $\pm$ 4 & 2.238 $\pm$ 0.009 & -28.579 $\pm$ 0.008 & 0.00159 $\pm$ 0.00002 & 0.284 $\pm$ 0.002 & 0.05\\
HD\,120620 & gBa & 217.2 $\pm$ 1.0 & 0.0 + 0.05 & 2456980 $\pm$ 100 & - $\pm$ - & 14.02 $\pm$ 0.16 & 33.56 $\pm$ 0.13 & 0.062 $\pm$ 0.002 & 0.280 $\pm$ 0.004 & 0.401\\
HD\,123585 & sgCH & 459.4 $\pm$ 1.5 & 0.0 + 0.05 & 2448419 $\pm$ 500 & - $\pm$ - & 11.7 $\pm$ 0.2 & 25.49 $\pm$ 0.17 & 0.077 $\pm$ 0.004 & 0.496 $\pm$ 0.008 & 0.49\\ 
HD\,127392 & sgCH & 1508 $\pm$ 2 & 0.093 $\pm$ 0.010 & 2448739 $\pm$ 700 & 1711 $\pm$ 10 & 8.25 $\pm$ 0.09 & -64.21 $\pm$ 0.07 & 0.087 $\pm$ 0.003 & 1.14 $\pm$ 0.02 & 0.17\\
HD\,141804 & sgCH & 2652 $\pm$ 95 & 0.0 + 0.05 & 2449979 $\pm$ 1500 & - $\pm$ - & 6.28 $\pm$ 0.16 & -49.3 $\pm$ 0.2 & 0.068 $\pm$ 0.007 & 1.531 $\pm$ 0.09 & 0.20\\ 
HD\,147609 & dBa & 1146.2 $\pm$ 1.5 & 0.058 $\pm$ 0.005 & 2455716 $\pm$ 130 & 123 $\pm$ 5 & 2.974 $\pm$ 0.016 & -16.277 $\pm$ 0.009 & 0.00310 $\pm$ 0.00005 & 0.3128 $\pm$ 0.0016 & 0.057\\
HD\,150862 & sgCH & 291 $\pm$ 4 & 0.28 $\pm$ 0.09 & 2448733 $\pm$ 100 & 299 $\pm$ 21 & 12 $\pm$ 2 & -82.1 $\pm$ 0.7 & 0.045 $\pm$ 0.003 & 0.305 $\pm$ 0.009 & 0.15\\
HD\,182274 & sgCH & 8231 $\pm$ 54 & 0.0 + 0.05 & 2451522 $\pm$ 3000 & - $\pm$ - & 2.99 $\pm$ 0.16 & -16.34 $\pm$ 0.19 & 0.0227 $\pm$ 0.005 & 2.26 $\pm$ 0.12 & 0.24\\
HD\,207585 & sgCH & 672 $\pm$ 2 & 0.0 + 0.05 & 2452172 $\pm$ 500 & - $\pm$ - & 10.4 $\pm$ 0.3 & -62.3 $\pm$ 0.2 & 0.078 $\pm$ 0.007 & 0.641 $\pm$ 0.018 & 0.73\\
HD\,216219 & gBa & 4117 $\pm$ 122 & 0.0 + 0.05 & 2456272 $\pm$ 2000 & - $\pm$ - & 2.98 $\pm$ 0.10 & -6.82 $\pm$ 0.10 & 0.0113 $\pm$ 0.0013 & 1.13 $\pm$ 0.05 & 0.383\\
HD\,221531 & dBa & 1399 $\pm$ 3 & 0.163 $\pm$ 0.007 & 2456940 $\pm$ 81 & 184 $\pm$ 4 & 6.44 $\pm$ 0.04 & 2.82 $\pm$ 0.02 & 0.0372 $\pm$ 0.0007 & 0.818 $\pm$ 0.006 & 0.14\\
\hline\hline
\rule[0mm]{0mm}{4mm}
Name & Type & Period [d] & e & T$_0$ [HJD] & $\omega$ [$^{\circ}$] & $K_{\rm{i}}$ [$\mathrm{km\,s}^{-1}$] & $\gamma$ [$\mathrm{km\,s}^{-1}$] & f(m) [M$_{\sun}$] & $a_{\rm{i}}\sin{i}$ [AU] & $\sigma$(O-C) [$\mathrm{km\,s}^{-1}$]\\
\hline
HD\,48565$_{\rm{in}}$ & Trip. & 73.344 $\pm$ 0.013 & 0.220 $\pm$ 0.007 & 2446943 $\pm$ 2 & 214 $\pm$ 3 & 9.17 $\pm$ 0.06 & - & 0.00542 $\pm$ 0.00010 & 0.0603 $\pm$ 0.0004 & 1.41\\
HD\,48565$_{\rm{out}}$ & Trip. & 10\,531 $\pm$ 210 & 0.37 $\pm$ 0.06 & 2464195 $\pm$ 120 & 222 $\pm$ 5 & 3.9 $\pm$ 0.2 & -21.01 $\pm$ 0.12 & 0.053 $\pm$ 0.003 & 3.5 $\pm$ 0.3 & 0.31\\
HD\,114520a & SB2 & 437.94 $\pm$ 0.18 & 0.523 $\pm$ 0.003 & 2456152 $\pm$ 2 & 249.7 $\pm$ 0.5 & 12.41 $\pm$ 0.06 & -9.07 $\pm$ 0.03 & 0.054 $\pm$ 0.002 & 0.426 $\pm$ 0.005 & 0.11\\
HD\,114520b & SB2 & " & " & " & " & -17.02 $\pm$ 0.12 & " & " & -0.138 $\pm$ 0.015 & 0.75\\
\hline\hline
\end{tabular}
\end{sidewaystable*}

\subsection{Single-lined spectroscopic binaries}\label{ssec:elements}
We derived orbital elements for 25 single-line spectroscopic binaries (SB1). Some of these binaries were already discussed by \citet{North00}, but their orbital elements were never published. However, five of them have spectroscopic orbits published and included in the ninth catalogue of spectroscopic binary orbits (SB9; \citealt{SB9}). As mentioned before, the orbits of the three stars flagged as gBa stars in the literature, HD\,22589, HD\,120620, and HD\,216219, were determined by \cite{Udry98} and \cite{Udry98II}. Our new orbital elements are in agreement with theirs within uncertainties. In addition, \cite{Griffin18} determined orbital elements for HD\,34654 from photoelectric radial velocities, and \cite{Willmarth16} determined orbital elements for HD\,95241. Our results are also in good agreement with theirs. 

In the following subsections, we focus on some individual systems with remarkable properties.

\subsection{HD~48565}\label{ssec:HD48565}

\begin{figure}[t]
\centering
\includegraphics[width=0.49\textwidth]{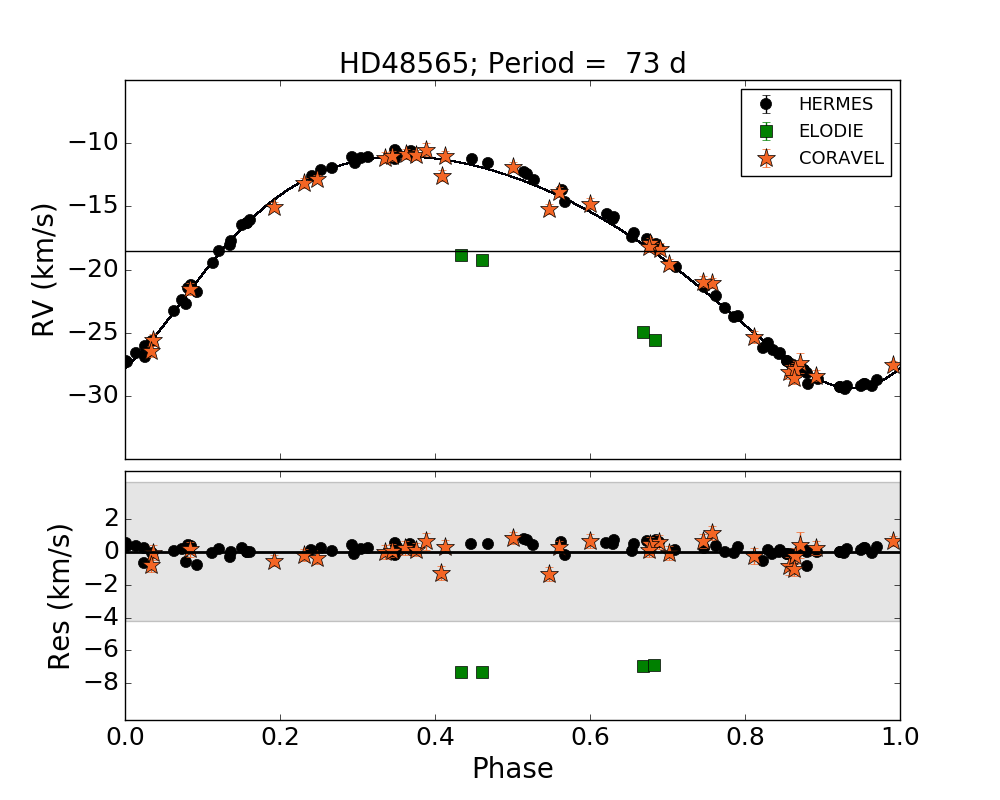}
\caption{\label{HD48565p} Phase-folded HERMES and CORAVEL radial velocities and best-fitting orbital model for the short orbit of the triple system HD\,48565. The lower panel shows the observed minus calculated (O-C) residuals of the fit and a shadowed region which corresponds to three times the standard deviation of the residuals.}
\end{figure}

\begin{figure}[t]
\centering
\includegraphics[width=0.49\textwidth]{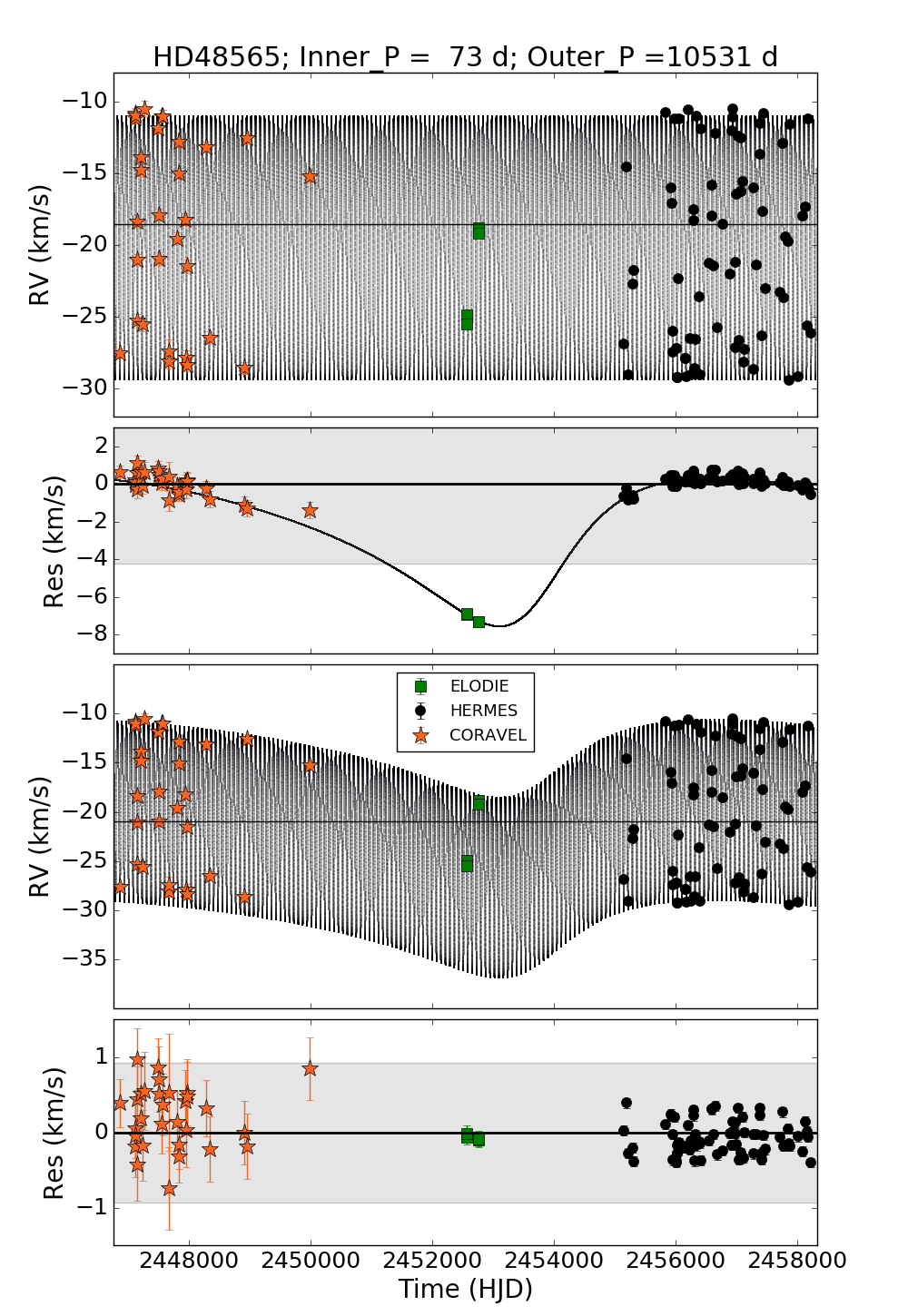}
\caption{\label{HD48565t} Same as Fig. \ref{HD48565p} but as a function of time. The best fitting orbital model to the residuals is included in the second panel. The combined fit of both orbits and its residuals are shown in the third and fourth panels, respectively.}
\end{figure}

HD\,48565 is an SB1 and the dominant signal of its radial-velocity curve can be best fit by a Keplerian orbit with a period of 73 days and an eccentricity of 0.22 (see upper panel of Figs. \ref{HD48565p} and \ref{HD48565t}). This period was already constrained by \cite{North00}, who also noticed a trend on the residuals and suggested the possible presence of a third star in the system, in a much longer orbit. 

Combining our HERMES observations with CORAVEL and ELODIE data, we confirm the presence of a trend in the residuals (see second panel of Fig. \ref{HD48565t}) and estimate the orbital elements of the outer orbit. We assume that the secondary orbit is only a small perturbation of the primary one so we can fit them independently. The best-fitting model of the residuals, with a period of more than 10\,500 days and an eccentricity of almost 0.4 is presented in the second panel of Fig. \ref{HD48565t}. The third panel of this figure shows the modelled radial-velocity curve, which results from combining the two Keplerian orbits, and the fourth panel shows the final residuals. To plot the third and fourth panels, we have accounted for light-time effects (LTE) caused by the motion of the inner pair around the center of mass of the triple. We followed Eq. 1 from \cite{TorresStefanik00}. The orbital parameters of the inner and outer orbits are included in the bottom part of Table \ref{table_dBa}. The systemic velocity of the system was obtained by fixing $\gamma_{\rm in}$ to zero. Given the ratio of the periods, it is likely that the triple system is dynamically stable (\citealt{Tokovinin14}, \citealt{Toonen16}).

\cite{North00} suggested that the inner orbit is too small to have hosted an AGB star, so that the WD must be in the outer orbit while the other two components are less evolved. However, the complexity of the evolution of hierarchical triple systems and the amount of dynamical processes that could have affected HD\,48565 make it difficult to describe the history of the system. For example, from Eq. 24 of \citealt{Toonen16}, we estimate that Kozai-Lidov cycles in the system are of the order of several thousand years, which are short enough to have an important effect on the evolution of the system. Note that for small relative inclinations, the amplitude of these cycles is zero, so they do not have to exist, but we do not have enough information to know if they do or not. Additionally, the derived mass functions are compatible with the WD being in both orbits. Since we cannot confidently say in which orbit the WD is, we are cautious about including HD\,48565 in certain parts of our further analysis.

\subsection{HD~114520}\label{ssec:HD114520}

\begin{figure}[t]
\centering
\includegraphics[width=0.49\textwidth]{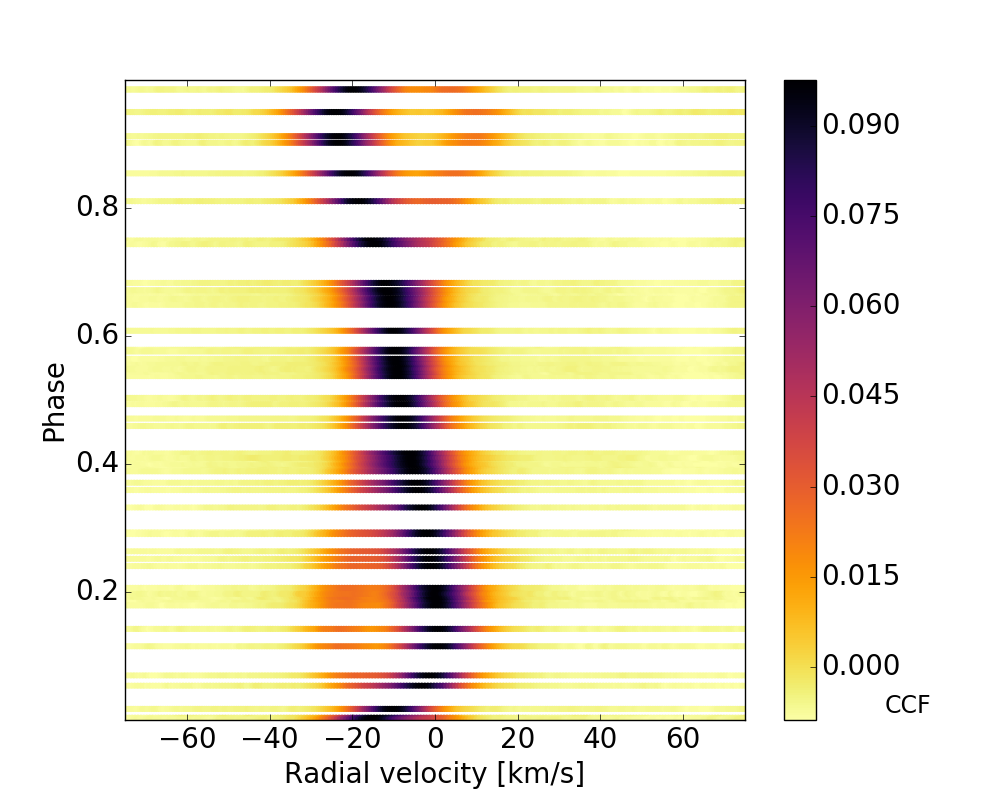}
\caption{\label{HD114520_CCF} Intensity map of the CCF profiles of HD\,114520 phase-folded using the orbital period of the SB2 ($P$ = 437.8 days).}
\end{figure}

\begin{figure}[t]
\centering
\includegraphics[width=0.49\textwidth]{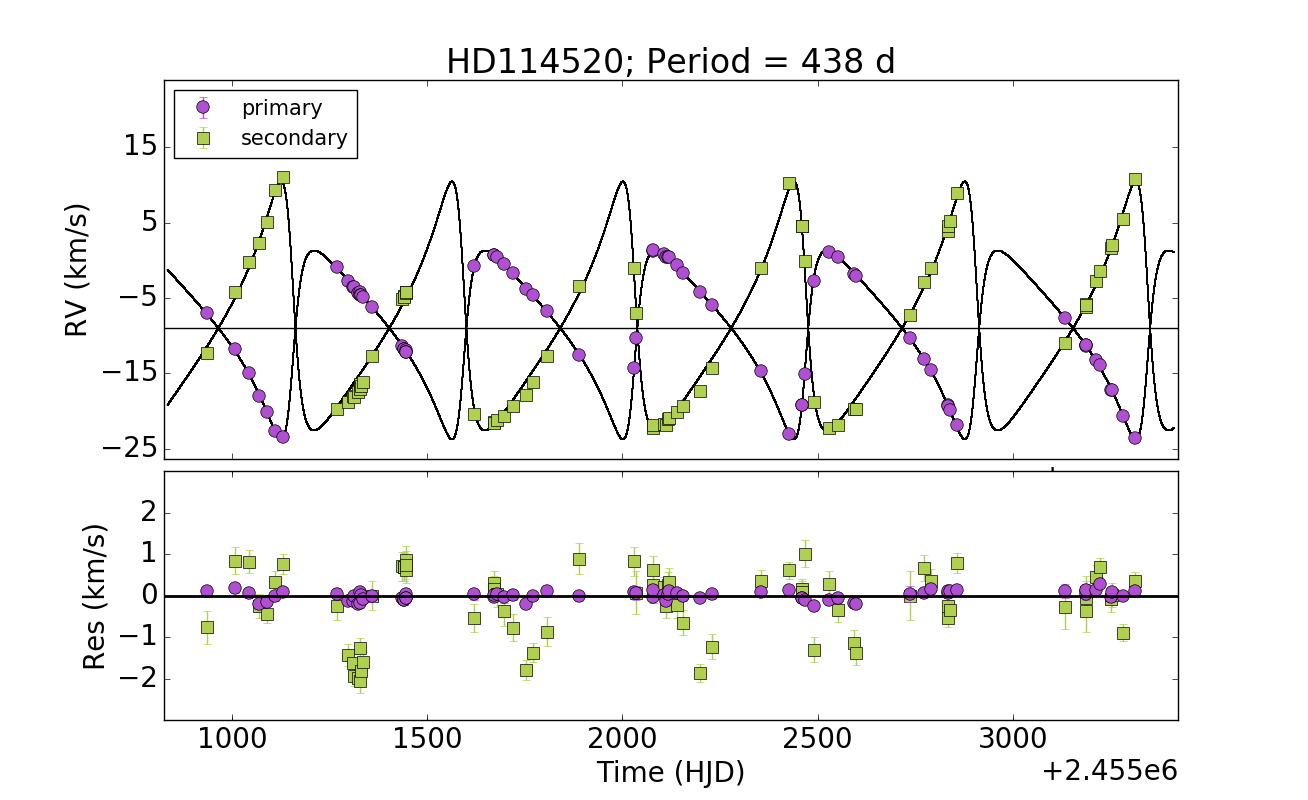}
\caption{\label{HD114520_orbit} Radial-velocity curves and orbital fit of the components of the SB2 in HD\,114520. The lower panel shows the O-C residuals of the fit of both components.}
\end{figure}

Although its binarity and the strength of its \ion{Sr}{II} line at 4077\,\text{\AA} had been reported before, HD\,114520 was first classified as a Ba dwarf by \cite{Gray11}. From high-resolution spectra, they reported an overabundance of some s-process elements, and noticed that the system was an SB2. They also detected an excess in the UV, that they associated with a WD companion. This makes HD\,114520 a triple system as well.

We have 81 high-resolution HERMES spectra of HD\,114520 obtained over six years, with which we cover several orbital cycles. In Fig. \ref{HD114520_CCF}, we show an intensity map with all the available CCF profiles phase-folded with the dominant period found in the data ($\sim$\,438\,days). At given phases, the CCF clearly shows a double peak confirming the SB2 nature of the system. The good orbital coverage and the high eccentricity of the system are also clear from this figure.

We obtained the radial-velocity data of the two components by fitting a double-Gaussian to the CCF profile of each HERMES spectrum. To obtain accurate orbital parameters, we excluded a few data points obtained at phases of zero or almost zero relative radial-velocity between the two stars. Disentangling the two contributions to the CCF at these phases was not easy and the poor quality of those radial-velocity points strongly affected the orbital fit. The best-fit solution to the two components is shown in Fig. \ref{HD114520_orbit} and the derived orbital parameters are included in Table \ref{table_dBa}. We can also calculate the mass ratio between the two stars as $q = m_2 / m_1 = K_1 / K_2 = 0.73$. The secondary is probably another main-sequence star of a cooler spectral-type.

The derived orbital period must correspond to the inner orbit, since \cite{Tokovinin06} and \cite{Tokovinin14}, among others, have shown the absence of outer periods shorter than $\sim$\,1000\,days in triple systems with F- or lower-type dwarfs. This means that the WD companion is in an outer and longer orbit. According to the stability criterion discussed by \cite{Tokovinin14}, the ratio between the long period and the short period must be higher than 4.7, so the outer orbit of HD\,114520 must have a period of at least 2058 days, although it could be much longer.

The residuals of the fit do not show any long trend that can help us confirm the presence of a third body, but notice that our data covers a period only slightly longer than the minimum stable outer period of this system. This contrasts with the more than 31 years of time coverage that we have for HD\,48565, thanks to the CORAVEL and HERMES monitoring effort. Additionally, the residuals are in phase with the orbit, which could hide the long-period and low-amplitude signal of a third body. Since we are not able to detect the orbit of the WD companion, we do not include this object in the subsequent analysis.

\subsection{HD~26455 and HD~177996}

\begin{figure}[t]
\centering
\includegraphics[width=0.24\textwidth]{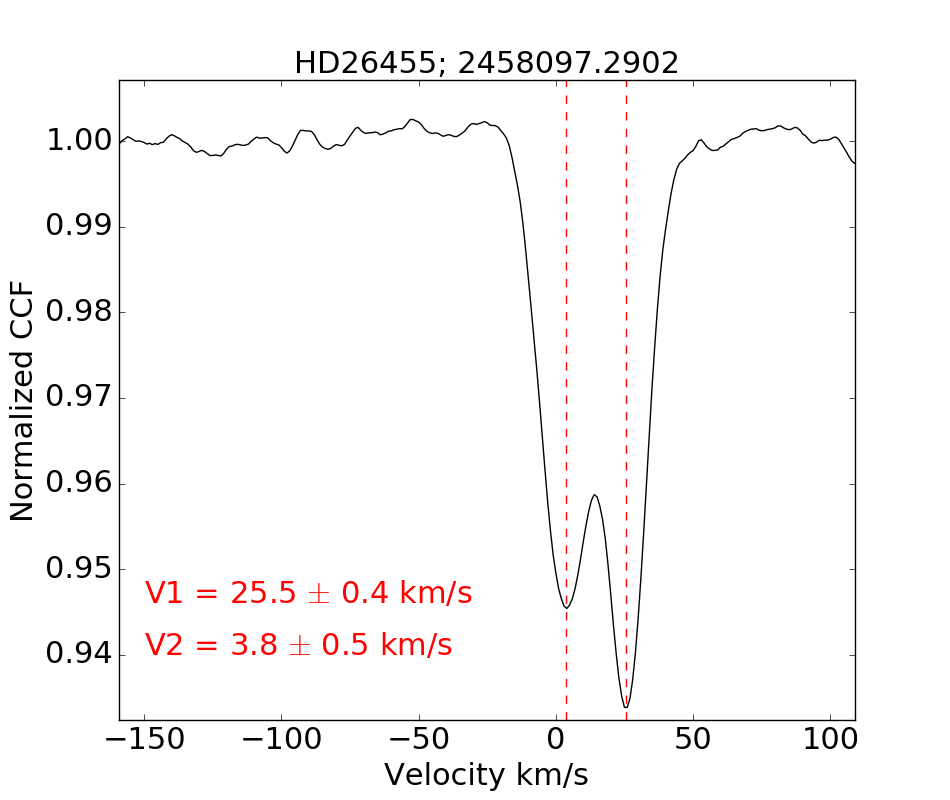}
\includegraphics[width=0.24\textwidth]{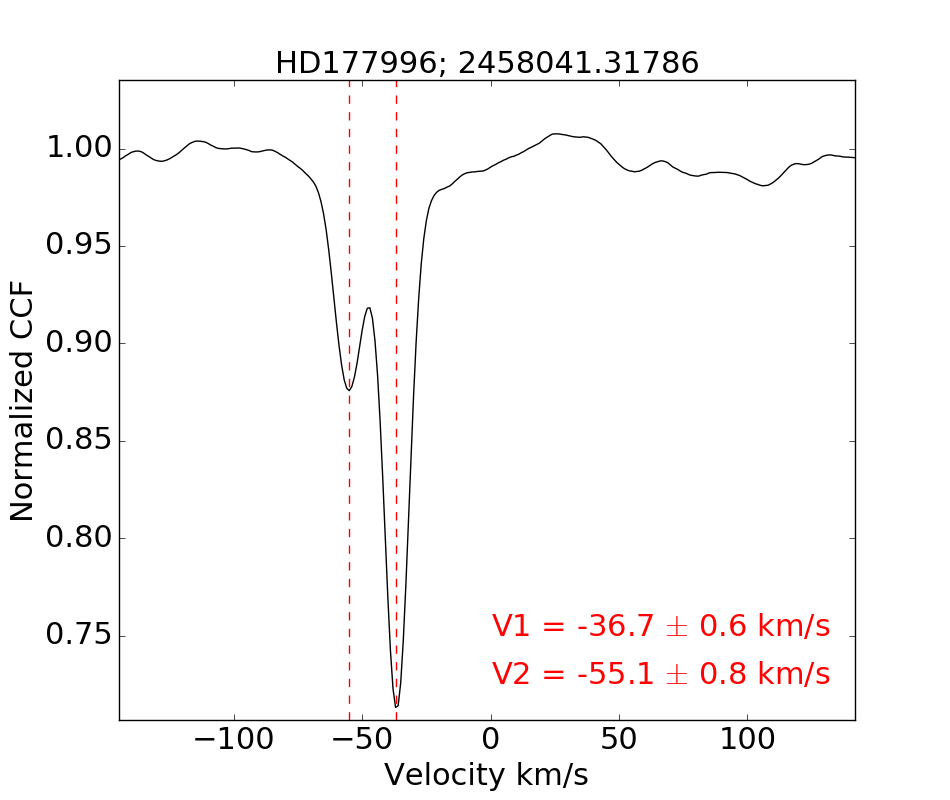}
\caption{\label{ccfs} Double-peaked CCF of two SALT-HRS spectra of HD\,26455 (left) and HD\,177996 (right).}
\end{figure}

These two systems show a double-peaked CCF (see Fig. \ref{ccfs}) in at least one of the available SALT spectra. The Ba star must be an SB2 like HD\,114520, but we do not have enough data yet to solve for the different components. For both systems, there is no indication that the second component of the SB2 spectra is the expected WD companion. This suggests that both of these systems must be triple systems as well.

\subsection{Binaries with incomplete orbital phase coverage and unconfirmed binaries}\label{ssec:others}

\begin{table}[t] 
\begin{small}
\caption{Spectroscopic binaries with no orbits yet available (marked in bold face and as "SB" in the last column) and objects with no variability detected. The second column (std(RV)) gives the standard deviation of the RV points and the third column (3$\sigma$(RV)) gives three times the average error bar considering all the RV points. We claim that a star is in a binary system when the number of the second column is higher than the one of the third.}
\label{uncovered}
\begin{center}
\begin{tabular}{lccc}
\hline\hline
\rule[0mm]{0mm}{3mm}
 & std(RV) & 3$\sigma$(RV) & \\
\rule[0mm]{0mm}{3mm}
Name & [$\mathrm{km\,s}^{-1}$] & [$\mathrm{km\,s}^{-1}$] & Comments\\
\hline
\textbf{BD-11$^\circ$3853} & 2.37 & 0.37 & SB\\ 
\textbf{CD-62$^\circ$1346} & 2.03 & 1.70 & SB\\
\textbf{HD\,2454$^{(1)}$} & 0.5863 & 0.1917 & SB\\
HD\,6434 & 0.44 & 0.95 & \\
HD\,9529 & 0.59 & 1.52 & \\
HD\,13555$^{(2)}$ & 0.14 & 0.33 & \\
\textbf{HD\,18853} & 2.43 & 1.03 & SB\\
\textbf{HD\,26455} & 5.15 & 1.63 & SB2\\
HD\,31732 & 0.30 & 1.23 & \\
HD\,35296 & 0.24 & 0.59 & \\
HD\,60532 & 0.15 & 0.31 & \\
\textbf{HD\,69578} & 17.99 & 1.65 & SB\\
HD\,82328 & 0.13 & 0.25 & \\
\textbf{HD\,92545} & 0.50 & 0.42 & SB\\
HD\,101581 & 0.10 & 0.88 & \\
HD\,103840 & 0.52 & 0.97 & \\
HD\,104342 & 0.58 & 1.17 & \\
HD\,105671 & 0.31 & 0.94 & \\
\textbf{HD\,109490} & 14.67 & 7.54 & SB\\
HD\,113402 & 1.06 & 2.91 & \\
HD\,117288 & 0.50 & 1.08 & \\
HD\,124850 & 0.42 & 0.51 & \\
HD\,130255 & 0.21 & 0.43 & \\
HD\,146800 & 0.21 & 0.99 & \\
HD\,170149 & 0.42 & 1.47 & \\
\textbf{HD\,177996} & 9.53 & 1.18 & SB2\\
\textbf{HD\,188985} & 1.03 & 0.69 & SB\\
HD\,202400 & 4.03 & 4.21 & \\
\textbf{HD\,205156} & 1.40 & 1.18 & SB\\
HD\,219899 & 0.36 & 0.96 & \\
HD\,220117 & 0.17 & 0.47 & \\
\textbf{HD\,222349} & 3.26 & 1.35 & SB\\
\textbf{HD\,224621} & 6.32 & 0.98 & SB\\
\hline\hline
\end{tabular}
\end{center}
\end{small}
\tablefoot{$^{(1)}$ std(RV) and 3$\sigma$(RV) have been computed using only HERMES data. Binarity is clear from this data set while the CORAVEL RV points are very scattered (see third panel of Fig. \ref{Fig:dBauncovered}). 
$^{(2)}$ std(RV) and 3$\sigma$(RV) have been computed excluding the second CORAVEL RV point, which we considered must be an outlier (see third panel of Fig. \ref{Fig:dBaothers}).}
\end{table}

\setlength{\tabcolsep}{5.5pt}
\renewcommand{\arraystretch}{1.2}
\begin{table*}[t] 
\begin{small}
\caption{Effective temperature ($T_{\rm eff}$), surface gravity (log$g$), metallicity ([Fe/H]), and microturbulence velocity ($\xi$) of the objects with a determined orbit. Column 7 gives the reference of the adopted stellar parameters when no HERMES spectra were available. Column 8 gives the line-of-sight extinction, and columns 9 and 10 list the derived luminosities and masses.
\textbf{References for the spectroscopic stellar parameters:} L\&B91: \citet{LuckBond91}; K+13: \citet{Kordopatis13}; K\&G14: \citet{Karinkuzhi14}; P\&J03: \citet{PereiraJunqueira03}; N+94: \citet{North94}.}
\label{params}
\begin{center}
\begin{tabular}{lccccccccc}
\hline\hline
\rule[0mm]{0mm}{4mm}
Name & Type & $T_{\rm eff}$ [K] & log$g$ [dex] & [Fe/H] [dex] & $\xi$ [$\mathrm{km\,s}^{-1}$] & Ref. & E(B-V) & L/L$_{\sun}$ & M/M$_{\sun}$\\
\hline
BD-10$^\circ$4311 & sgCH & 5858 $\pm$ 50 & 3.9 $\pm$ 0.3 & -0.65 $\pm$ 0.10 & 1.25 $\pm$ 0.07 & This work & 0.042 $\pm$ 0.015 & 1.99 $\pm$ 0.05 & 0.80 $\pm$ 0.10 \\
BD-18$^\circ$255 & dBa & 6800 $\pm$ 400 & 4.3 $\pm$ 0.3 & -0.18 $\pm$ 0.05 & 2.4 $\pm$ 0.5 & L\&B91 & 0.015 $\pm$ 0.005 & 4.2 $\pm$ 1.2 & 1.30 $\pm$ 0.15\\
BD+18$^\circ$5215 & dBa & 6452 $\pm$ 50 & 4.4 $\pm$ 0.4 & -0.50 $\pm$ 0.09 & 0.82 $\pm$ 0.09 & This work & 0.014 $\pm$ 0.007 & 1.86 $\pm$ 0.11 & 1.10 $\pm$ 0.10\\
HD\,15306 & dBa & 6898 $\pm$ 50 & 3.8 $\pm$ 0.2 & -0.40 $\pm$ 0.14 & 0.9 $\pm$ 0.3 & This work & 0.037 $\pm$ 0.007 & 8.7 $\pm$ 2.3 & 1.30 $\pm$ 0.10 \\
HD\,22589 & gBa & 5672 $\pm$ 56 & 3.9 $\pm$ 0.2 & -0.07 $\pm$ 0.09 & 1.20 $\pm$ 0.04 & This work & 0.048 $\pm$ 0.008 & 6.9 $\pm$ 0.8 & 1.30 $\pm$ 0.10 \\
HD\,24864 & dBa & 6337 $\pm$ 95 & 3.82 $\pm$ 0.18 & 0.0 $\pm$ 0.11 & 1.0 $\pm$ 1.0 & K+13 & 0.0 + 0.005 & 9.2 $\pm$ 0.6 & 1.38 $\pm$ 0.06\\
HD\,34654 & dBa & 6174 $\pm$ 50 & 4.4 $\pm$ 0.3 & -0.09 $\pm$ 0.07 & 1.36 $\pm$ 0.04 & This work & 0.012 $\pm$ 0.008 & 1.8 $\pm$ 0.3 & 1.19 $\pm$ 0.05 \\
HD\,48565 & dBa & 6030 $\pm$ 100 & 3.8 $\pm$ 0.3 & -0.6 $\pm$ 0.3 & 1.3 $\pm$ 0.3 & K\&G14 & 0.018 $\pm$ 0.010 & 2.8 $\pm$ 0.3 & 0.92 $\pm$ 0.05 \\
HD\,50264 & dBa & 5800 $\pm$ 100 & 4.2 $\pm$ 0.2 & -0.34 $\pm$ 0.08 & 1.0 $\pm$ 0.3 & P\&J03 & 0.002 $\pm$ 0.010 & 0.89 $\pm$ 0.08 & 0.90 $\pm$ 0.10\\
HD\,76225 & dBa & 6340 $\pm$ 50 & 3.9 $\pm$ 0.2 & -0.37 $\pm$ 0.08 & 1.30 $\pm$ 0.06 & This work & 0.053 $\pm$ 0.010 & 5.9 $\pm$ 0.7 & 1.21 $\pm$ 0.06 \\
HD\,87080 & sgCH & 5483 $\pm$ 50 & 3.6 $\pm$ 0.3 & -0.60 $\pm$ 0.08 & 1.20 $\pm$ 0.04 & This work & 0.028 $\pm$ 0.012 & 12.5 $\pm$ 0.8 & 1.6 $\pm$ 0.2\\
HD\,89948 & sgCH & 6000 $\pm$ 400 & 4.0 $\pm$ 0.3 & -0.13 $\pm$ 0.05 & 1.8 $\pm$ 0.5 & L\&B91 & 0.0 $\pm$ 0.003 & 1.36 $\pm$ 0.07 & 0.95 $\pm$ 0.08\\
HD\,95241 & dBa & 5837 $\pm$ 50 & 3.6 $\pm$ 0.2 & -0.37 $\pm$ 0.07 & 1.36 $\pm$ 0.04 & This work & 0.0 + 0.008 & 7.8 $\pm$ 0.8 & 1.30 $\pm$ 0.10 \\
HD\,98991 & dBa & 6388 $\pm$ 50 & 2.7 $\pm$ 0.4 & -0.44 $\pm$ 0.08 & 0.07$^{(1)}$ & This work & 0.0 + 0.014 & 14.7 $\pm$ 1.0 & 1.45 $\pm$ 0.08 \\
HD\,106191 & dBa & 5946 $\pm$ 50 & 4.3 $\pm$ 0.4 & -0.29 $\pm$ 0.10 & 1.21 $\pm$ 0.06 & This work & 0.010 $\pm$ 0.010 & 2.0 $\pm$ 0.2 & 0.95 $\pm$ 0.05 \\
HD\,107574 & dBa & 6340 $\pm$ 100 & 3.6 $\pm$ 0.5 & -0.80 $\pm$ 0.25 & 1.8 $\pm$ 0.3 & N+94 & 0.0 + 0.005 & 5.6 $\pm$ 0.2 & 1.11 $\pm$ 0.05 \\
HD\,120620 & gBa & 4831 $\pm$ 50 & 3.0 $\pm$ 0.3 & -0.29 $\pm$ 0.10 & 1.11 $\pm$ 0.05 & This work & 0.055 $\pm$ 0.013 & 12.2 $\pm$ 1.6 & 1.2 $\pm$ 0.2 \\
HD\,123585 & sgCH & 6047 $\pm$ 100 & 3.5 $\pm$ 0.5 & -0.50 $\pm$ 0.25 & 1.8 $\pm$ 0.3 & N+94 & 0.054 $\pm$ 0.012 & 3.5 $\pm$ 1.0 & 1.00 $\pm$ 0.10\\ 
HD\,127392 & sgCH & 5600 $\pm$ 300 & 3.9 $\pm$ 0.3 & -0.52 $\pm$ 0.05 & 2.6 $\pm$ 0.5 & L\&B91 & 0.011 $\pm$ 0.011 & 0.98 $\pm$ 0.13 & 0.8 $\pm$ 0.3\\
HD\,141804 & sgCH & 6000 $\pm$ 400 & 3.5 $\pm$ 0.3 & -0.41 $\pm$ 0.05 & 1.7 $\pm$ 0.5 & L\&B91 & 0.0 + 0.003 & 1.77 $\pm$ 0.07 & 0.9 $\pm$ 0.2 \\ 
HD\,147609 & dBa & 6411 $\pm$ 50 & 3.9 $\pm$ 0.5 & -0.23 $\pm$ 0.09 & 1.26 $\pm$ 0.07 & This work & 0.044 $\pm$ 0.007 & 10.2 $\pm$ 1.3 & 1.40 $\pm$ 0.10\\
HD\,150862 & sgCH & 6182 $\pm$ 50 & 4.1 $\pm$ 0.3 & -0.33 $\pm$ 0.08 & 0.94 $\pm$ 0.06 & This work & 0.006 $\pm$ 0.006 & 2.2 $\pm$ 0.2 & 1.04 $\pm$ 0.05\\
HD\,182274 & sgCH & 6327 $\pm$ 50 & 4.3 $\pm$ 0.2 & -0.32 $\pm$ 0.09 & 1.05 $\pm$ 0.06 & This work & 0.012 $\pm$ 0.007 & 2.06 $\pm$ 0.14 & 1.09 $\pm$ 0.05 \\
HD\,207585 & sgCH & 5400 $\pm$ 300 & 3.3 $\pm$ 0.3 & -0.57 $\pm$ 0.05 & 1.8 $\pm$ 0.5 & L\&B91 & 0.0 + 0.003 & 3.8 $\pm$ 0.8 & 0.96 $\pm$ 0.15\\
HD\,216219 & gBa & 5900 $\pm$ 50 & 3.6 $\pm$ 0.3 & -0.17 $\pm$ 0.08 & 1.42 $\pm$ 0.04 & This work & 0.042 $\pm$ 0.009 & 9.9 $\pm$ 0.8 & 1.45 $\pm$ 0.10\\
HD\,221531 & dBa & 6460 $\pm$ 100 & 4.2 $\pm$ 0.5 & -0.30 $\pm$ 0.25 & 2.4 $\pm$ 0.3 & N+94 & 0.0 + 0.004 & 4.7 $\pm$ 1.4 & 1.20 $\pm$ 0.10\\
\hline\hline
\end{tabular}
\end{center}
\end{small}
\tablefoot{$^{(1)}$ Uncertain value due to the low signal-to-noise ratio of the spectrum around Fe lines.}
\end{table*}

We could not determine orbits for 33 out of 60 objects in our sample. However, binary motion is clearly detected for 13 of them. Some of these were already mentioned as spectroscopic binaries in \cite{North00}, but we include them again in the present paper for the sake of completeness. To decide whether a star is member of a binary system or not, we use the following criterion: when the standard deviation of the observations, std(RV), is higher than three times the average errorbar, $\sigma$(RV), we consider the star to present variability. Table \ref{uncovered} summarises the results. The confirmed binaries are marked in boldface and their radial velocities are shown in Fig. \ref{Fig:dBauncovered} of Appendix \ref{Sect:AppendixdBa}.

Additionally, Fig. \ref{Fig:dBaothers} in Appendix \ref{Sect:AppendixdBa}, shows the radial velocity data of the 20 objects for which we cannot prove binary motion yet. In some cases, this is due to the lack of data (see HD\,105671 in Fig. \ref{Fig:dBaothers} as an example). In others, the radial-velocity changes hint at binary motion, but the measured RV dispersion is still too small for a firm conclusion (e.g. HD\,220117 in Fig. \ref{Fig:dBaothers}).

\section{The Hertzsprung–Russell diagram}\label{sec:hrd}

\begin{figure*}
\centering
\includegraphics[width=0.99\textwidth]{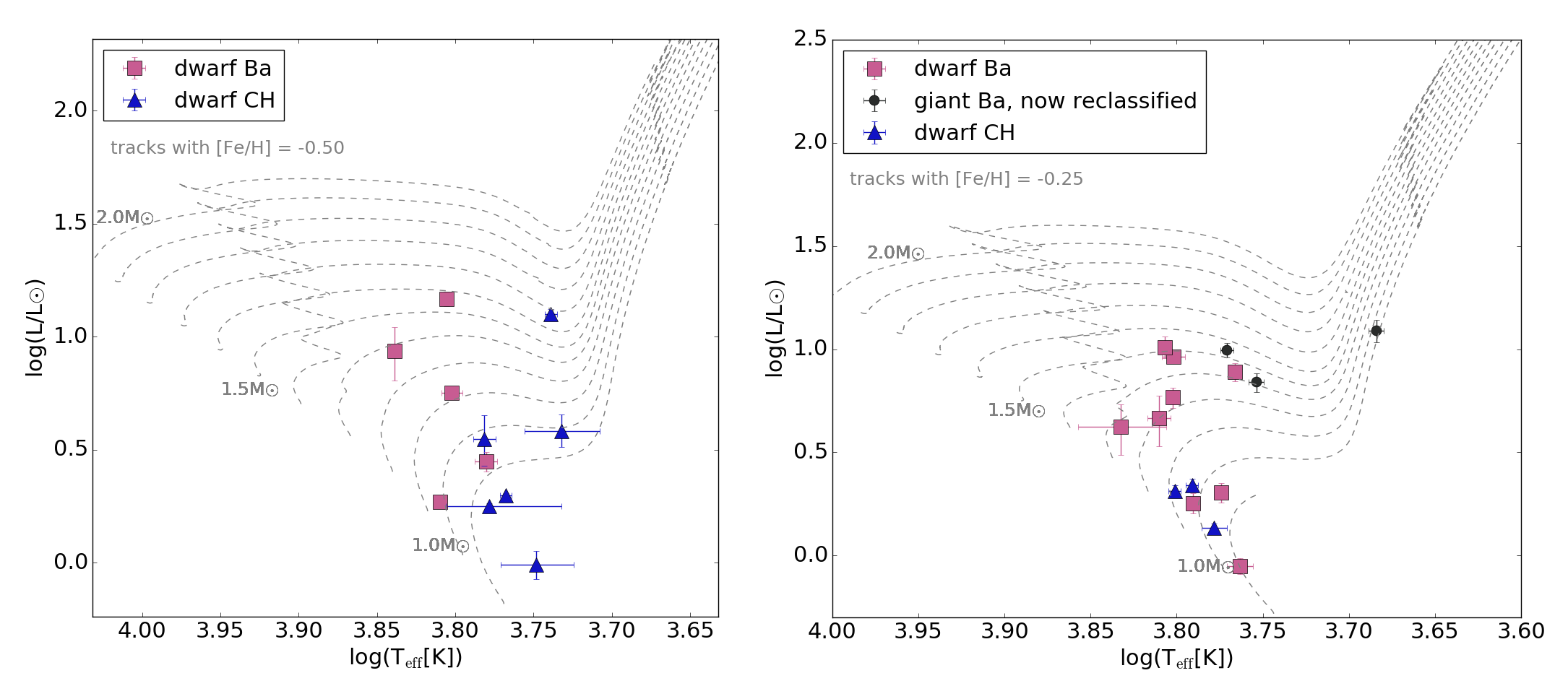}
\caption{\label{HRD} HRD of the dwarf and subgiant Ba (pink squares) and CH (blue triangles) stars for which we could derive orbital parameters in this work. The three stars initially classified as giants are plotted with black circles. We have distributed the objects in two HRDs to plot them with evolutionary tracks of the most appropriate metallicity ($\rm{[Fe/H]}=-0.50$ in the left panel and $\rm{[Fe/H]}=-0.25$ one on the right panel). Evolutionary tracks have been computed with the STAREVOL code (\citealt{Siess00}; \citealt{SiessArnould08}).}
\end{figure*}

We developed a method to obtain atmospheric parameters and luminosities of a sample of more than 400 Ba stars by modelling their spectral energy distributions (SEDs; \citealt{Escorza17}). SEDs are mainly sensitive to the effective temperature, and provide only a limited amount of information about other parameters, especially about metallicity. In this work, we derived the stellar parameters of our main sequence and subgiant stars from HERMES high-resolution spectra whenever available. We used the \textsc{bacchus} pipeline \citep{Masseron16}, which uses Turbospectrum (\citealt{Turbospectrum98}; \citealt{Turbospectrum12}), a 1D local thermodynamical equilibrium (LTE) spectral-synthesis code. An additional check based on the equilibrium of excitation and ionisation iron abundances derived from \ion{Fe}{I} and \ion{Fe}{II} lines was performed to ensure the consistency of the derived atmospheric parameters (see \citealt{Karinkuzhi18} for more details). The \ion{Fe}{I} and \ion{Fe}{II} lines that we used are listed in \cite{Jorissen19}. The derived effective temperature ($T_{\rm eff}$), surface gravity (log$g$), metallicity ([Fe/H]), and microturbulence velocity ($\xi$) are presented in Table \ref{params}. In this manuscript, metallicity is expressed with respect to the solar value as given by \cite{Asplund09}. We did this analysis for the 25 stars with a determined SB1 orbit and for HD\,48565, since there is signature of only one star in its spectra, but we did not include HD\,114520. For targets with no HERMES spectra available, or when the signal-to-noise ratio of the available spectra was not enough to determine accurate stellar parameters, we used spectroscopic parameters from the literature. The relevant references are also included in Table \ref{params}.

Once the stellar parameters were derived, we used the SED-fitting tool described in \cite{Escorza17} to find the best-fitting MARCS model atmosphere \citep{Gustafsson2008} to the photometry of each target and to obtain the total line-of-sight extinction ($E(B-V)$). We used the parameter ranges obtained from the spectroscopic analysis (see Table \ref{params}) to limit $T_{\rm eff}$ and log$g$, and we fixed the metallicity to the closest available in the MARCS grid (0.0, -0.25, -0.5 or -0.75). $E(B-V)$ was the only parameter kept completely free. We integrated the best-fitting SED models over all wavelengths and used distances from \cite{Bailer-Jones18}, based on \textit{Gaia}-DR2 parallaxes \citep{Lindegren18}, to compute accurate luminosities. Note that the \textit{Gaia}-DR2 parallaxes were derived from a single-star astrometric solution and all our stars are binaries. However, our orbital periods are different from one year and we do not expect this to have a big impact. Table \ref{params} includes the derived extinction and luminosities. The luminosity errors come from the distance errors and from the error of the integrated flux. To obtain the latter, we integrated 1000 additional well-fitting SED models and used the standard deviation of these new fluxes as 1$\sigma$-uncertainty on the flux. We chose these 1000 models by combining stellar parameters randomly chosen from normal distributions centred around the spectroscopic values and within the uncertainties.

Figure \ref{HRD} shows the position on the HRD of all the dwarf and subgiant Ba (pink squares) and CH (blue triangles) stars for which we could constrain an orbit. In these plots, we keep the classification found in the literature for each star. However, as mentioned before, the distinction between dBa and sgCH stars is not very clear, and some targets have been classified as both in different studies (see Table \ref{info}). In order to decide in which group to include a star with two classifications in the literature, we used the following criterion: when the star had [Fe/H] lower that -0.5, we classified it as a CH star and when it was higher, as a Ba star. This is a classical criterion, generally used for their giant counterparts since CH stars were defined as metal-deficient Ba stars (see, for example, \citealt{LuckBond91} or \citealt{Vanture92_I, Vanture92_II, Vanture92_III}) with $\rm{[Fe/H]}$ values ranging from $-0.50$ to $-1.7$ \citep{Wallerstein98}. We stress, however, that this classification based on metallicity is is not fully satisfactory (e.g. \citealt{Yamashita75}), and should be revised in the future. For example, HD\,89948 with $\rm{[Fe/H]}=-0.13$ is referred to as a CH star in the literature, while HD\,48565 with $\rm{[Fe/H]}=-0.6$ is defined as a Ba star!

Following this criterion, HD\,87080 appears now among main sequence CH stars, while BD-18$^\circ$255 and HD\,50264 appear among the Ba dwarfs. If only one classification was found in the literature, we kept it independently of the metallicity. The three stars initially classified as gBa (HD\,22589, HD\,120620, and HD\,216219) in the catalogue by \cite{Lu83} are represented in Fig. \ref{HRD} with black circles. Their position in the HRD is more compliant with subgiants as they have not ascended the RGB yet. From now on, we will add these three stars to our dwarf and subgiant Ba star sample. Figure \ref{HRD} is divided in two panels. The left panel shows an HRD with all the stars in our sample with $\rm{[Fe/H]}\lesssim-0.40$ and STAREVOL (\citealt{Siess00}; \citealt{SiessArnould08}) evolutionary tracks computed with $\rm{[Fe/H]}=-0.50$. The HRD on the right panel shows stars with $\rm{[Fe/H]}\gtrsim-0.40$ and STAREVOL evolutionary tracks computed with $\rm{[Fe/H]}=-0.25$.

\begin{figure}
\centering
\includegraphics[width=0.49\textwidth]{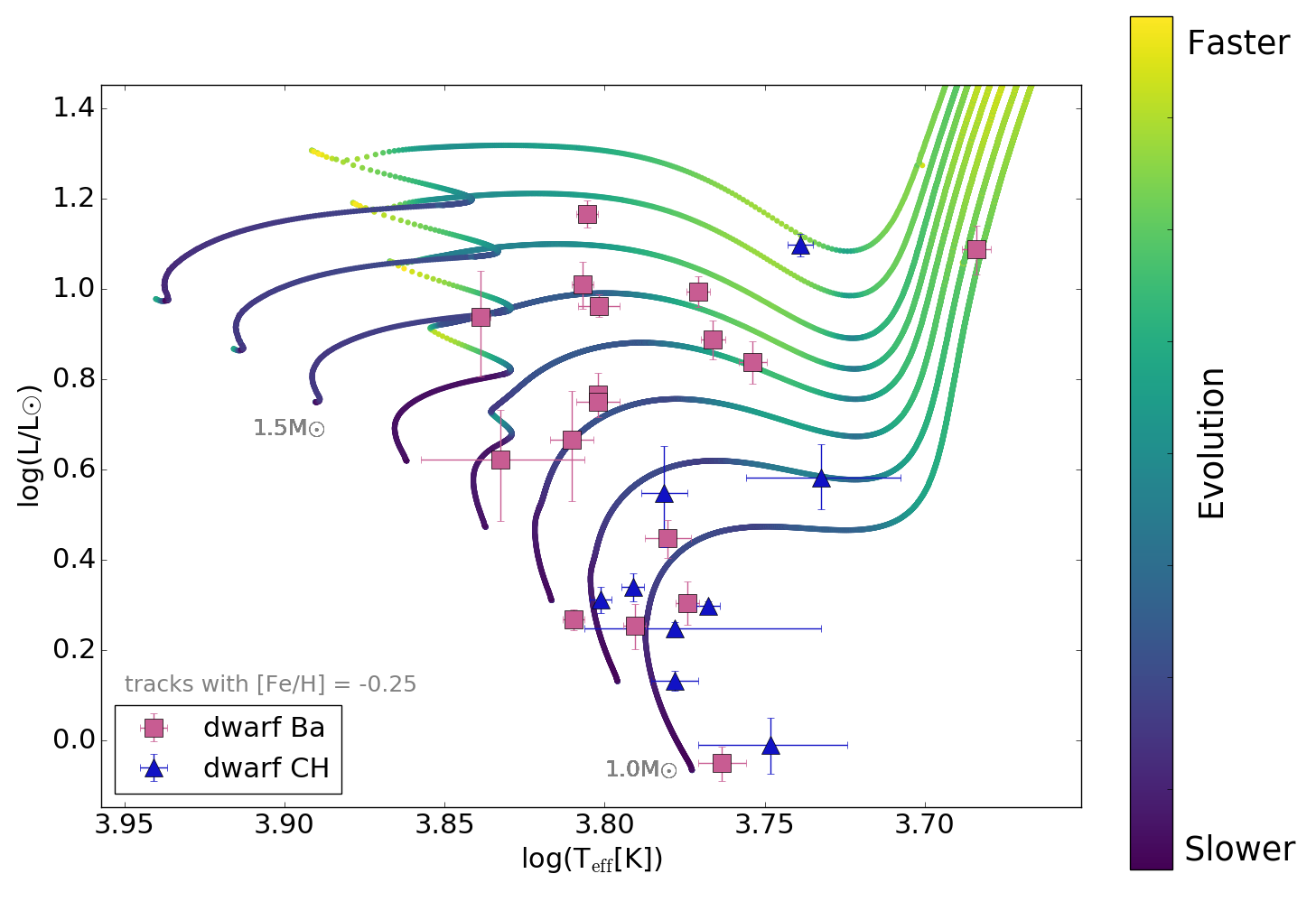}
\caption{\label{Hgap} Same as the right panel of Fig. \ref{HRD} but with stars of all metallicities included. The colour of the tracks is proportional to an evolutionary velocity computed as the distance between two points divided by the age difference between them. Dark blue colours trace much slower evolution than light green ones.}
\end{figure}

A noticeable feature of our HR diagrams is the location of some of our most massive objects with respect to the Hertzsprung gap. A considerable fraction of our objects appear in regions of the HRD, where evolution is fast and not many stars are expected to be observed. To illustrate this more clearly, in Fig. \ref{Hgap}, we have coloured our evolutionary tracks of $\rm{[Fe/H]}=-0.25$ according to their apparent evolutionary speed. We have computed the latter as the distance between two points on a track divided by the age difference between them. Stellar evolution on the dark blue parts of a track is much slower than on the light green parts. Although this is a simple approach, it is good enough to show that several dBa stars occupy rapid phases of evolution.

Finally, to obtain individual masses for our primary stars, we used the same method as in \cite{Escorza17}. We derived the mass by comparing the position of the stars on the HRD with a grid of evolutionary models computed with the STAREVOL code (\citealt{Siess00}; \citealt{SiessArnould08}). The grid covers a wide range of masses, from 0.9\,M$_{\sun}$ to 4.0\,M$_{\sun}$ with a mass step of 0.1\,M$_{\sun}$, and four different metallicities, $\rm{[Fe/H]}=0.0$, $-0.25$, $-0.5$ and $-1.0$. To compute these evolutionary models, we included a mass-dependent formulation for the overshooting on top of the convective core, following \cite{ClaretTorres18}, and considered a wind prescription as described by \cite{SchroderCuntz07} until the beginning of the AGB phase. We also included some overshooting at the base of the convective envelope, following the exponential decay expression of \cite{Herwig97} with $f_{\rm{over}} = 0.1$. Finally, we used a grey atmosphere surface boundary condition. The derived masses are presented in Table \ref{params}. The uncertainty on the mass is obtained by deriving the upper and lower mass limits given the error bars of the spectroscopic parameters.

\begin{figure*}
\centering
\includegraphics[width=0.80\textwidth]{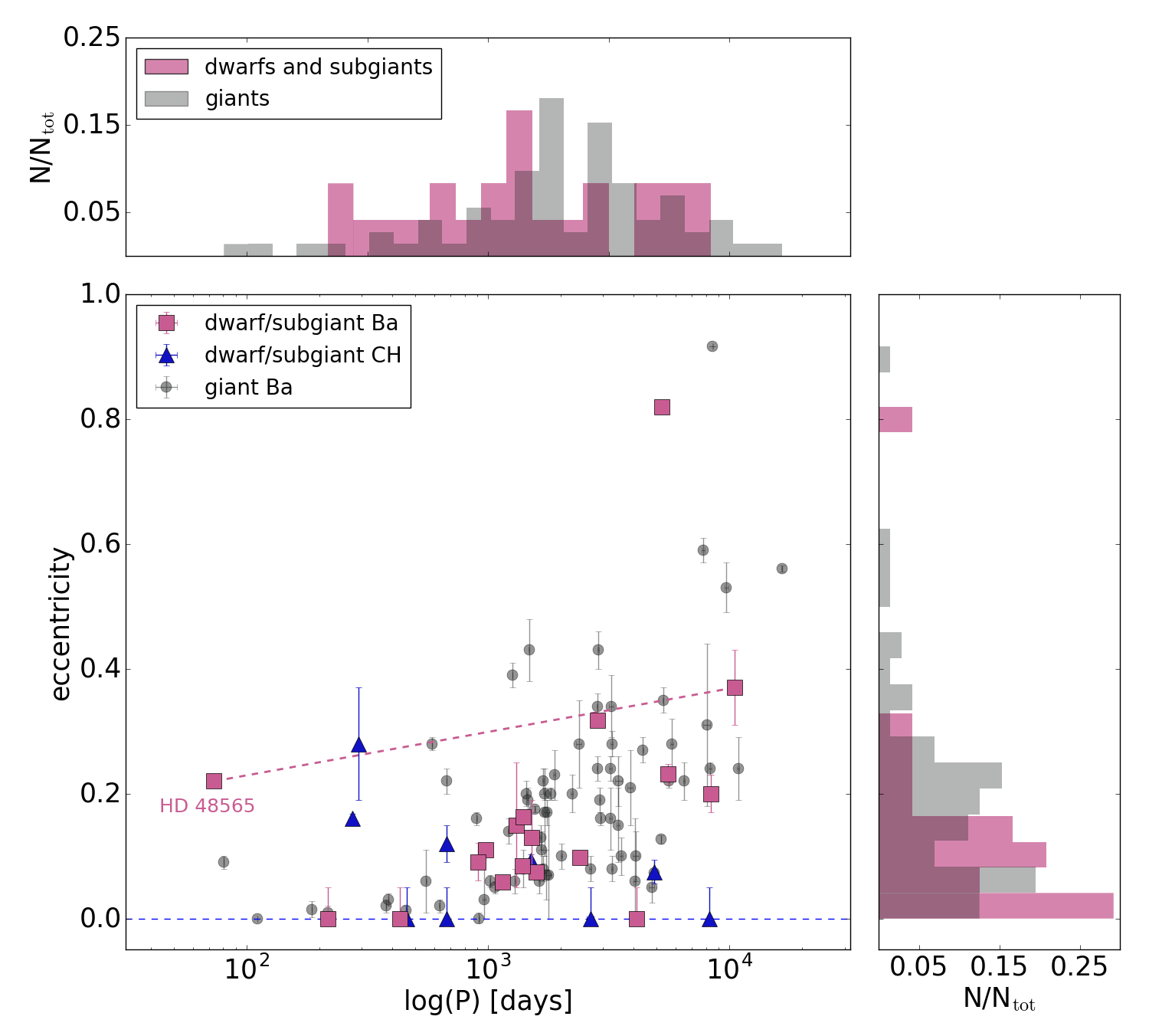}
\caption{\label{e-logP} Eccentricity-period diagram of dwarf and subgiant Ba stars (pink squares), including HD\,22589, HD\,120620, and HD\,216219, initially called giants, and of dwarf and subgiant CH stars (blue triangles). The two orbits of the triple system HD\,48565 are linked with a dashed line. Orbits of giant Ba are included as grey circles for comparison \citep{Jorissen19}. The top and right panels show normalised histograms of $logP$ and $e$, respectively.}
\end{figure*}

\section{Discussion}\label{sec:disc}
\subsection{Eccentricity-period diagram}\label{ssec:e-logP}

Figure \ref{e-logP} presents the eccentricity-period ($e-logP$) diagram of Ba stars at different evolutionary stages and metallicities. The pink squares are the stars flagged as Ba dwarfs and subgiants including HD\,22589, HD\,120620, and HD\,216219 and the blue triangles are the dwarf and subgiant CH stars. The two orbits of the triple system HD\,48565 are linked with a dashed line since we cannot be sure which orbit hosts the WD. The grey circles and crosses represent, respectively, Ba and CH giant orbits from the literature \citep{Jorissen16, Jorissen19} that we include for comparison. The top and on the right panels of Fig.~\ref{e-logP} draws the histograms of the period and eccentricity distributions, respectively. In these diagrams, the giant Ba and CH stars are represented in grey and the dwarf and the subgiant Ba and CH stars are
represented together in pink. Figure 9 shows that:

\begin{itemize}
\item The ranges of observed periods and eccentricities are similar for all type of objects.\\

\item At short periods ($P < 1000$~d), we observe both circular orbits and a few remarkably eccentric ones. Among the new orbits, two CH dwarfs stand out: HD\,87080 ($P = 274.3 \pm 1.9$~d, $e = 0.161 \pm 0.007$) and HD\,150862 ($P = 291 \pm 4$~d, $e = 0.28 \pm 0.09$). We do not have many radial-velocity points for these objects, but the eccentric orbital model fits clearly better, and the HERMES point confirms the CORAVEL eccentric orbit in both cases (see Figs. \ref{HD87080} and \ref{HD150862}).

The inner orbit of HD\,48565 stands out as well among the short period binaries. This is likely a result of dynamical interaction in the triple system, as it was discussed in Sect. \ref{ssec:HD48565}.

\item At longer periods ($P > 1000$~d), all barium giants seem to be eccentric, but among the new systems, we find three with zero eccentricity.

\item Finally, we have giant systems with very long periods ($P > 10 000$~d) and large eccentricities. Only one Ba dwarf (HD\,95241) populates that region of the diagram, but note that among our "uncovered" orbits, we have some confirmed binaries which will have very long periods as well.
\end{itemize}

Overall the $e-logP$ diagram of the dwarf and subgiant stars does not show specific features to distinguish them from their giant counterparts, apart from the presence of a couple of large-eccentricity systems at short periods among the CH dwarfs.

\subsection{dBa vs. sgCH}\label{ssec:location}
Figure \ref{HRD} corroborates the finding of our earlier work \citep{Escorza17}: there is no evolutionary distinction between dwarf Ba stars and subgiant CH stars. Some "dBa" stars are not real dwarfs and many "sgCH" stars are located on the main sequence and are not subgiants. Additionally, we do not see a compelling metallicity difference between the two classes of stars. Those stars which are classified in the literature as dwarf Ba stars have, on average, a higher metallicity than the ones classified as subgiants CH stars, but we find outliers among both subsamples. 

These two classes of stars look very similar in many aspects, but there is a noticeable difference between their location on the HRD. The dwarf CH stars are located on the bottom right part of the main sequence, so they have a lower temperature and are less massive than the dwarf Ba stars. CH stars are characterised by strong G bands due to the CH molecule \citep{Keenan42} that might not be detectable anymore in the spectra of hotter main-sequence stars. This might be causing a bias in the classification of s-process-enhanced main-sequence stars making CH dwarfs appear as the low-mass analogues of Ba dwarfs. These families of stars would benefit from a new and systematic spectroscopic classification, but this is beyond the scope of this study. We do not see this effect on the giant counterparts, Ba and CH giants occupy the same region on the HRD (see Fig. 7 of \citealt{Escorza17}).

\subsection{Mass distribution}\label{ssec:massd}
\begin{figure}
\centering
\includegraphics[width=0.49\textwidth]{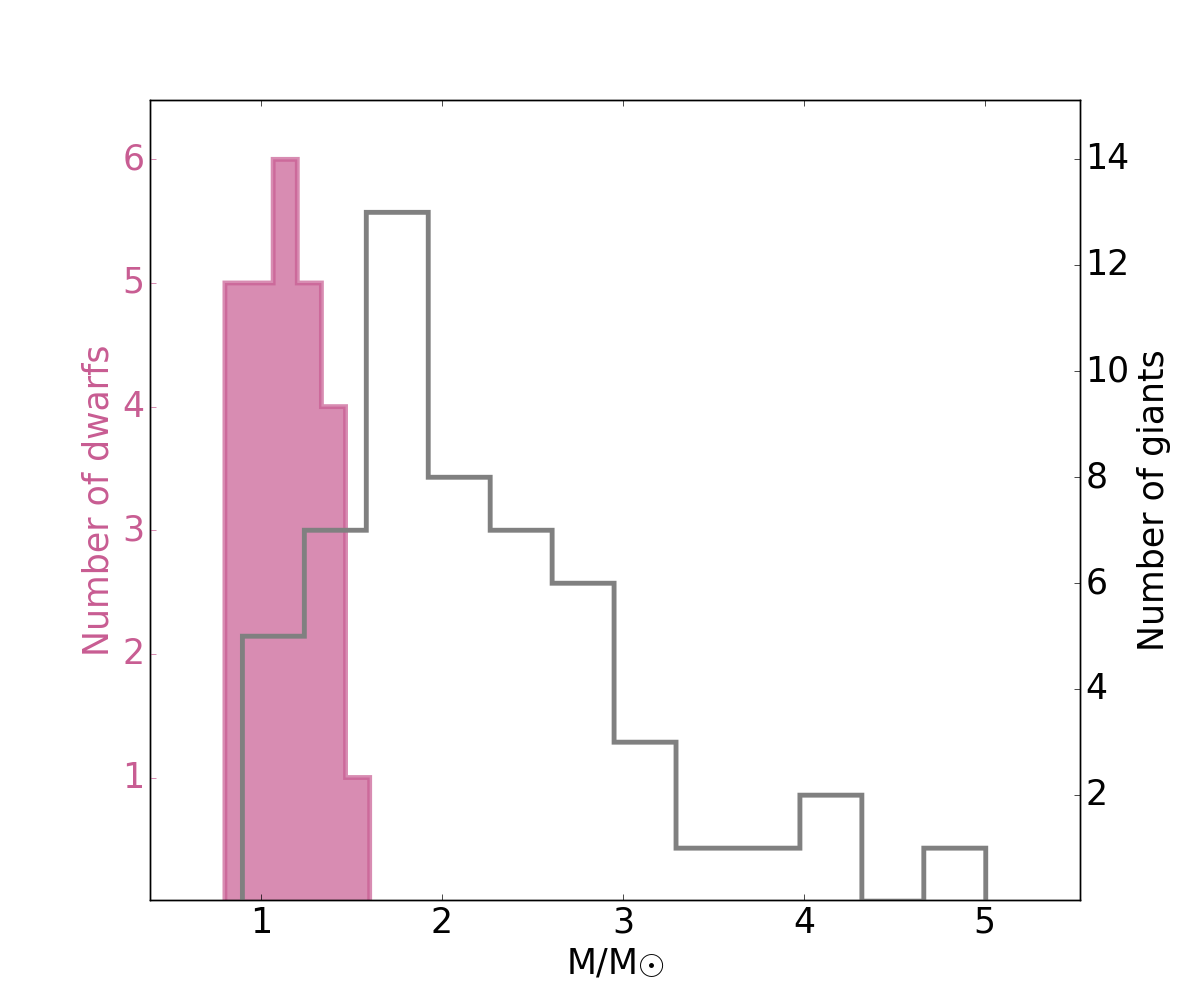}
\caption{\label{md} Mass distribution of dwarf and subgiant Ba and CH stars (solid pink) and of Ba giants (grey) from \cite{Jorissen19}.}
\end{figure}

Figure \ref{md} presents the mass distribution of Ba and CH stars before (pink coloured histogram) and after (black histogram) they ascend the RGB. Only objects with determined orbital solutions are included in Fig. \ref{md} (same systems than in Fig. \ref{e-logP}). The masses of the giants come from \cite{Jorissen19}. The mass distribution of Ba dwarf and subgiant stars peaks at significantly lower value than the mass distribution of Ba giants, which peaks around 2\,M$_{\sun}$. Hence, the Ba and CH stars that we discuss in this paper will not evolve into the prototypical Ba giants that are observed, only into the least massive ones. The reason for this is likely an observational bias against hotter dwarf Ba stars. A-type stars rotate typically fast, hence the lines appear broadened in their spectra, which makes detecting and quantifying the s-process overabundances difficult. Moreover, in this region of the main sequence, a wide variety of stellar processes can cause stellar variability (e.g. $\delta$ Scuti, $\gamma$ Doradus and roAp stars) and chemical inhomogeneities (e.g. Am and Ap stars). The effect of these could wash out any signature of past mass-transfer in the spectra. It is likely that some of the delta Scuti stars with suspected WD companions studied by \cite{Murphy18} gained enriched material from a former AGB companion, but a systematic high-resolution spectral survey at short wavelength is needed to confirm this hypothesis.

\subsection{Mass function and companion masses}\label{ssec:mf}
The mass function relates the masses of the two components in a binary ($m_1$ being the mass of the Ba star and $m_2$ being the mass of the WD companion in our case) and the inclination of the system (sin\,$i$) with the parameters derived from the spectroscopic orbit (period, $P$; eccentricity, $e$; and semi-amplitude, $K_1$) as follows: 

\begin{equation}\label{eq:fm}
f(m)=\frac{m_2^3}{(m_1+m_2)^2}\sin^3 i=1.0361 \cdot 10^{-7}\cdot(1-e^2)^{3/2}K_1^3 P \,\,\,\,\,\, [\rm{M}_{\sun}]
\end{equation}

\noindent with $P$ expressed in days and $K_1$ in $\mathrm{km\,s}^{-1}$.

We derived the mass functions of our binaries in Sect. \ref{sec:orbits} (column 9 of Table \ref{table_dBa}) and the masses of the primary stars in Sect. \ref{sec:hrd} (column 9 of Table \ref{params}). We are only missing information about the orbital inclination to derive the mass of the WD companions. Following the methodology used in \cite{PJ2000}, \cite{Pourbaix03} and \cite{Jancart05}, among others, we combined our new orbital solutions with Hipparcos astrometric data to derive the astrometric orbit of our spectroscopic binaries. This had been done for some of our targets before \citep{PJ2000}, however, the statistical tests introduced later by \cite{Pourbaix01} and \cite{Jancart05} were not applied at the time. We repeated the reprocessing of the Hipparcos astrometric data, and only four of our Ba and CH stars passed the statistical tests. The obtained parallaxes, inclinations and companion masses are presented in Table \ref{WD}. Since the orbital periods of these targets are far from one year, one does not expect disagreement in the parallaxes with respect to the single star solution. The obtained parallaxes are in good agreement with the Hipparcos \citep{HippCat97} and \textit{Gaia}-DR2 values \citep{Lindegren18}. Additionally, the derived WD masses are compatible with current estimates for field WD masses
\citep{Kleinman13} and with the mass distribution of WD companion of Ba giants obtained by \cite{Jorissen19}.

\begin{table}[t] 
\begin{small}
\caption{Parallaxes and orbital inclinations from the combination of spectroscopic parameters and Hipparcos astrometric data and companion masses of the Ba and CH stars that passed the tests presented in \cite{Pourbaix01} and \cite{Jancart05}. }
\label{WD}
\begin{center}
\begin{tabular}{lccc}
\hline
\rule[0mm]{0mm}{4mm}
ID & $\varpi$ [mas] & Inclination [$^{\circ}$] & Companion mass [M$_{\sun}$]\\
\hline
HD\,34654 & 21.5 $\pm$ 1.0 & 80 $\pm$ 4 & 0.621 $\pm$ 0.018\\
HD\,50264 & 14.1 $\pm$ 1.1 & 109 $\pm$ 5 & 0.60 $\pm$ 0.05\\ 
HD\,89948 & 23.9 $\pm$ 0.8 & 102 $\pm$ 3 & 0.54 $\pm$ 0.03\\
HD\,123585 & 9.5 $\pm$ 1.7 & 64 $\pm$ 13 & 0.66 $\pm$ 0.11\\
\hline
\end{tabular}
\end{center}
\end{small}
\end{table}

Only after the third data release of the \textit{Gaia} satellite, which will include binary astrometric solutions, will we be able to obtain companion masses for our other systems. In the meantime, several assumptions can be made about the unknown parameters in order to model the observed cumulative distribution of the mass functions. The mass distribution of the primary stars (m$_1$) is described in our model as a Gaussian centred at a mass $\mu_1$ and with standard deviation $\sigma_1$. We allow these two parameters to vary within ranges compatible with Fig. \ref{md}. Additionally, it is safe to assume that the secondary stars are WDs, so their mass distribution is also described with a Gaussian characterised by $\mu_2$ and $\sigma_2$, which are also free parameters within ranges compatible with WD masses from \citet{Kleinman13}. Finally, we need a distribution of orbital inclinations. If we assume that orbital planes are randomly oriented in space and take into account projection effects, we can use the following distribution:

\begin{equation}
i =  \arccos (z)
\end{equation}

\noindent where $z$ is the length of the projection of the unit normal vector along the line of sight and is uniformly distributed in the range [-1, 1].

\begin{figure}
\centering
\includegraphics[width=0.49\textwidth]{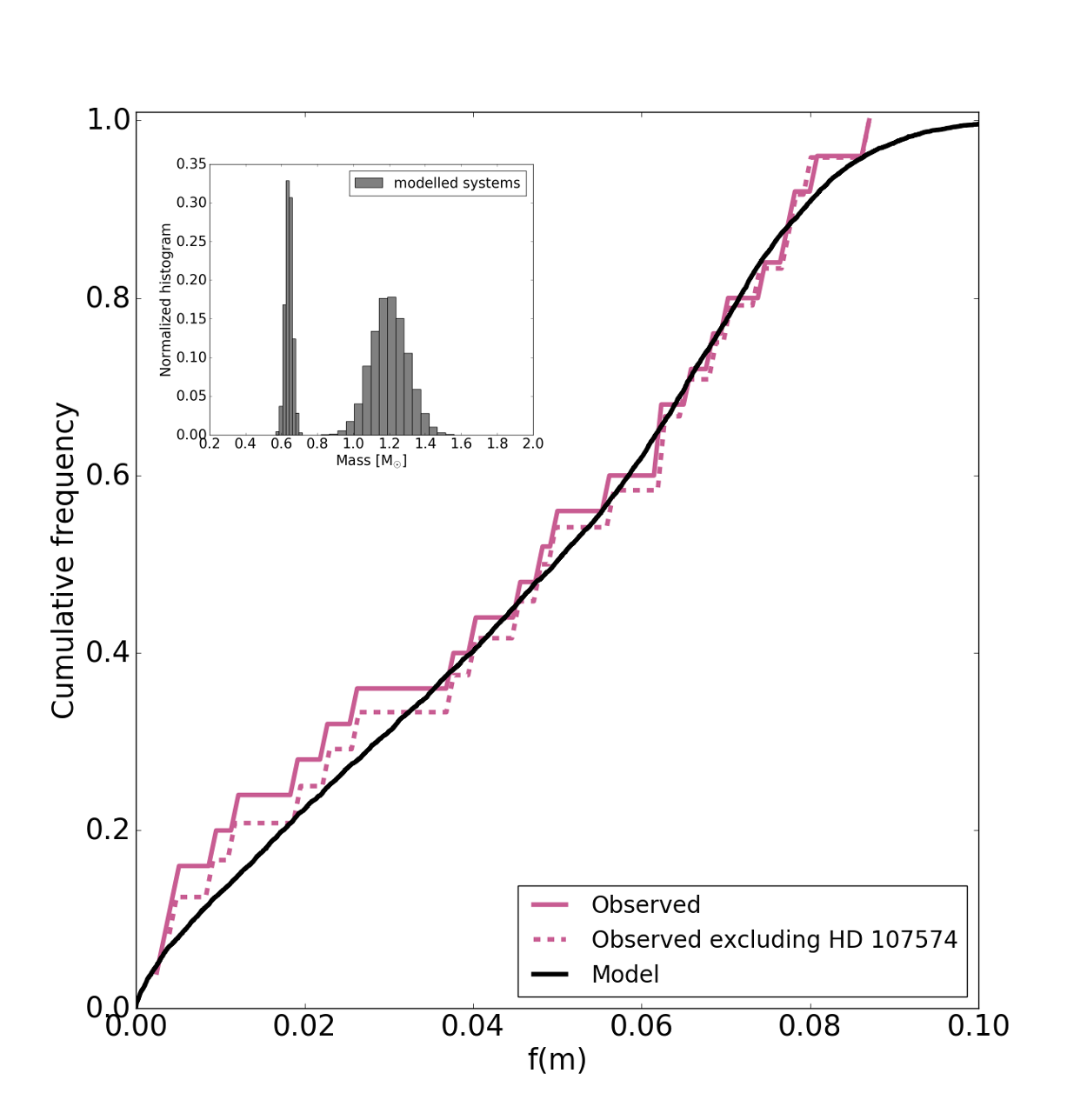}
\caption{\label{massf} Cumulative distribution of the mass functions of Ba and CH dwarf and subgiant stars (continuous pink line) and best fitting modelled to the distribution (dashed pink line). Giant Ba stars have been included in grey for comparison.}
\end{figure}

We generated 10\,000 random orbits and compared the modelled cumulative distribution with our observations to constrain the parameters that describe the mass distribution of the two stars in these systems. Figure \ref{massf} shows the comparison of the observed distribution (solid pink line) and the best fitting model (solid black line). This has been found for a primary mass distribution centred at 1.2~M$_{\sun}$ with $\sigma_1$ = 0.1~M$_{\sun}$ and a secondary (WD) mass distribution centred at 0.64~M$_{\sun}$ with $\sigma_2$ = 0.02~M$_{\sun}$. The distributions are shown in the panel on the top left corner of Fig. \ref{massf}. There is a small excess of low mass functions with respect to the best fitting model, which seems to be mostly caused by HD\,107574, a target with a remarkably low mass function compared with the others (see Table \ref{params}). The dashed pink line in Fig. \ref{massf} shows the observed cumulative distribution of mass functions when we exclude this object, and it seems to agree better with our best fitting model. However, we do not have a reason to exclude it from our sample. Assuming our value for $m_1$ is correct, the small mass function could be due to a very small companion mass, or to a very small orbital inclination (of about 16$^{\circ}$ to host a 0.6~M$_{\sun}$ WD), but we can not distinguish between these two possibilities with the available information.

\cite{Webbink86} already suggested that the mass-function distribution of Ba stars can be fitted by a sample of orbits with a very narrow distribution of $Q$, where $f(m) = Q \sin^3 i$ and several studies have confirmed this for Ba giants (e.g. \citealt{Jorissen98}, \citealt{Jorissen19}). Our model (black solid line in Fig. \ref{massf}) corresponds to a Gaussian distribution of $Q$ with $\mu_Q = 0.077$~M$_{\sun}$ and $\sigma_Q = 0.01$~M$_{\sun}$. The small sigma of the $Q$ distribution suggests that $m_1$ and $m_2$ could be correlated.

\begin{figure}
\centering
\includegraphics[width=0.49\textwidth]{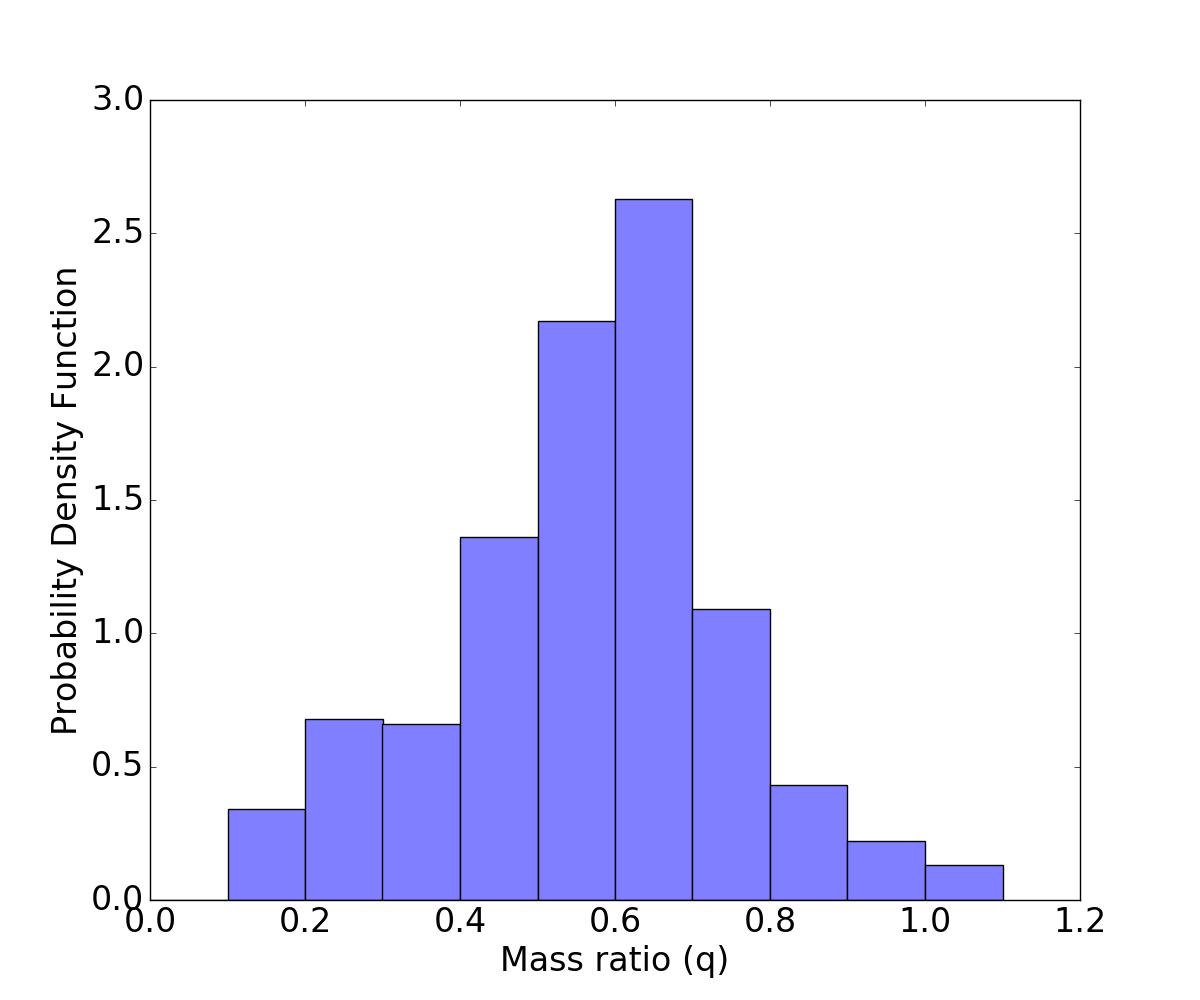}
\caption{\label{q} Probability density function (PDF) of mass ratios of our sample of Ba and CH dwarf and subgiant stars.}
\end{figure}

Finally, $Q$ can also be expressed as follows: 

\begin{equation}\label{eq:Q}
Q = m_1 \frac{q^3}{(1+q)^2}\,\,\,\,\,\, [\rm{M}_{\sun}],
\end{equation}

\noindent where $q = m_2 / m_1$ is the mass ratio between the two stars. Applying the Lucy-Richardson inversion method (e.g. \citealt{Boffin92}; \citealt{Cerf94}) to the distribution of $Q$ values, we can also obtain the distribution of $q$ (Fig. \ref{q}).

\subsection{Comparison with evolutionary models}\label{ssec:models}
Since the first $e-logP$ diagrams of Ba giants were obtained, a lot of research was devoted to understanding them. The interaction between the former AGB star and its less-evolved companion shapes the orbits in a way that can not be well reproduced by evolutionary models (e.g., \citealt{Pols03}, \citealt{BonacicMarinovic08}). However, Ba stars might have gone through a second stage of binary interaction when the dwarf ascended the RGB, which can also affect the eccentricity and the period of Ba giants.

In order to explore the evolutionary link between our Ba dwarfs and the observed low-mass Ba giants, we computed a grid of standard binary evolution models with the BINSTAR code \citep{Siess13}. We used input parameters inspired by our sample of main-sequence Ba and CH stars. The systems that we modelled were formed by a primary main-sequence star with $M$~=~1.5~M$_{\sun}$ and metallicity $\rm{[Fe/H]}=-0.25$, and a cool WD companion of $M$~=~0.6~M$_{\sun}$. We used seven initial orbital periods: 100, 300, 600, 1000, 2000, 3000, and 10\,000 days; and four eccentricities: 0.2, 0.4, 0.6, and 0.8. We only followed the evolution of the primary star, for which we used a mass-loss prescription as described by \cite{SchroderCuntz07} and a grey atmosphere as surface boundary condition. We considered "standard" binary interactions assuming that the stars are in solid rotation and that angular momentum evolution is governed by the effect of tides and mass loss. We considered that the wind carries away the specific orbital angular momentum of the star as in the Jeans mode. In \cite{Escorza17} we showed that Ba giants accumulate in the He-clump, so we allowed the primary star to evolve until the onset of core He-burning and compared the results with known orbits of giants.

\begin{figure}[t]
\begin{center}
\includegraphics[width=0.49\textwidth]{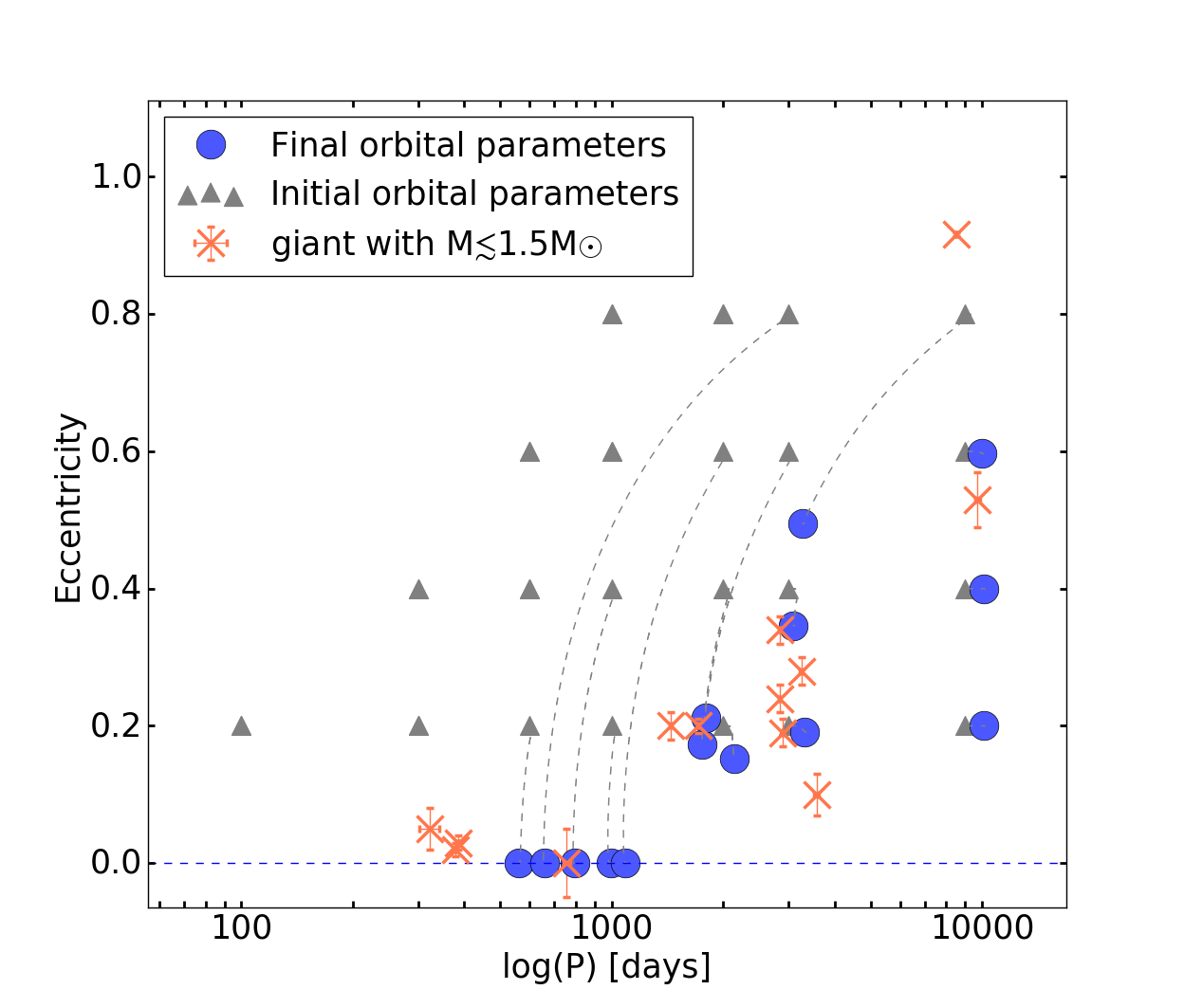}
\caption{\label{Fig:elogPth} Observed and modelled orbits of Ba giants after the binary interaction along the RGB phase. The triangles indicate the initial period an eccentricity of each BINSTAR model and the blue dots indicate the final orbital parameters at the time of core-He burning onset. The evolution of the models that reached the core-He burning phase is represented by a dashed gray line. Orange crosses are observed orbits of low mass ($\lesssim$1.5~M$_{\sun}$) Ba and CH giants from \cite{Jorissen19}.}
\end{center}
\end{figure}

Fig. \ref{Fig:elogPth} shows the final orbits of the models that reached the core He-burning phase (blue circles) together with observations of Ba giants (orange crosses) that have $M$~=~1.5~M$_{\sun}$ or lower. The grey triangles are the initial orbital parameters of the models. When the primary reaches the core He-burning phase, the initial and final orbital parameters are linked by a dashed line that shows the evolution of the orbit. Triangles that are not connected to an evolutionary track correspond to models in which the primary star leaves the RGB after losing most of its mass via Roche-lobe overflow (RLOF) and does not become one of the giant Ba systems that we observe. These models do not have a corresponding final orbit (blue circle) in the figure.

When we compare the observations with the model predictions for this primary mass, we find a few outlying objects. Three giant systems have periods shorter than those allowed by standard binary models of this mass (P\,$<$\,700~d). There is also one Ba giant with a very long period and a very high eccentricity that our models cannot reproduce either. As the model with $P_{\rm init}$~=~10\,000 days and $e_{\rm init}$~=~0.8 shows, it is very difficult to keep such a high eccentricity after a stage of tidal interaction in the RGB. Another object mentioned in \cite{Jorissen19} but without a well-constrained orbit yet, HD\,134698, will probably occupy the same region of the $e-logP$ diagram (the preliminary estimated period and eccentricity are 10005 days and 0.95). Binary interaction mechanisms, that we do not include in our standard binary calculations, need to be considered to reproduce all the observed Ba giant orbits.

\section{Summary and conclusions}

We combined radial velocity data from four different instruments to analyse a sample of 60 objects classified in the literature as dwarf Ba or subgiant CH stars. To our knowledge, this is the largest systematic radial-velocity survey of dwarf and subgiant Ba and related stars. We determined accurate orbital parameters for 27 binaries. Twenty-five of them are SB1s with a polluted primary and a suspected WD companion. Additionally, HD\,48565 is a triple system, for which we constrained both the inner and outer orbit, and HD\,114520 is an SB2. The latter must be a triple system as well since the two stars that contribute to the spectra are main-sequence stars enhanced in s-process elements \citep{Gray11}. We determined the orbital parameters of the SB2 and concluded that the WD must be in an outer orbit of at least 2058~days, but we did not detect its signature in the residuals of the orbital fit as we did for HD\,48565. In addition, we found two more SB2s, HD\,26455 and HD\,177996, but we could not constrain their orbits due to a lack of data. Among the remaining objects of the sample, we confirmed radial velocity variability of 11 of them, but again, we did not have enough data to determine their orbital parameters. The monitoring of these long period binaries will continue. The other 20 objects do not present clear variability according to the criterion adopted in Sect. \ref{ssec:others}.

We determined stellar parameters from HERMES high-quality spectra or adopted literature values coming from high-resolution spectroscopy to locate the primary stars of our SB1s on the HRD. We used distances derived by \cite{Bailer-Jones18} from \textit{Gaia} DR2 data to obtain accurate luminosities. From the comparison of the location of these stars on the HRD with STAREVOL evolutionary tracks, we have also derived their masses. Our spectroscopic metallicities show no clear distinction between stars classified in the literature as dBa or as sgCH stars and our HRD does not agree with the difference in evolutionary stage that their names suggest. However, we observe that main-sequence CH stars are cooler and less massive than main-sequence Ba stars. The HRD also shows that several targets occupy regions of fast stellar evolution where it is difficult to observe many stars.

When we compare our sample with a sample of Ba giants, we see that the mass distribution of main sequence and subgiants stars peaks at much lower mass than the distribution of masses of their more evolved analogues. This means that the sample of dwarf and subgiant stars that we studied does not represent the precursor sample of the observed Ba giants. We associate this with an observational bias against the detection of hotter post-mass-transfer main-sequence stars.

From our orbital analysis, we concluded that the periods and eccentricities of Ba and CH dwarfs lie on the same region of the $e$-log$P$ diagram as the orbital elements of Ba giants. They also present high eccentricities at short periods like many other families of post-interaction binaries. Another product of the orbital analysis is the mass function of the systems. We combined our spectroscopic orbital elements with Hipparcos astrometric data and could obtain the orbital inclination of four of our systems. Using the mass functions, the primary masses that we determined and these inclinations, we derived the absolute mass of the four WD companions. The obtained values are consistent with field WD masses. The combination of spectroscopic orbital elements and astrometric data is very powerful and, with \textit{Gaia} DR3, this method will allow us to put more stringent observational constraints to the formation and evolution of Ba stars.

Since we cannot determine absolute WD masses for all the systems yet, we also modelled the mass function distribution of our spectroscopic binaries, concluding that all these systems can be represented by a population of binary systems formed by primary main-sequence stars and WDs with very narrow and Gaussian mass distributions and orbital planes randomly oriented on the sky. The mass distribution obtained for the primaries peaked at 1.2~M$_{\sun}$, which is in good agreement with the masses that we obtained from independent observations. The secondary mass distribution is an expected distribution of WD masses as well.

Finally, we used the BINSTAR binary evolution code to study the evolution of the orbital elements of our main-sequence stars along the RGB until they become the low-mass Ba giants that \cite{Jorissen19} observed.
Our models can explain the majority of the giant orbits with primary mass $\lesssim 1.5$~M$_{\sun}$, but we find four unexplained systems, three with short periods and modest eccentricities and one with a long period and a very high eccentricity. This might indicate that additional binary interaction mechanisms are also needed to explain the orbital evolution of low-mass Ba star systems along the RGB, independently of the previous interaction during the AGB phase of the now WD companion.\\

\begin{acknowledgements}
This research has been funded by the Fonds voor Wetenschappelijk Onderzoek Vlaanderen (FWO) under contract ZKD1501-00-W01 and by the Belgian Science Policy Office under contract BR$/$143$/$A2$/$STARLAB.
A.E. is grateful to the staff members of the IvS (KU Leuven) and IAA (ULB) for the enriching discussions and valuable feedback and to all observers of the HERMES consortium for their time dedicated to the Mercator-HERMES long-term monitoring program. She is also grateful to Dr. Silvia Toonen for the discussions about triple systems.
D.K. acknowledges the financial support from SERB-DST, through the file number PDF$/$2017$/$002338.
L.S. and D.P are Senior FNRS Research Associates.
CJ acknowledges funding from the European Research Council (ERC) under the European Union's Horizon 2020 research and innovation programme (grant agreement N$^\circ$670519: MAMSIE).
B.M. acknowledges support from the National Research Foundation (NRF) of South Africa.
J.M. acknowledges the funding from the National Science Centre, Poland, through grant OPUS 2017$/$27$/$B$/$ST9$/$01940. 
This work has made use of data from the European Space Agency (ESA) mission \textit{Gaia} (\url{https://www.cosmos.esa.int/gaia}), processed by the \textit{Gaia} Data Processing and Analysis Consortium (DPAC, \url{https://www.cosmos.esa.int/web/gaia/dpac/consortium}). Funding for the DPAC has been provided by national institutions, in particular the institutions participating in the \textit{Gaia} Multilateral Agreement. Some of the observations reported in this paper were obtained with the Southern African Large Telescope (SALT). Polish participation in SALT is funded by grant No. MNiSW DIR$/$WK$/$2016$/$07.
This research has made use of the SIMBAD database, operated at CDS, Strasbourg, France. 
\end{acknowledgements}

\bibstyle{aa} 
\bibliography{literature} 

\clearpage

\appendix

\onecolumn

\section{Sample}\label{App:sampletable}
\renewcommand{\arraystretch}{1.2}
\LTcapwidth=\textwidth
\begin{longtable}{lccccccc}
\caption{Information about the initial sample considered for this paper including classification of the objects in the literature (column 3), number of available data points from each instrument (columns 4 to 7) and binarity status after this work (column 8). Column 7 "OTHER" lists the number of available CORALIE radial-velocity measurements for all targets but for HD\,48565, for which we have no CORALIE data but four ELODIE radial-velocity points.\\
\textbf{References for column 3:} 1: \cite{LuckBond91}; 2: \cite{Houk&Cowley75}; 3: \cite{Houk78}; 4: \cite{North95}; 5: \cite{North94}; 6: \cite{Tomkin89}; 7: \cite{Edvardsson93}; 8: \cite{Lu83}; 9: \cite{Udry98}; 10: \cite{Gray11}; 11: \citep{PereiraJunqueira03}; 12: \cite{Lu91}; 13: \cite{McClureWoodsworth90}\\
\textbf{Binarity type in column 8:} SB-O: Spectroscopic binary with determined orbital parameters in Table \ref{table_dBa}; SB: Confirmed spectroscopic binary without available orbit (see Table \ref{uncovered}); SB2: Binary with a double-peaked CCF; TS: Confirmed triple system; NV: No variability detected according to the criterion described in Sect. \ref{ssec:others}.
\label{info}}\\
\hline\hline
\rule[0mm]{0mm}{4mm}
ID used in this paper & Second ID & Type (Ref.) & CORAVEL & HERMES & SALT HRS & OTHER & Binarity \\
\hline
\endfirsthead
\hline\hline
\rule[0mm]{0mm}{4mm}
ID used in this paper & Second ID & Type (Ref.) & CORAVEL & HERMES & SALT HRS & OTHER & Binarity \\
\hline
\endhead
BD-10$^\circ$4311 & HIP\,80356 & sgCH(1) & 29 & 20 & 0 & 1 & SB-O\\ 
BD-11$^\circ$3853 & HIP\,73444 & sgCH(1) & 16 & 113 & 0 & 2 & SB\\ 
BD-18$^\circ$255 & TYC\,5852-1110-1 & dBa(2,3,4), sgCH(1) & 14 & 36 & 0 & 0 & SB-O\\ 
BD+18$^\circ$5215 & TYC\,1724-1717-1 & dBa(5) & 28 & 3 & 0 & 0 & SB-O\\ 
CD-62$^\circ$1346 & HIP\,104151 & sgCH(1) & 4 & 0 & 8 & 0 & SB\\ 
HD\,2454 & BD+09$^\circ$47 & dBa(6) & 46 & 54 & 0 & 0 & SB\\
HD\,6434 & HIP\,5054 & dBa(7) & 11 & 0 & 1 & 0 & NV\\
HD\,9529 & CD-71$^\circ$71 & dBa(2,3) & 7 & 0 & 6 & 0 & NV\\
HD\,13555 & BD+20$^\circ$348 & dBa(7) & 8 & 52 & 0 & 0 & NV\\
HD\,15306 & BD-01$^\circ$340 & dBa(5) & 26 & 28 & 0 & 3 & SB-O\\
HD\,18853 & HIP\,13842 & dBa(2) & 3 & 0 & 8 & 5 & SB\\
HD\,22589 & BD-07$^\circ$642 & gBa(8,9) & 20 & 30 & 0 & 0 & SB-O\\
HD\,24864 & CD-54$^\circ$750 & dBa(2) & 10 & 0 & 3 & 0 & SB-O\\
HD\,26455 & CD-53$^\circ$858 & dBa(2) & 1 & 0 & 4 & 5 & SB2\\
HD\,31732 & HIP\,22814 & dBa(2) & 4 & 0 & 2 & 0 & NV\\
HD\,34654 & HIP\,25222 & dBa(10) & 0 & 42 & 0 & 0 & SB-O\\
HD\,35296 & BD+17$^\circ$920 & dBa(7) & 16 & 58 & 0 & 0 & NV\\
HD\,48565 & BD+20$^\circ$1552 & dBa(5) & 30 & 76 & 0 & 4 & TS\\
HD\,50264 & HIP\,32894 & dBa(8), sgCH(11) & 9 & 1 & 1 & 0 & SB-O\\
HD\,60532 & BD-21$^\circ$2007 & dBa(7) & 6 & 40 & 0 & 0 & NV\\
HD\,69578 & HIP\,40208 & dBa(2) & 8 & 0 & 3 & 0 & SB\\
HD\,76225 & HIP\,43703 & dBa(5) & 3 & 22 & 0 & 4 & SB-O\\
HD\,82328 & BD+52$^\circ$1401 & dBa(7) & 5 & 79 & 0 & 0 & NV\\
HD\,87080 & HIP\,49166 & dBa(8), sgCH(11) & 9 & 1 & 0 & 0 & SB-O\\
HD\,89948 & HIP\,50805 & sgCH(1) & 18 & 1 & 0 & 0 & SB-O\\ 
HD\,92545 & BD-11$^\circ$2929 & dBa(5) & 19 & 72 & 0 & 0 & SB\\
HD\,95241 & BD+43$^\circ$2068 & dBa(7) & 9 & 61 & 0 & 0 & SB-O\\
HD\,98991 & BD-17$^\circ$3367 & dBa(7) & 8 & 39 & 0 & 0 & SB-O\\
HD\,101581 & HIP\,56998 & suspected dBa(12) & 9 & 0 & 0 & 0 & NV\\
HD\,103840 & HIP\,58290 & suspected dBa(12) & 10 & 0 & 0 & 0 & NV\\
HD\,104342 & HIP\,58582 & suspected dBa(12) & 7 & 0 & 0 & 0 & NV\\
HD\,105671 & HIP\,59296 & suspected dBa(12) & 6 & 0 & 0 & 0 & NV\\
HD\,106191 & BD-14$^\circ$3478 & dBa(5) & 29 & 5 & 0 & 0 & SB-O\\
HD\,107574 & HIP\,60299 & dBa(5) & 21 & 55 & 0 & 0 & SB-O\\
HD\,109490 & CD-43$^\circ$7765 & dBa(3) & 3 & 0 & 0 & 0 & SB\\
HD\,113402 & HIP\,63812 & dBa(5) & 8 & 0 & 0 & 0 & NV\\
HD\,114520 & BD+22$^\circ$2550 & dBa(10) & 0 & 85 & 0 & 0 & SB2\\
HD\,117288 & HIP\,65870 & suspected dBa(12) & 9 & 0 & 0 & 0 & NV\\
HD\,120620 & BD-03$^\circ$3537 & gBa(8,9) & 28 & 3 & 0 & 0 & SB-O\\
HD\,123585 & HIP\,69176 & sgCH(1) & 11 & 0 & 0 & 0 & SB-O\\
HD\,124850 & BD-05$^\circ$3843 & dBa(7) & 18 & 86 & 0 & 0 & NV\\
HD\,127392 & HIP\,71058 & sgCH(1) & 10 & 1 & 0 & 0 & SB-O\\
HD\,130255 & BD+01$^\circ$2980 & sgCH(13) & 26 & 68 & 0 & 0 & NV\\
HD\,141804 & CD-53$^\circ$6286 & sgCH(1) & 8 & 0 & 0 & 0 & SB-O\\
HD\,146800 & HIP\,80043 & suspected dBa(12) & 9 & 0 & 0 & 0 & NV\\
HD\,147609 & BD+27$^\circ$2631 & dBa(5) & 21 & 41 & 0 & 0 & SB-O\\
HD\,150862 & CD-24$^\circ$12805 & sgCH(1) & 12 & 1 & 0 & 0 & SB-O\\
HD\,170149 & HIP\,90674 & suspected dBa(12) & 8 & 0 & 2 & 0 & NV\\
HD\,177996 & HIP\,94050 & suspected dBa(12) & 16 & 0 & 2 & 0 & SB2\\
HD\,182274 & BD+19$^\circ$3996 & sgCH(1) & 19 & 11 & 0 & 0 & SB-O\\
HD\,188985 & HIP\,98431 & dBa(5) & 2 & 0 & 2 & 0 & SB\\
HD\,202400 & HIP\,105294 & dBa(5) & 5 & 0 & 9 & 0 & NV\\
HD\,205156 & HIP\,106560 & suspected dBa(12) & 17 & 0 & 2 & 0 & SB\\
HD\,207585 & HIP\,107818 & sgCH(1) & 16 & 2 & 2 & 0 & SB-O\\
HD\,216219 & BD+17$^\circ$4818 & gBa(8,9,12) & 29 & 1 & 0 & 0 & SB-O\\
HD\,219899 & HIP\,115183 & suspected dBa(12) & 8 & 0 & 0 & 0 & NV\\
HD\,220117 & BD+37$^\circ$4817 & dBa(7) & 11 & 69 & 0 & 0 & NV\\
HD\,221531 & BD-12$^\circ$6514 & dBa(5) & 29 & 40 & 0 & 0 & SB-O\\
HD\,222349 & CD-57$^\circ$8842 & dBa(5) & 4 & 0 & 4 & 1 & SB\\
HD\,224621 & HIP\,118266 & sgCH(1) & 4 & 0 & 0 & 0 & SB\\
\hline\hline
\end{longtable}

\clearpage
\twocolumn
\newpage
\section{Orbital solutions of the Ba dwarfs}
\label{Sect:AppendixdBa}

\begin{figure}[ht]
\centering
\includegraphics[width=0.49\textwidth]{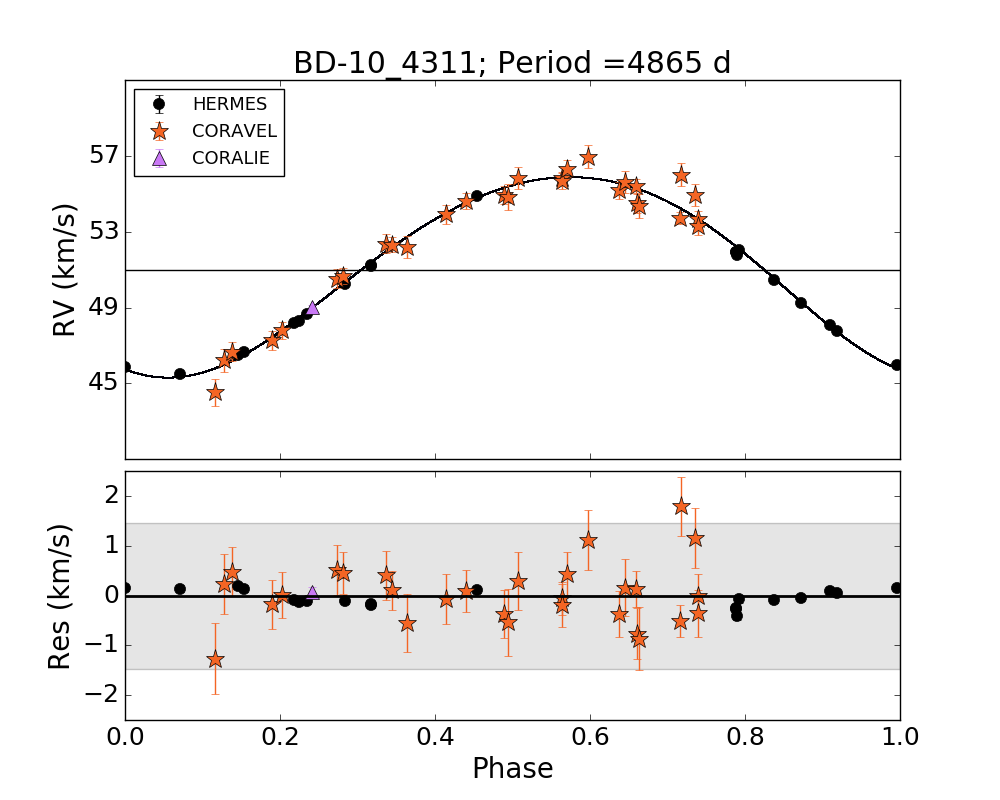}
\caption{\label{BD-10_4311} BD-10$^\circ$4311}
\end{figure}

\begin{figure}[ht]
\centering
\includegraphics[width=0.49\textwidth]{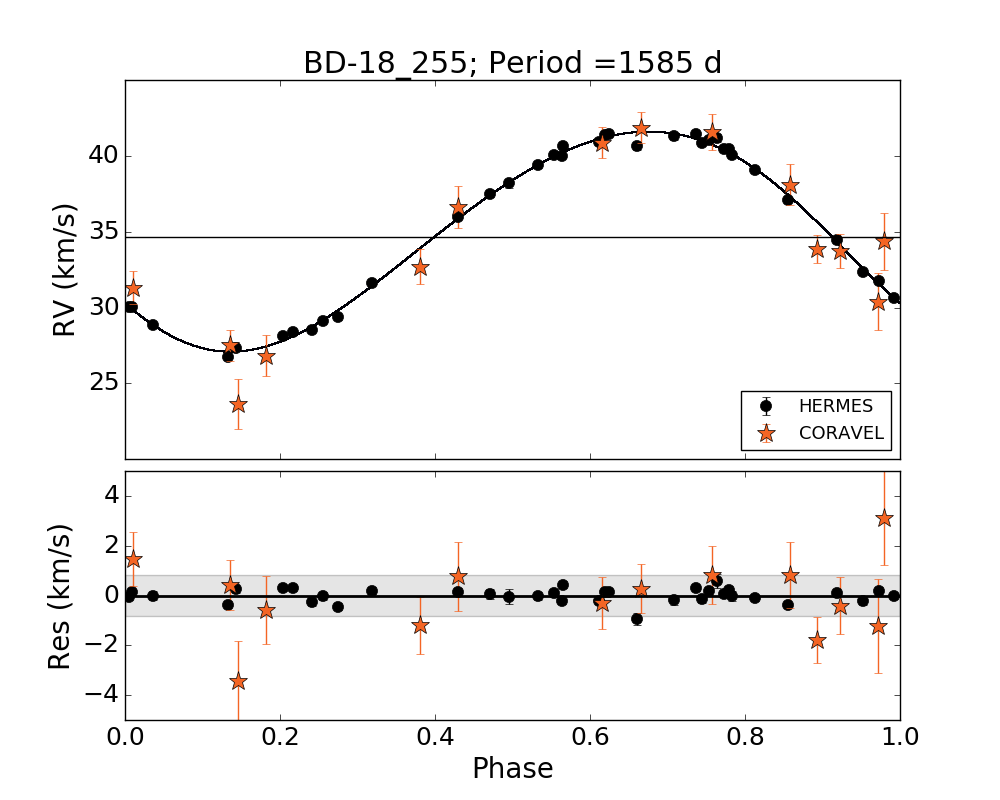}
\caption{\label{BD-18_255}BD-18$^\circ$255}
\end{figure}

\begin{figure}[ht]
\centering
\includegraphics[width=0.49\textwidth]{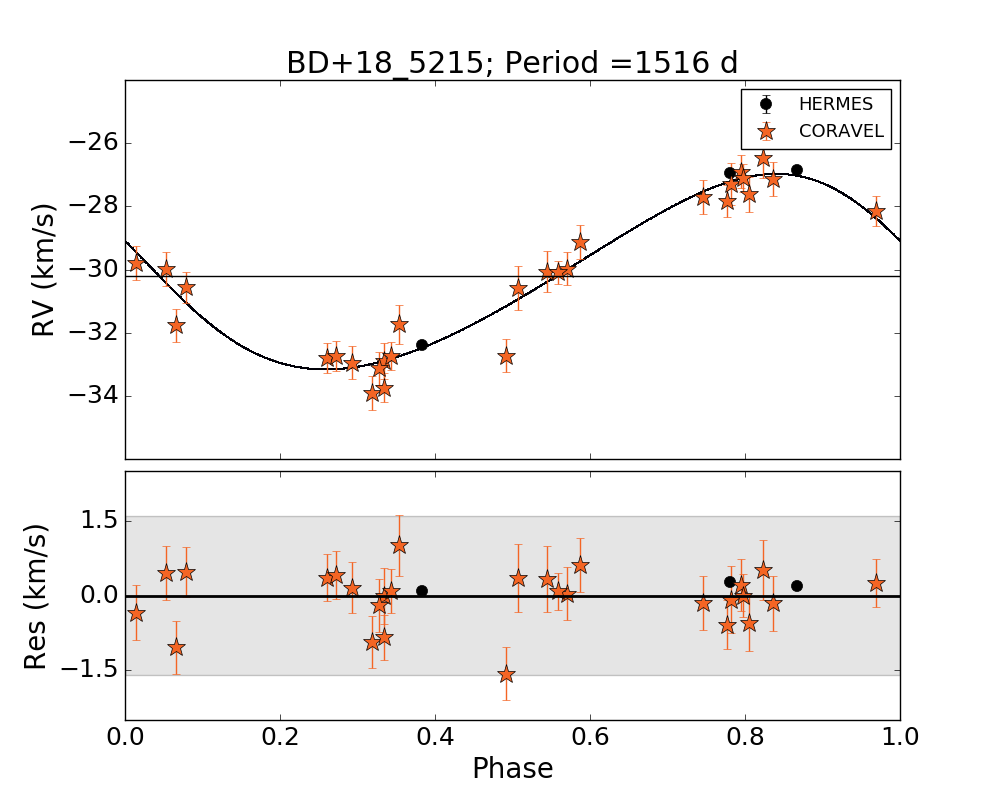}
\caption{\label{BD+18_5215} BD+18$^\circ$5215}
\end{figure}

\begin{figure}[ht]
\centering
\includegraphics[width=0.49\textwidth]{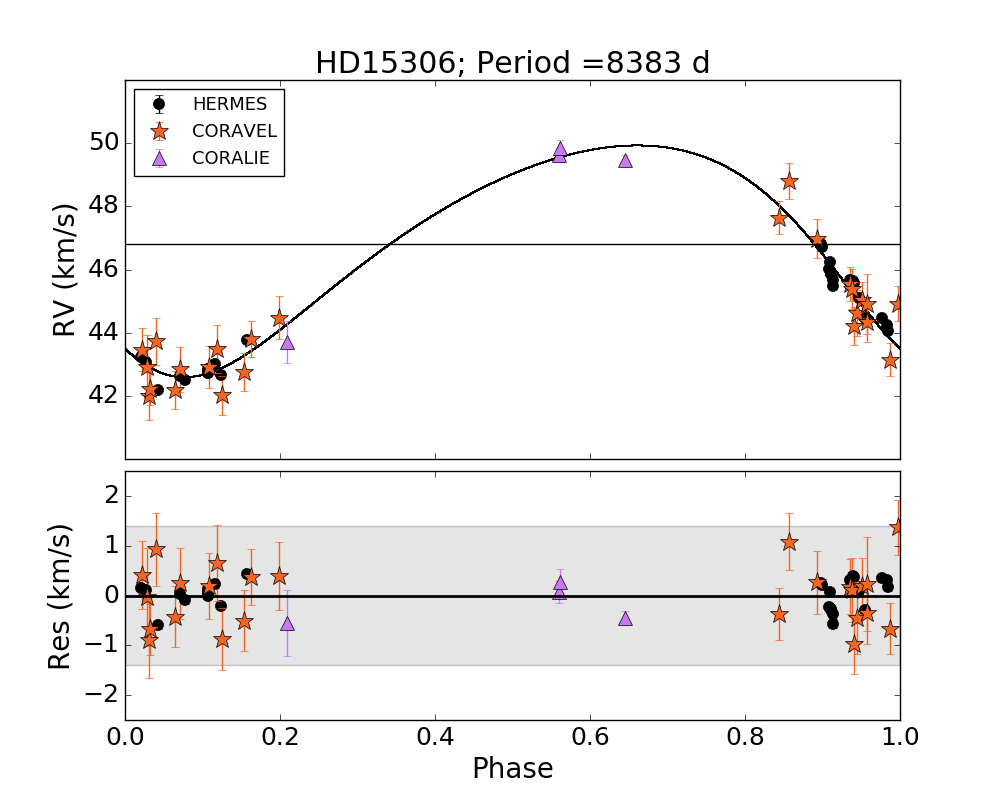}
\caption{\label{HD15306} HD\,15306}
\end{figure}

\begin{figure}[ht]
\centering
\includegraphics[width=0.49\textwidth]{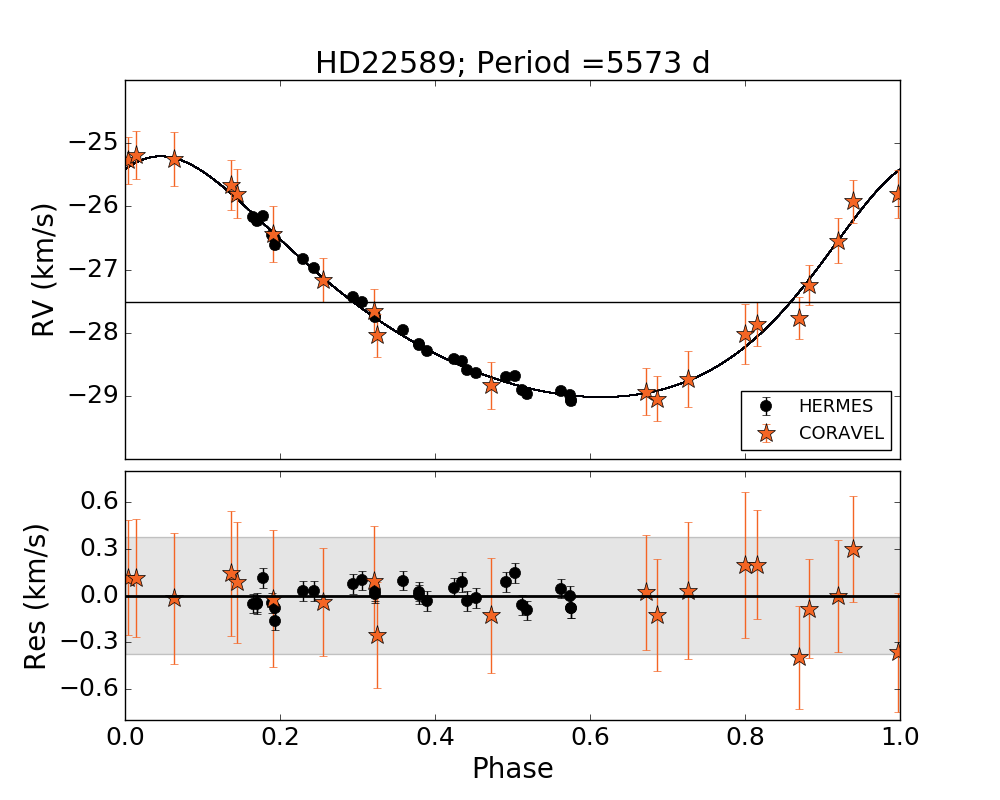}
\caption{\label{HD22589} HD\,22589}
\end{figure}

\begin{figure}[ht]
\centering
\includegraphics[width=0.49\textwidth]{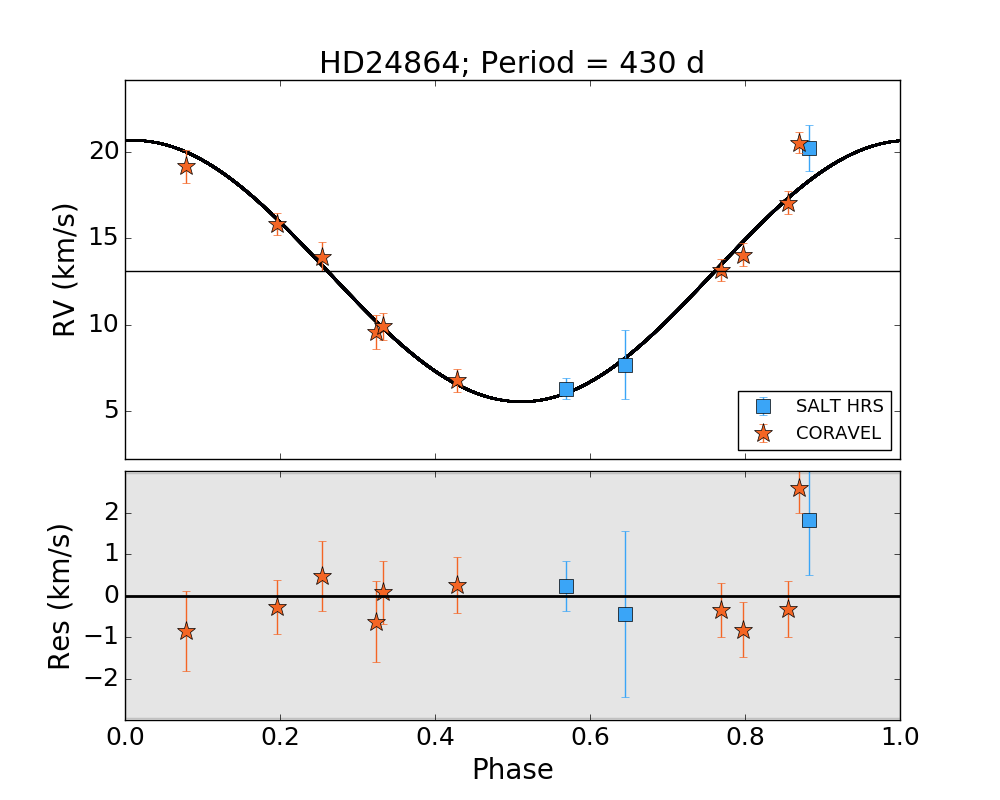}
\caption{\label{HD24864} HD\,24864}
\end{figure}

\begin{figure}[ht]
\centering
\includegraphics[width=0.49\textwidth]{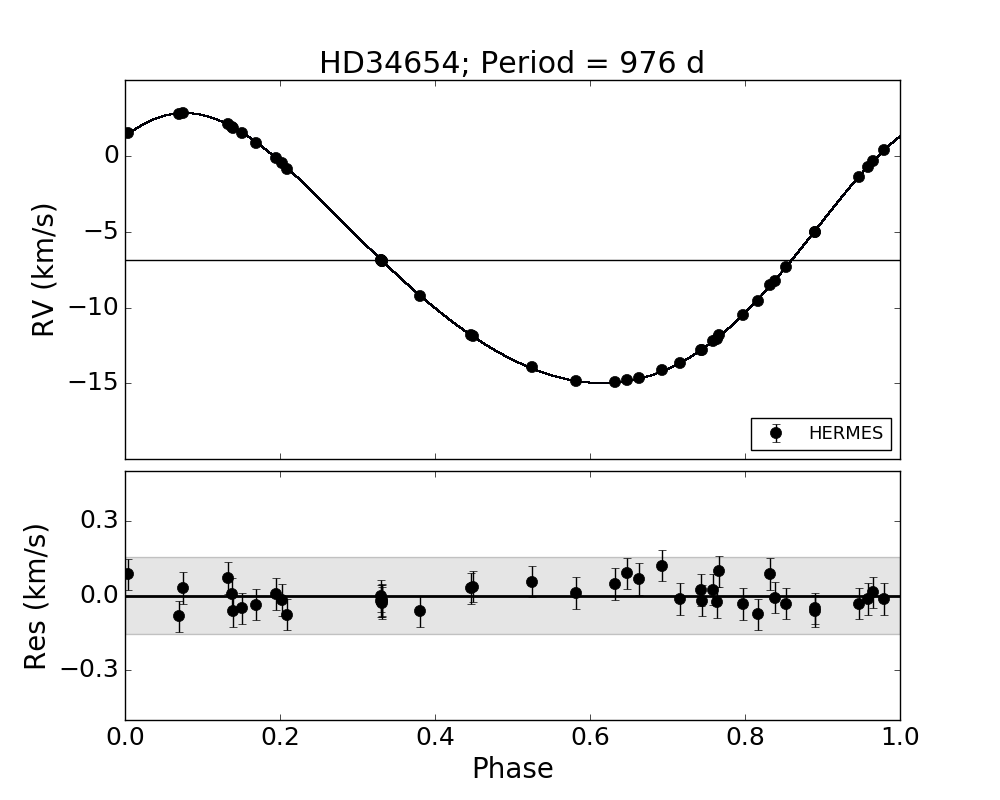}
\caption{\label{HD34654} HD\,34654}
\end{figure}

\begin{figure}[ht]
\centering
\includegraphics[width=0.49\textwidth]{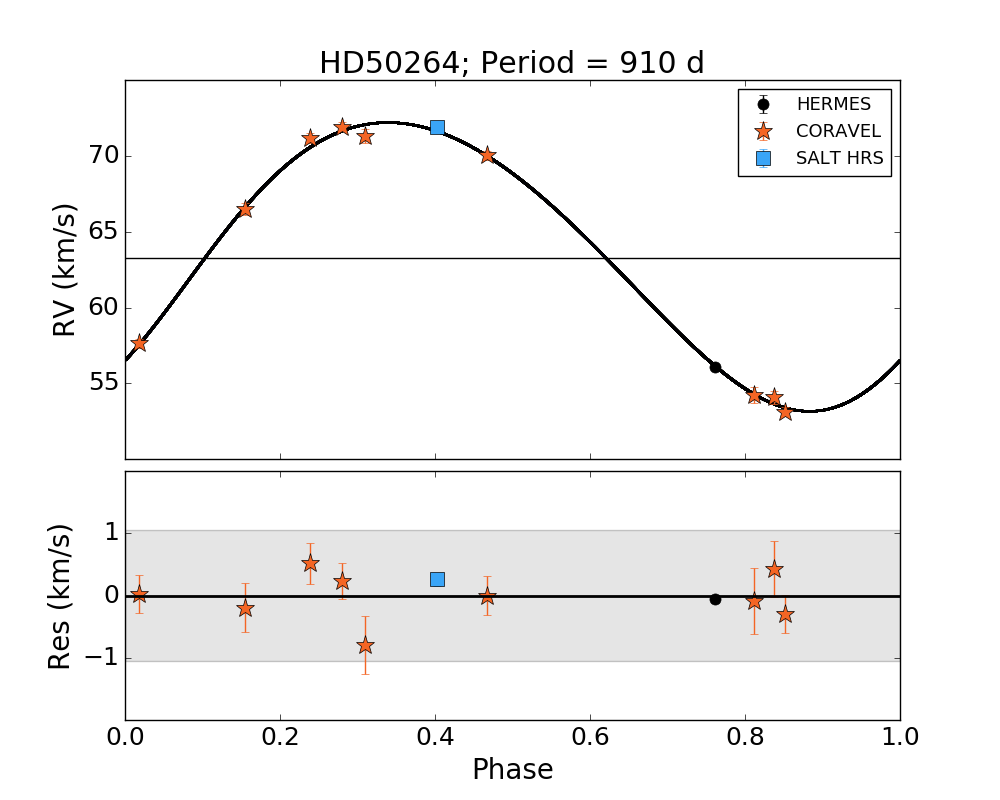}
\caption{\label{HD50264} HD\,50264}
\end{figure}

\begin{figure}[ht]
\centering
\includegraphics[width=0.49\textwidth]{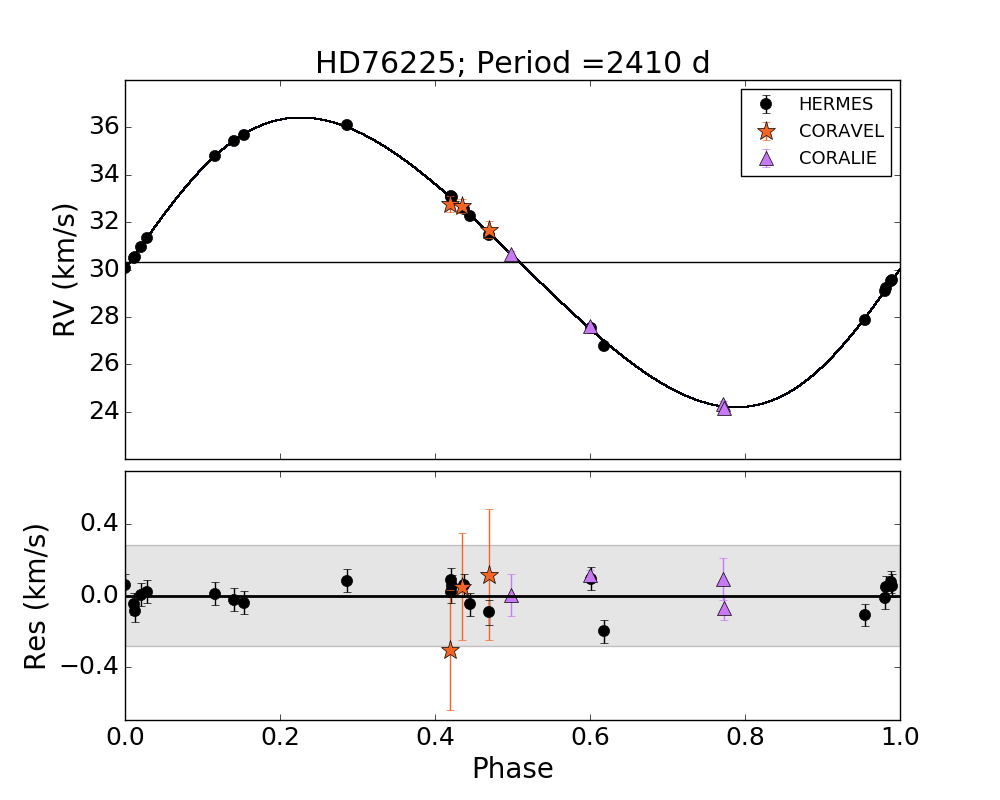}
\caption{\label{HD76225} HD\,76225}
\end{figure}

\begin{figure}[ht]
\centering
\includegraphics[width=0.49\textwidth]{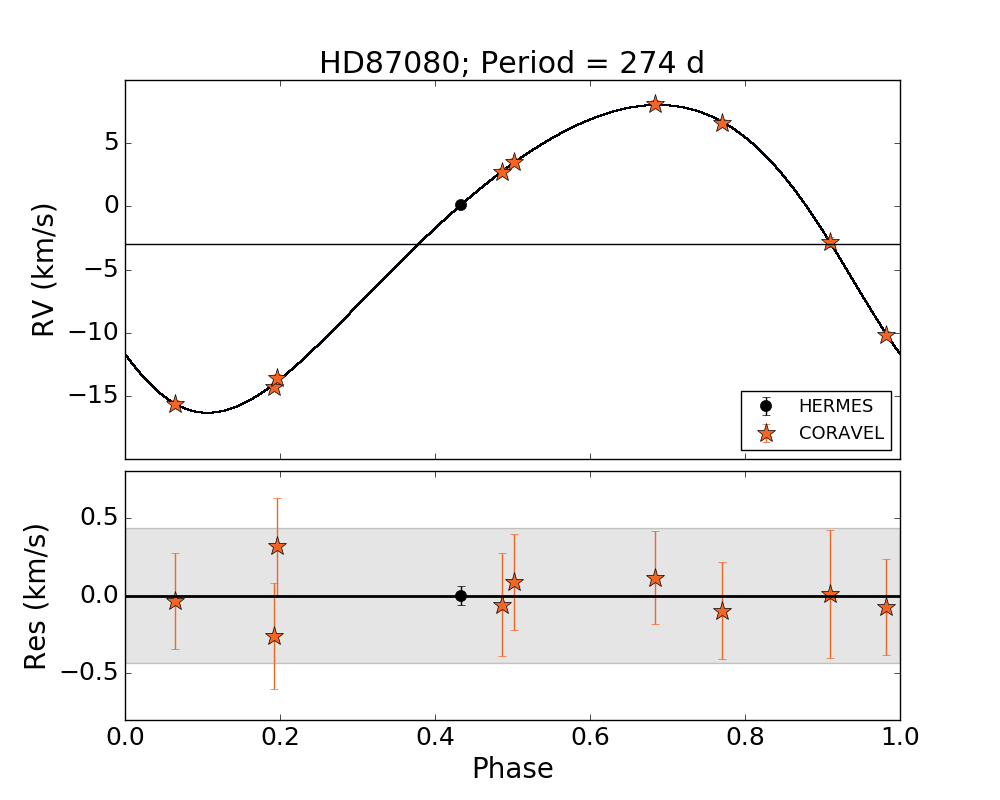}
\caption{\label{HD87080} HD\,87080}
\end{figure}

\begin{figure}[ht]
\centering
\includegraphics[width=0.49\textwidth]{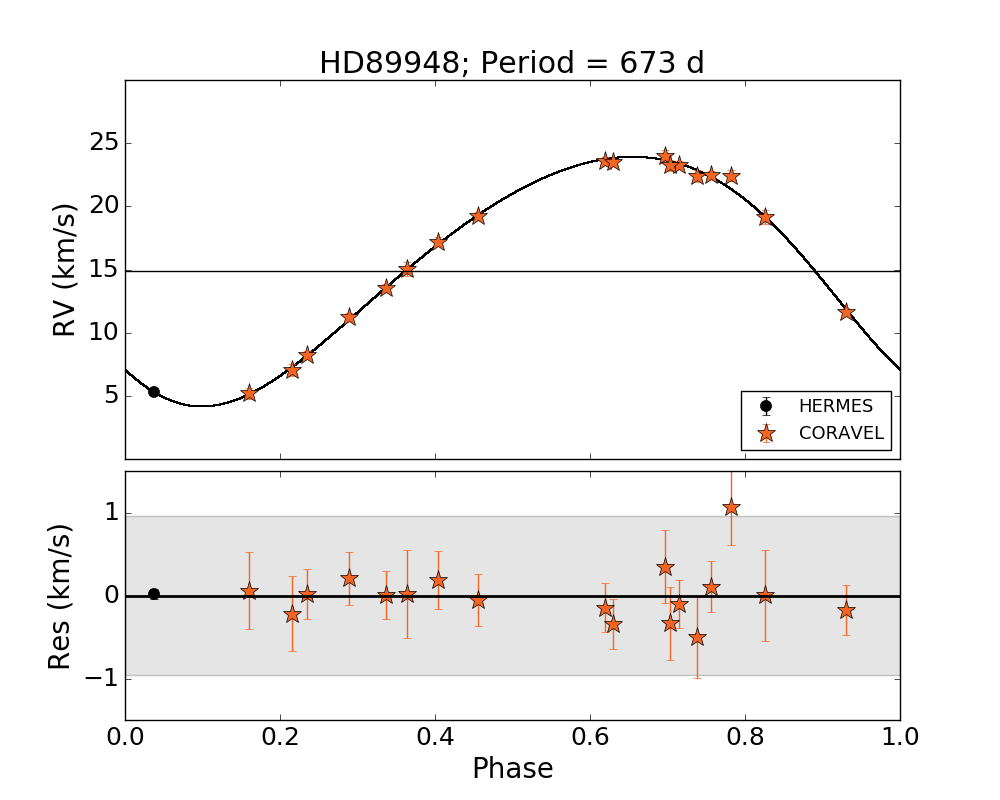}
\caption{\label{HD89948} HD\,89948}
\end{figure}

\begin{figure}[ht]
\centering
\includegraphics[width=0.49\textwidth]{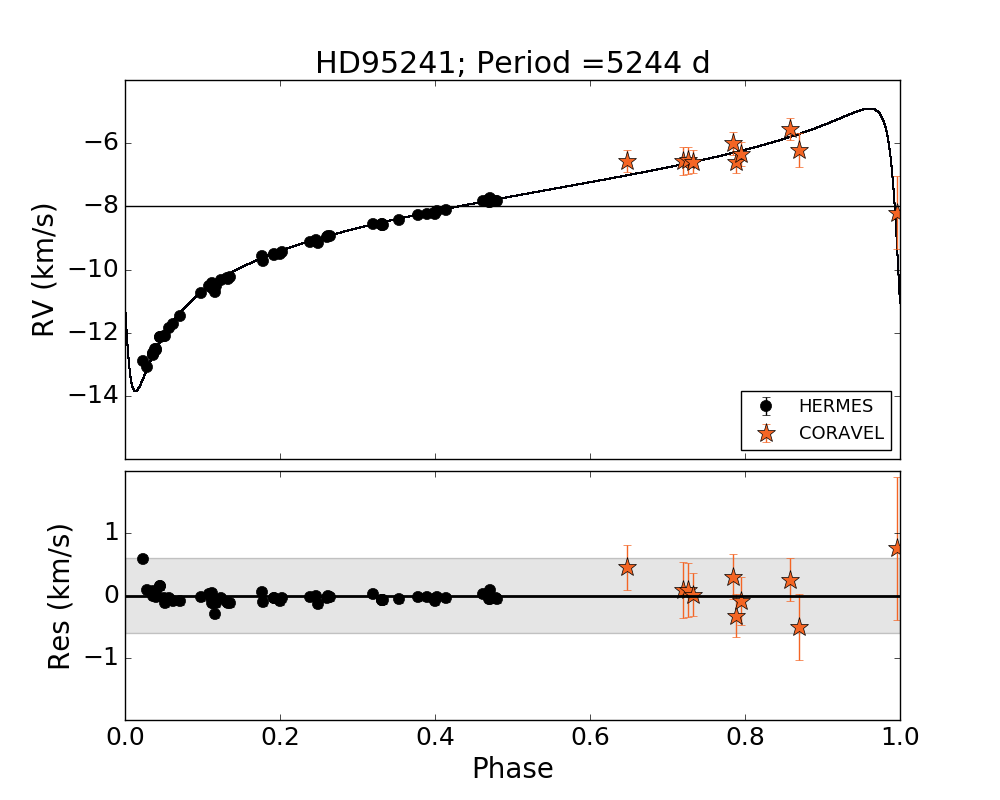}
\caption{\label{HD95241} HD\,95241}
\end{figure}

\begin{figure}[ht]
\centering
\includegraphics[width=0.49\textwidth]{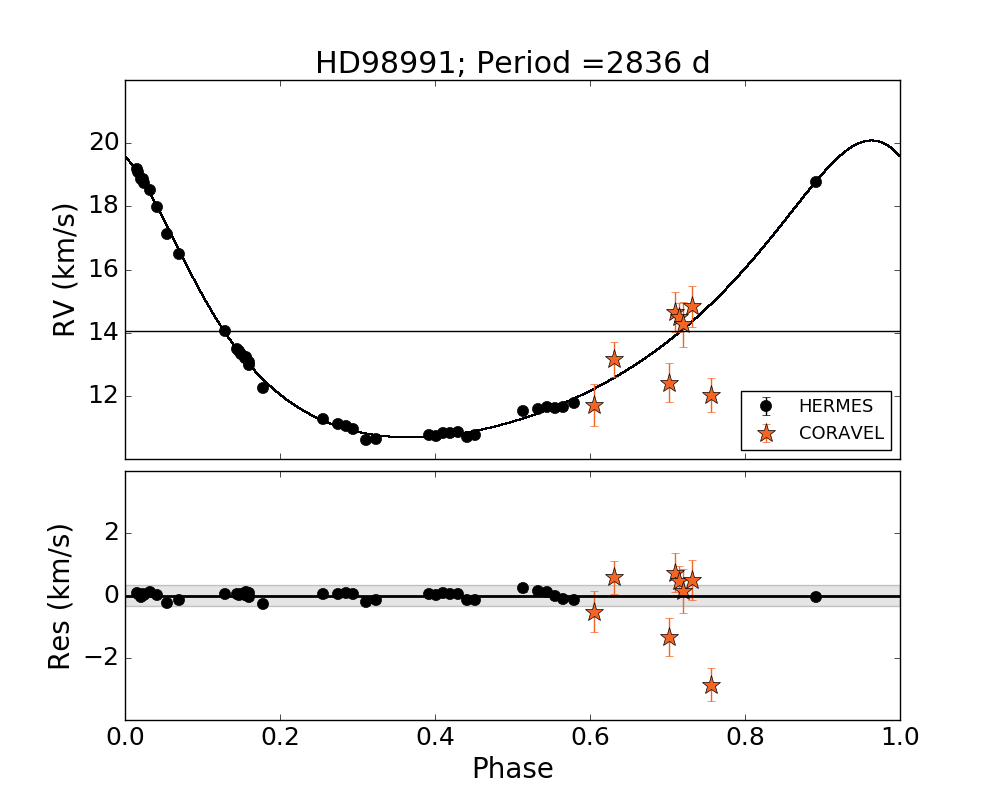}
\caption{\label{HD98991} HD\,98991}
\end{figure}

\begin{figure}[ht]
\centering
\includegraphics[width=0.49\textwidth]{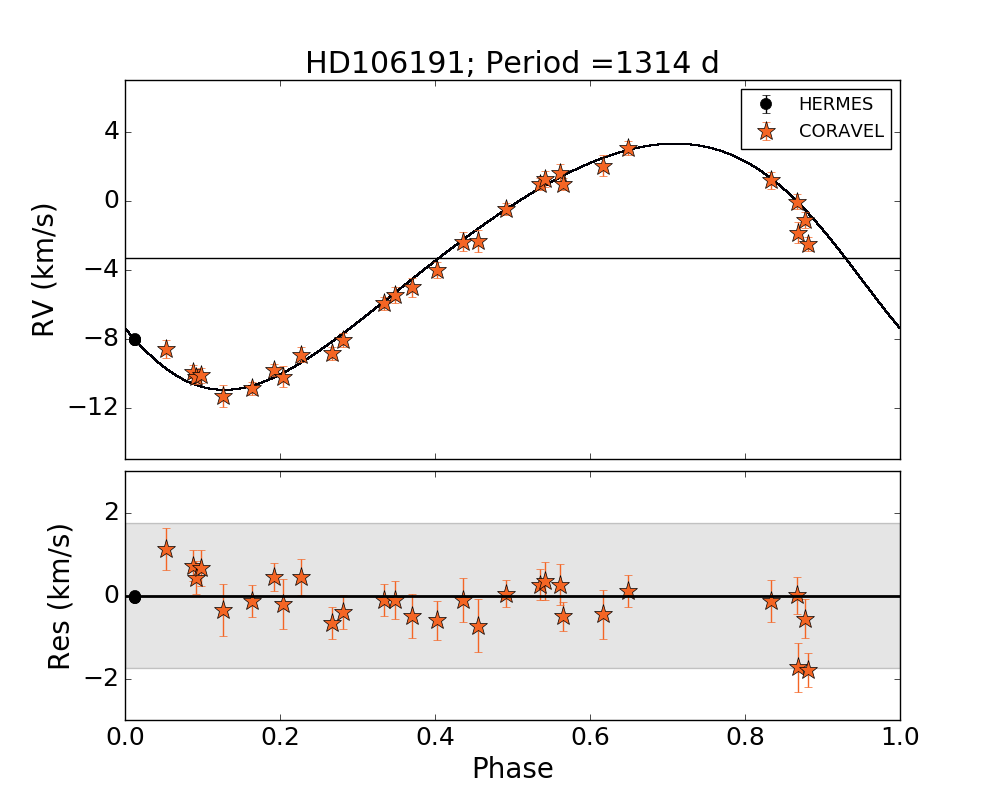}
\caption{\label{HD106191} HD\,106191}
\end{figure}

\begin{figure}[ht]
\centering
\includegraphics[width=0.49\textwidth]{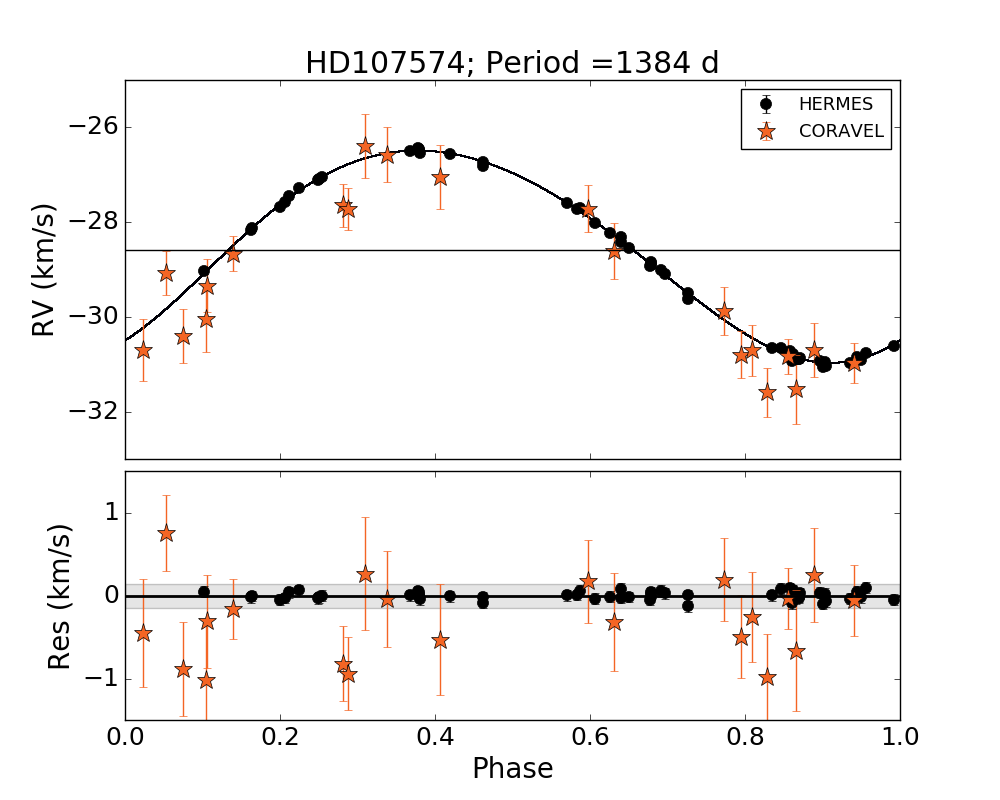}
\caption{\label{HD107574} HD\,107574}
\end{figure}

\begin{figure}[ht]
\centering
\includegraphics[width=0.49\textwidth]{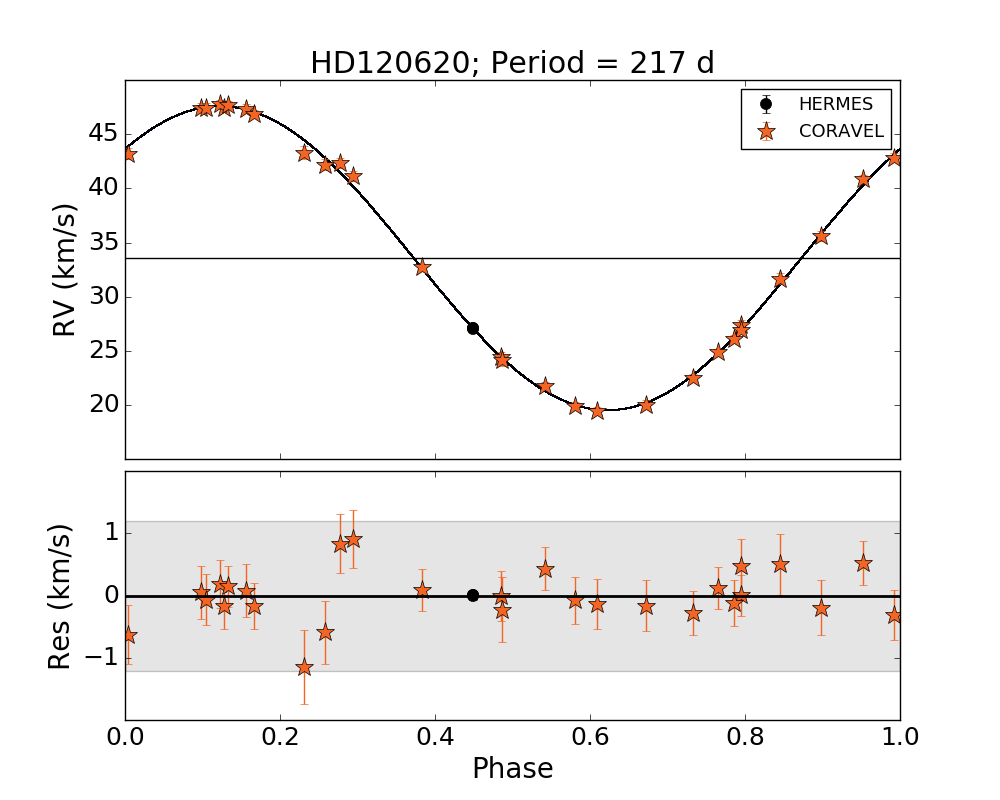}
\caption{\label{HD120620} HD\,120620}
\end{figure}

\begin{figure}[ht]
\centering
\includegraphics[width=0.49\textwidth]{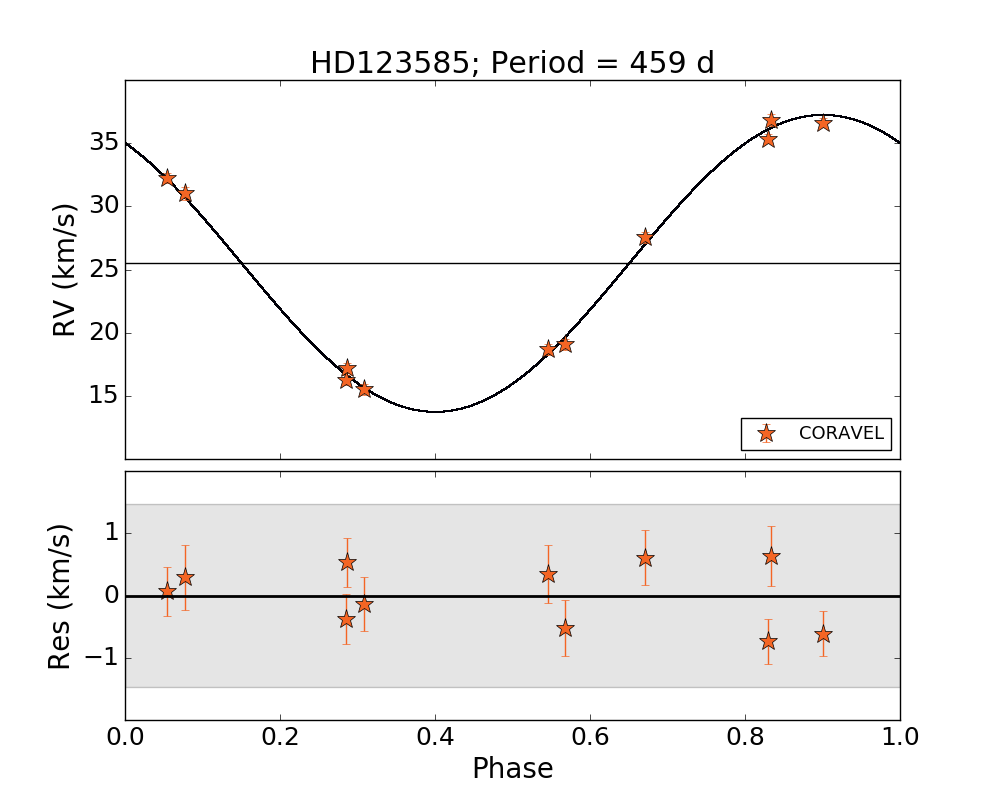}
\caption{\label{HD123585} HD\,123585}
\end{figure}

\begin{figure}[ht]
\centering
\includegraphics[width=0.49\textwidth]{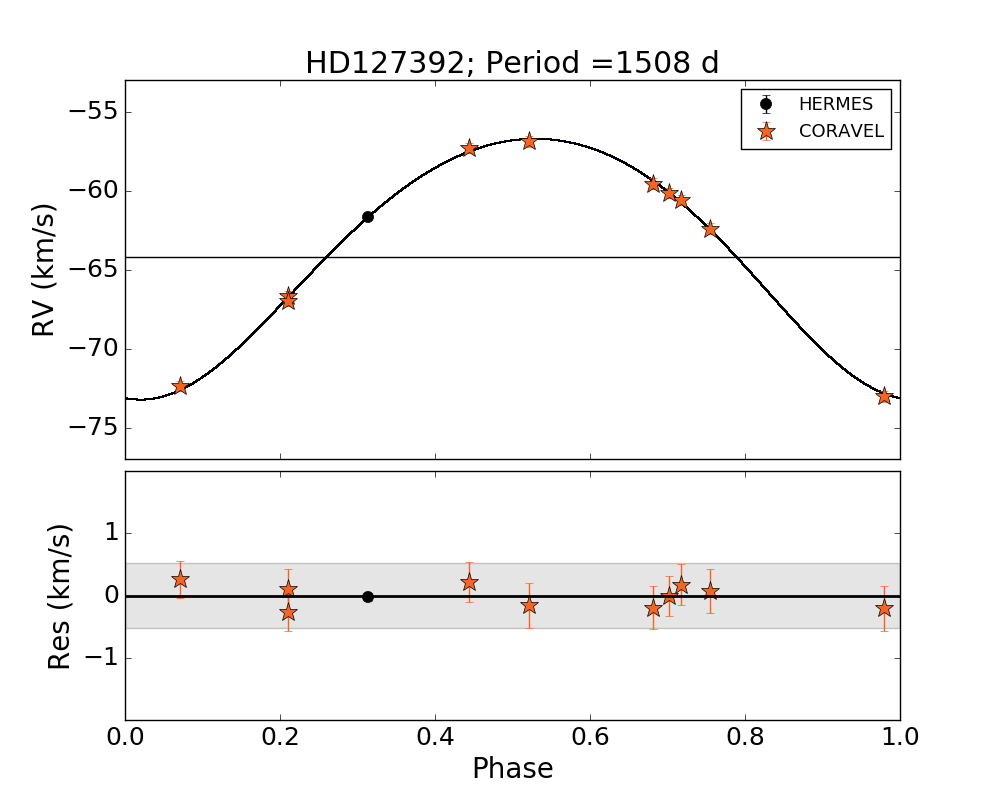}
\caption{\label{HD127392} HD\,127392}
\end{figure}

\begin{figure}[ht]
\centering
\includegraphics[width=0.49\textwidth]{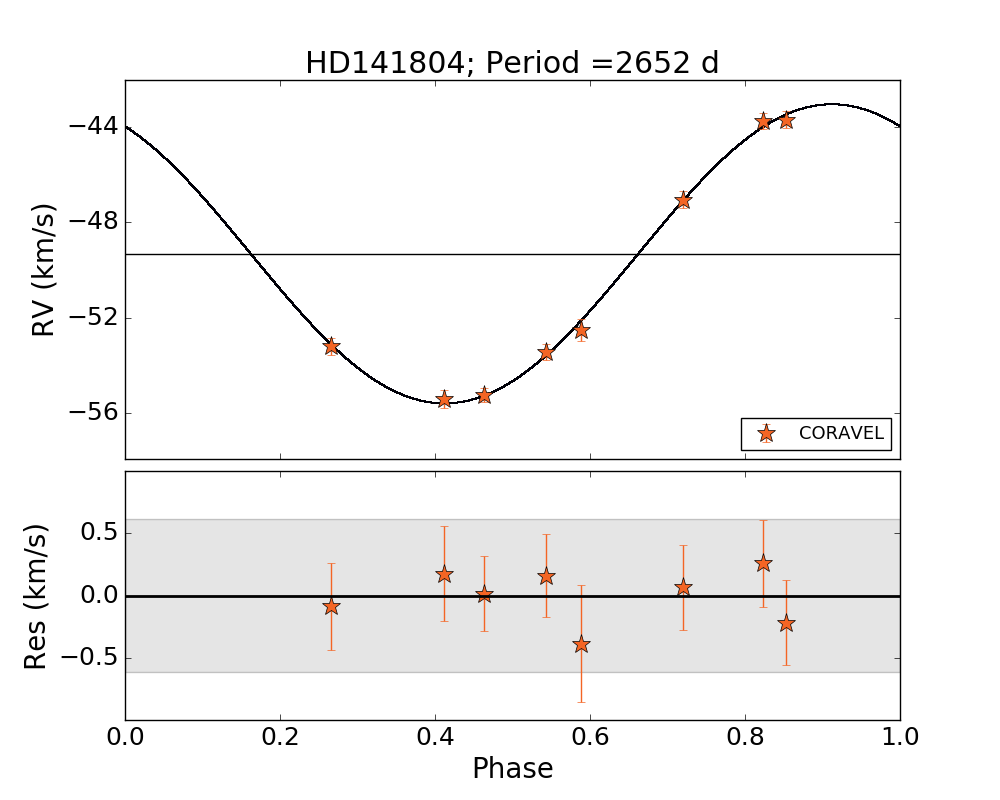}
\caption{\label{HD141804} HD\,141804}
\end{figure}

\begin{figure}[ht]
\centering
\includegraphics[width=0.49\textwidth]{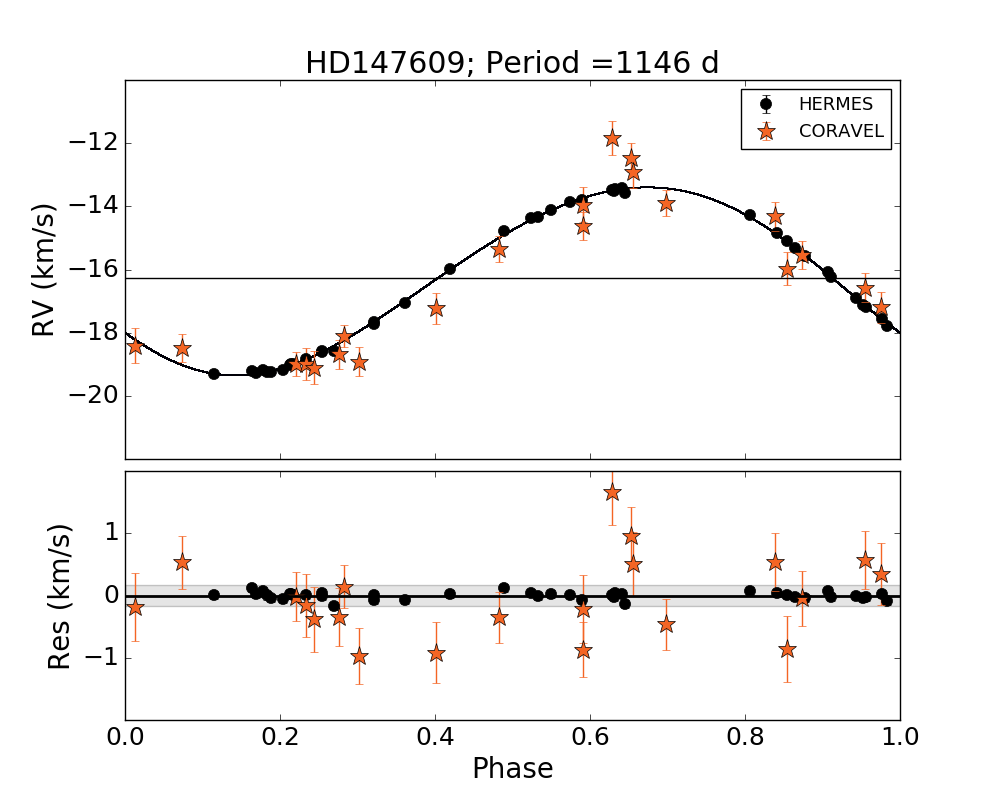}
\caption{\label{HD147609} HD\,147609}
\end{figure}

\begin{figure}[ht]
\centering
\includegraphics[width=0.49\textwidth]{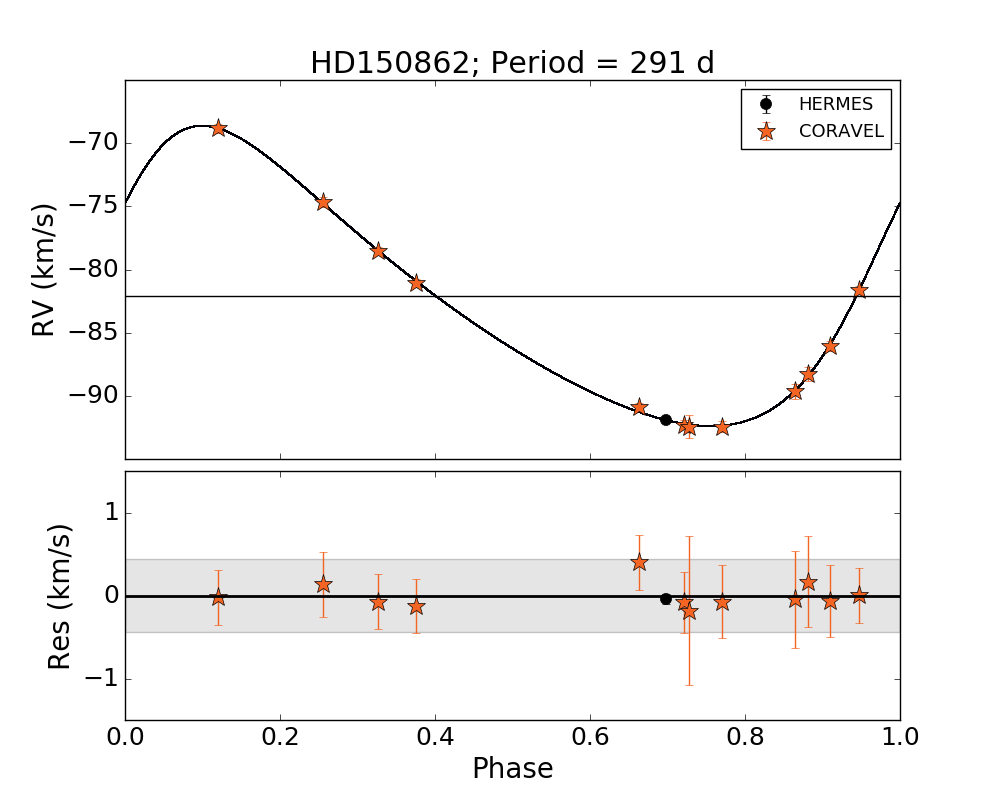}
\caption{\label{HD150862} HD\,150862}
\end{figure}

\begin{figure}[ht]
\centering
\includegraphics[width=0.49\textwidth]{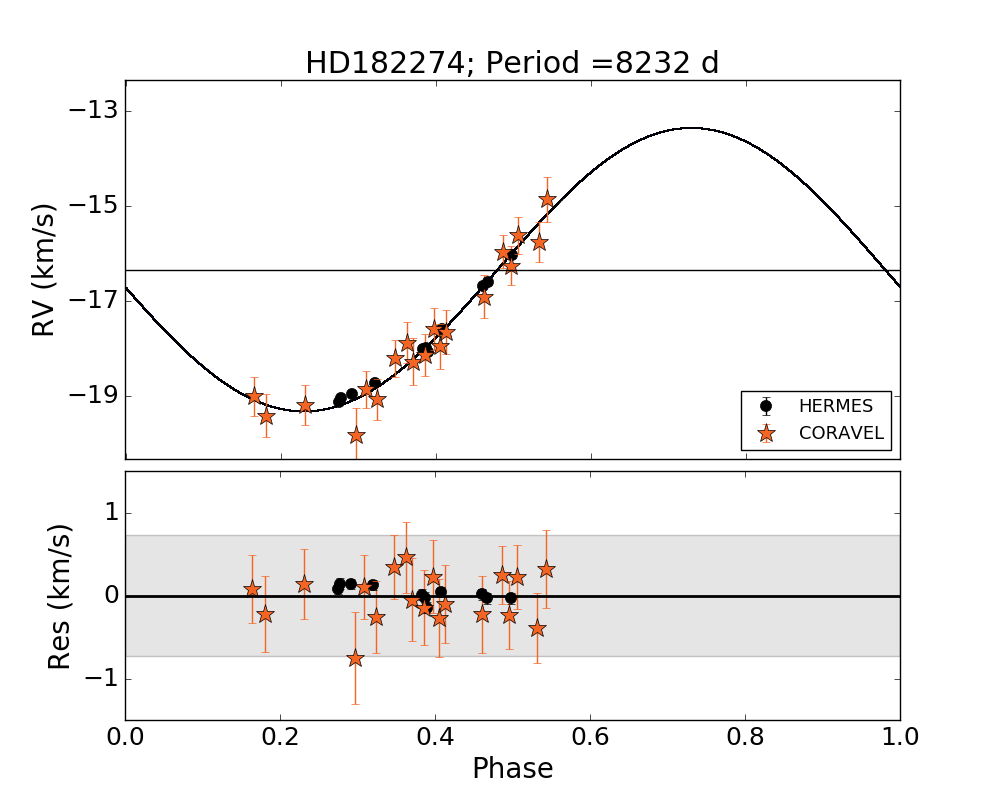}
\caption{\label{HD182274} HD\,182274}
\end{figure}

\begin{figure}[ht]
\centering
\includegraphics[width=0.49\textwidth]{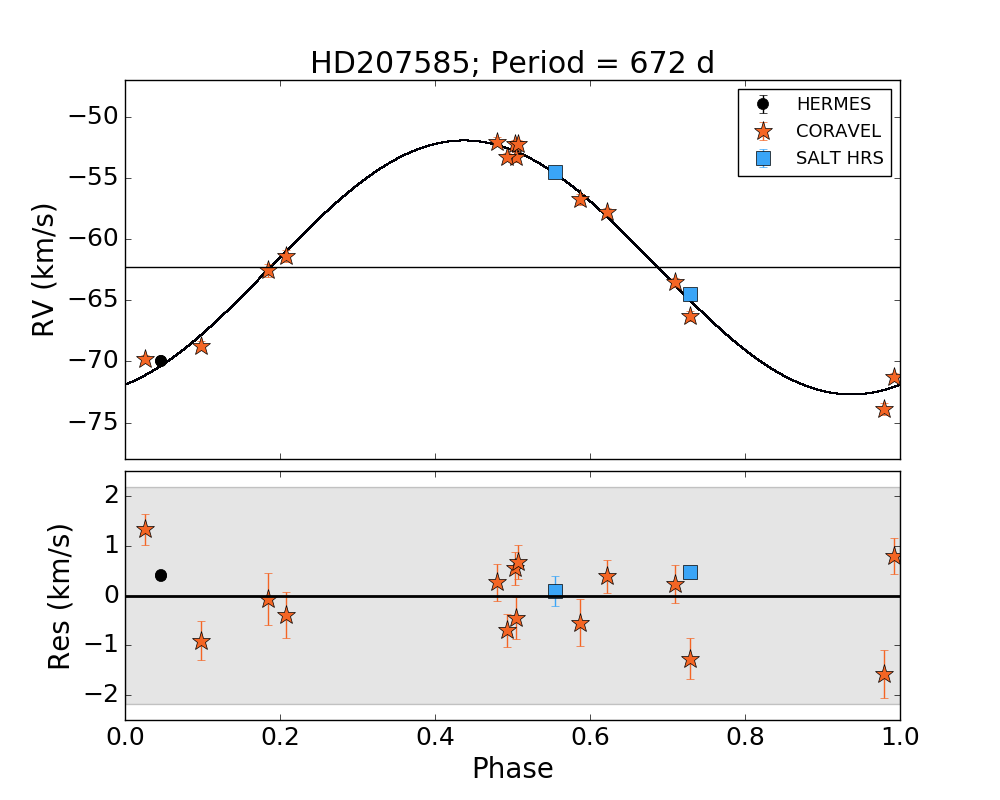}
\caption{\label{HD207585} HD\,207585}
\end{figure}

\begin{figure}[ht]
\centering
\includegraphics[width=0.49\textwidth]{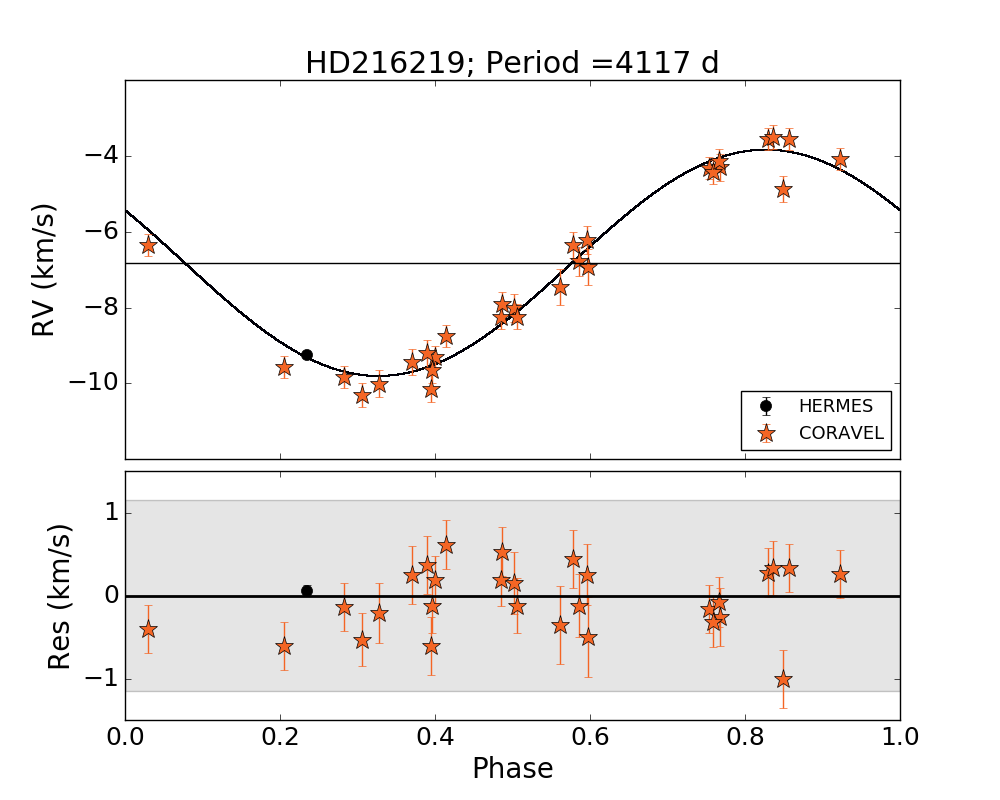}
\caption{\label{HD216219} HD\,216219}
\end{figure}

\begin{figure}[ht]
\centering
\includegraphics[width=0.49\textwidth]{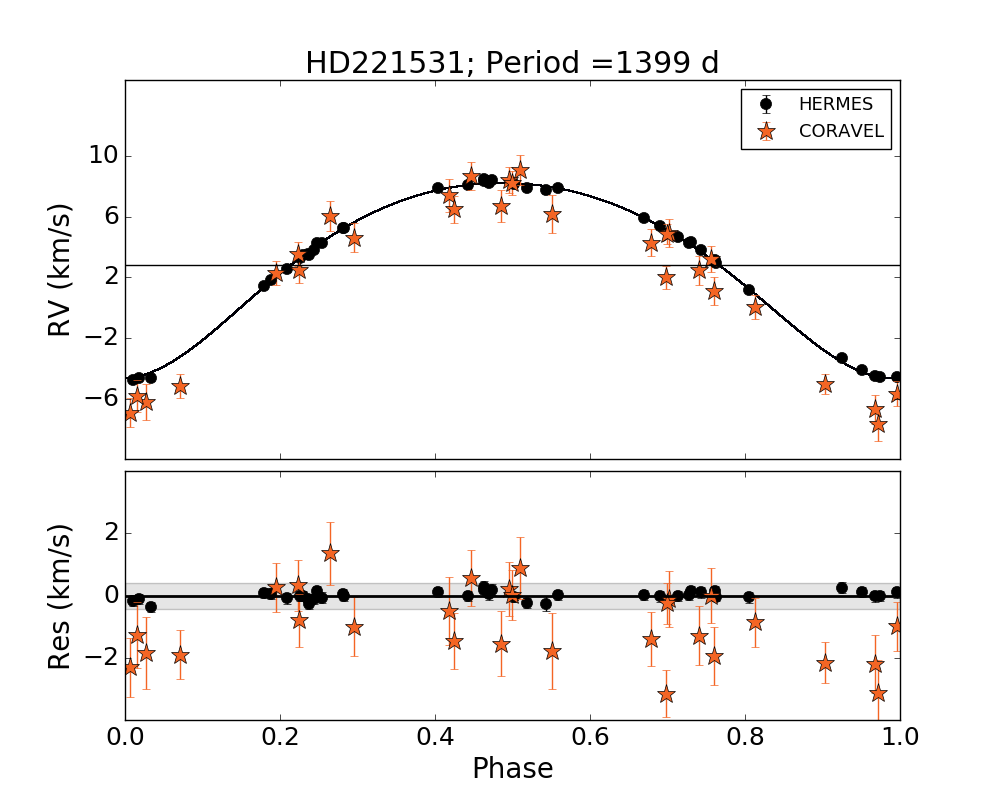}
\caption{\label{HD221531} HD\,221531}
\end{figure}

Figures from \ref{BD-10_4311} to \ref{HD221531} show the RV curves and the best-fitting solutions of the 25 new orbits obtained in this paper. The upper panel of each graph includes, when available for each specific target, HERMES (black dots), CORAVEL (orange stars), CORALIE (purple triangles), and SALT (blue squares) radial velocity data and the best fitting model of the orbit (black solid line). The lower panel includes the O-C residuals of the fit and a shadowed region which corresponds to three times the standard deviation of the residuals.

\begin{figure*}[b]
\includegraphics[width=0.33\textwidth,clip]{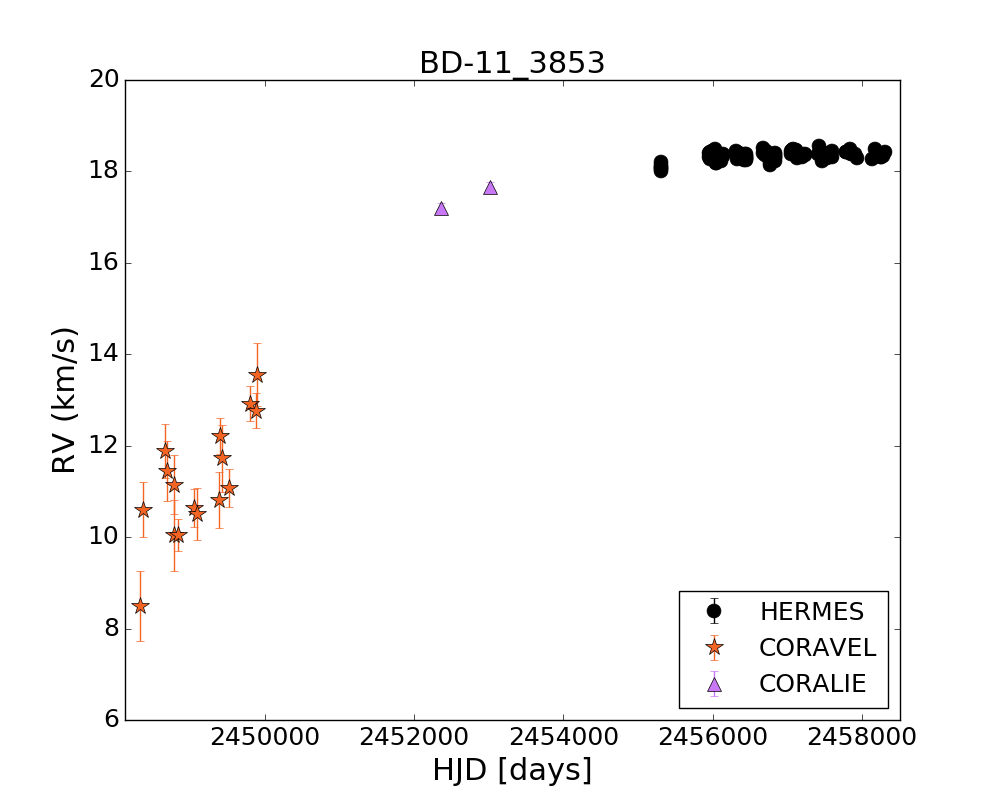}
\includegraphics[width=0.33\textwidth,clip]{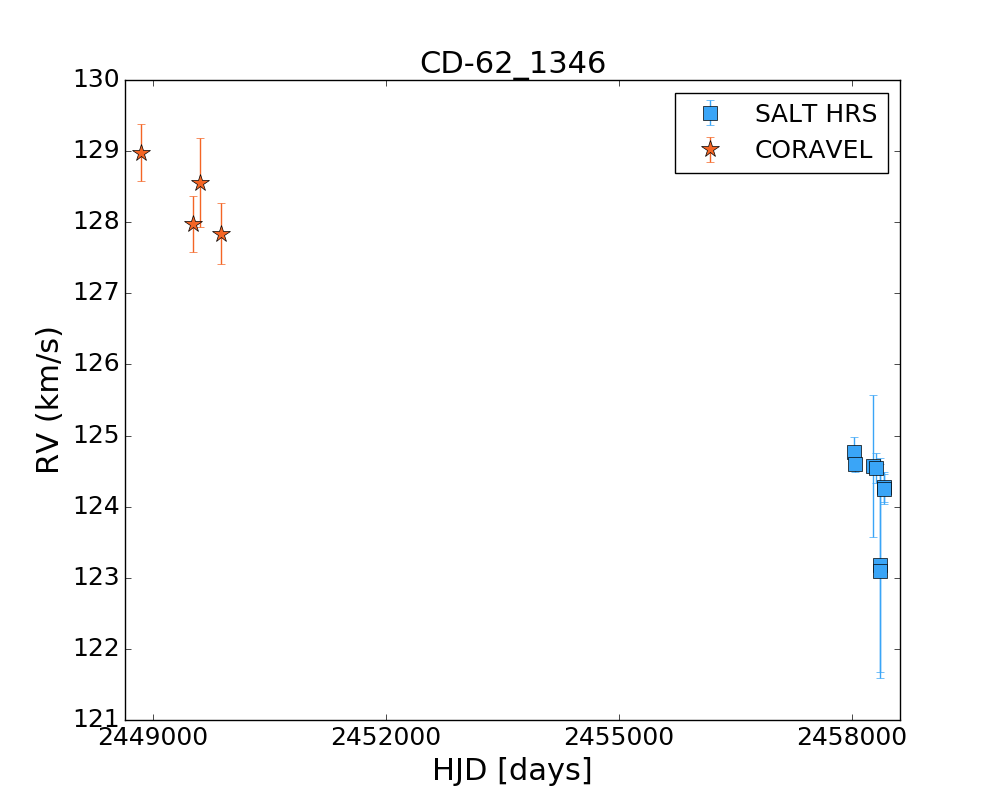}
\includegraphics[width=0.33\textwidth,clip]{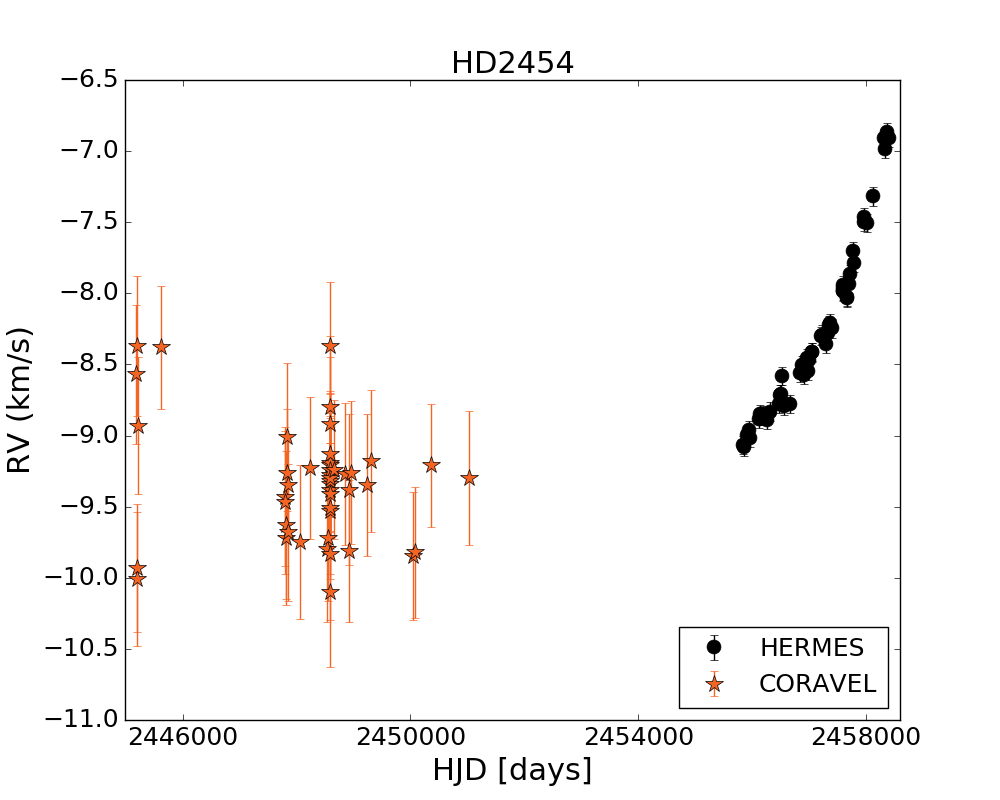}
\includegraphics[width=0.33\textwidth,clip]{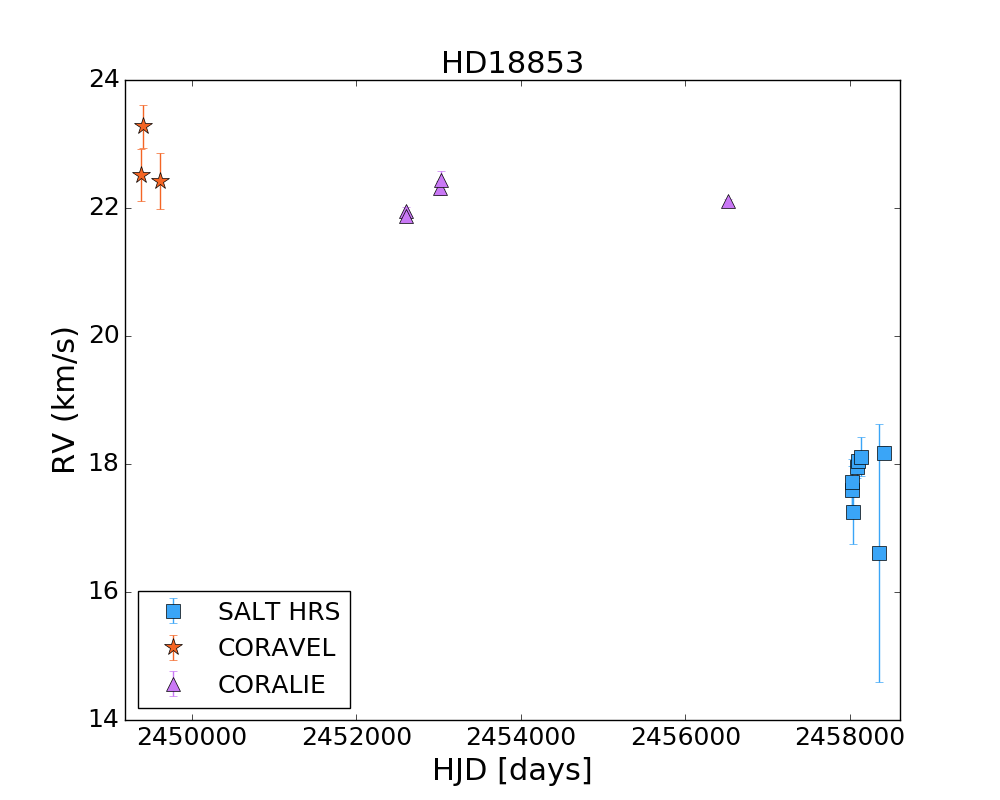}
\includegraphics[width=0.33\textwidth,clip]{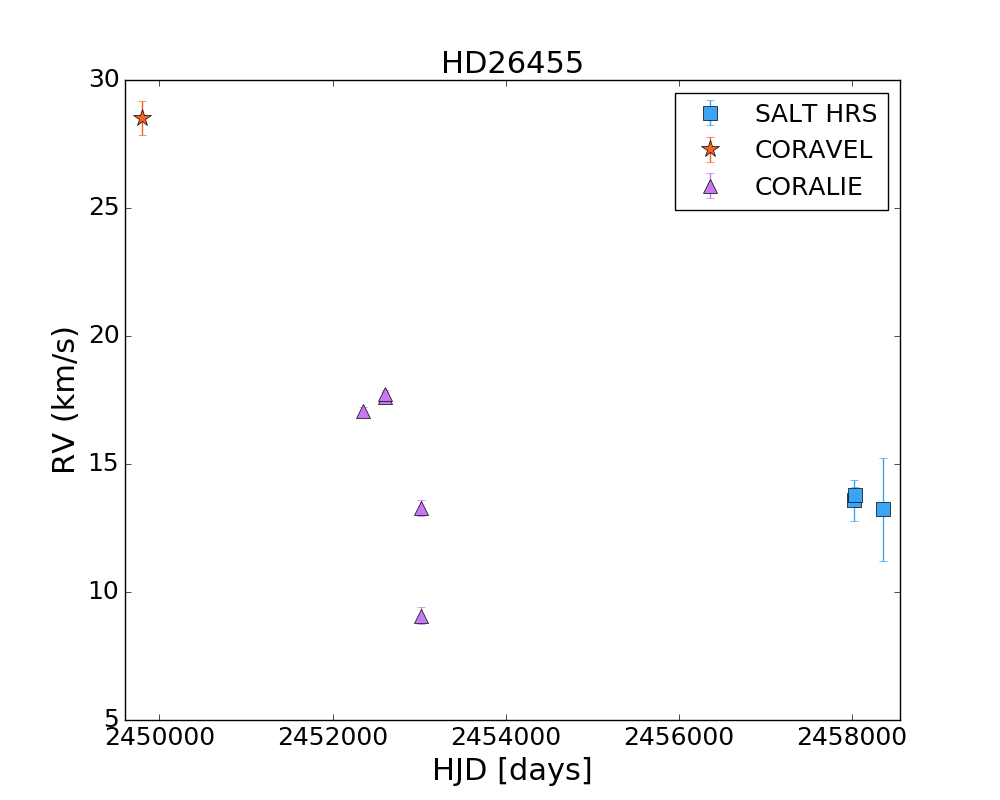}
\includegraphics[width=0.33\textwidth,clip]{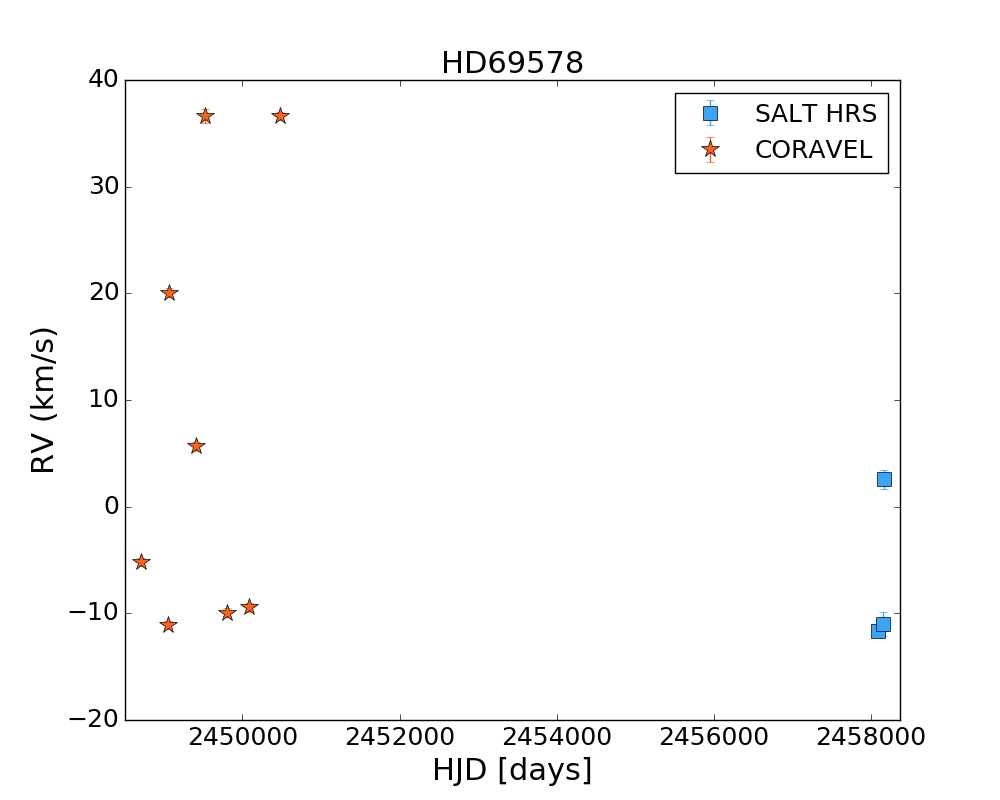}
\includegraphics[width=0.33\textwidth,clip]{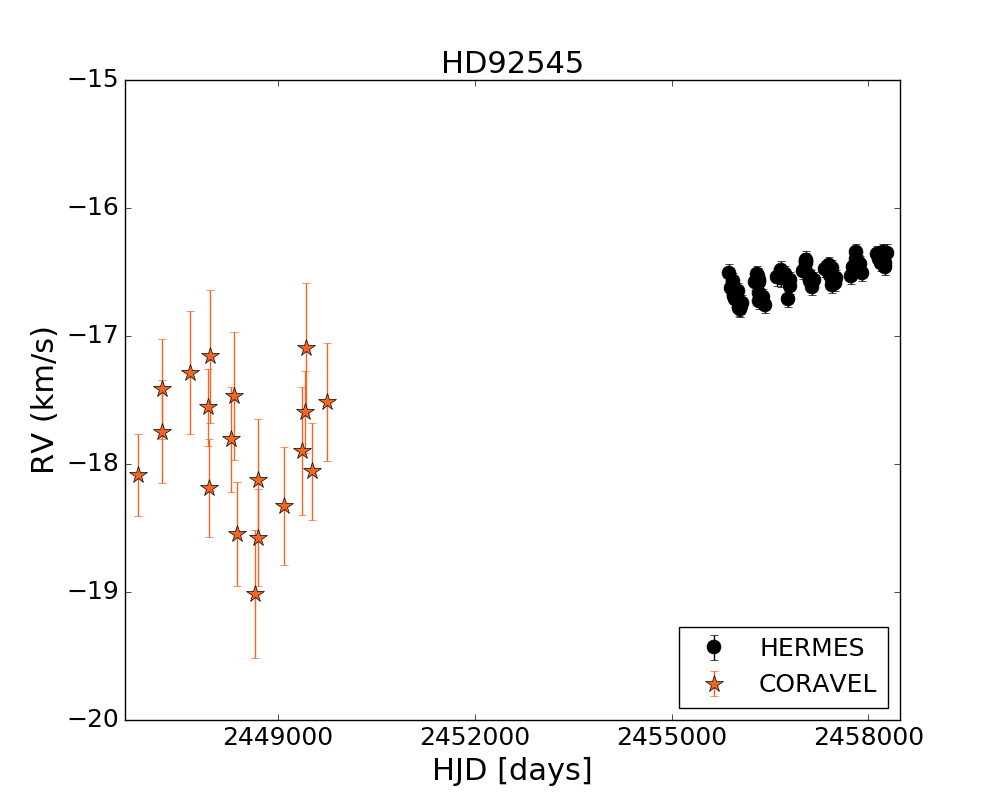}
\includegraphics[width=0.33\textwidth,clip]{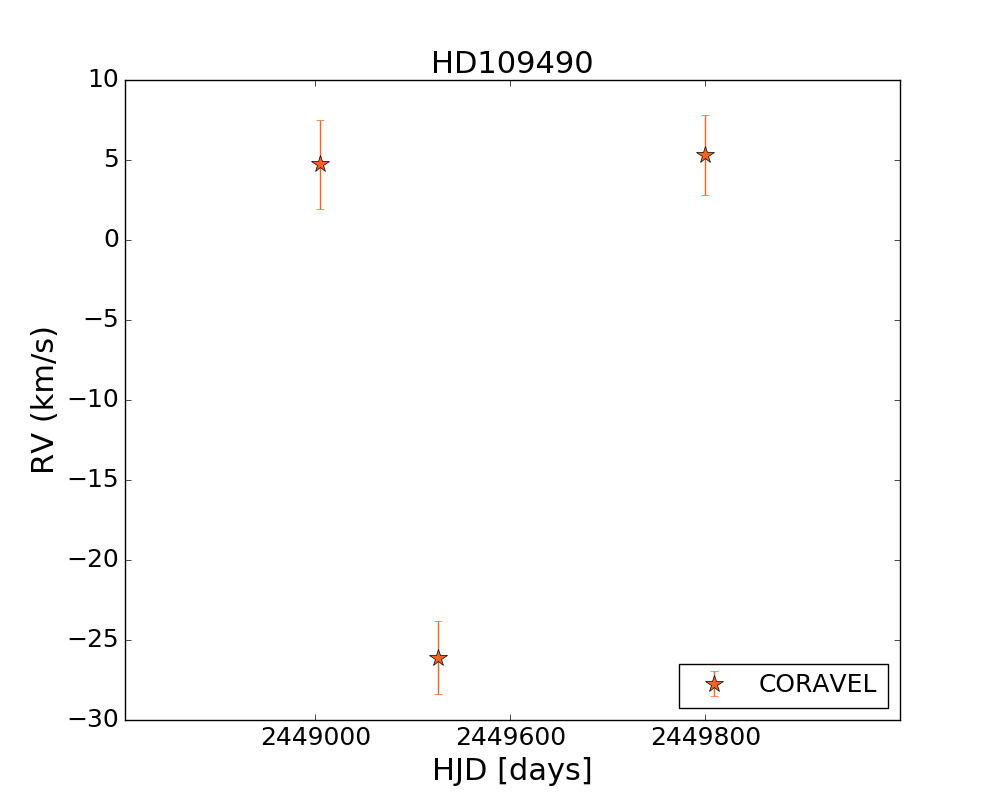}
\includegraphics[width=0.33\textwidth,clip]{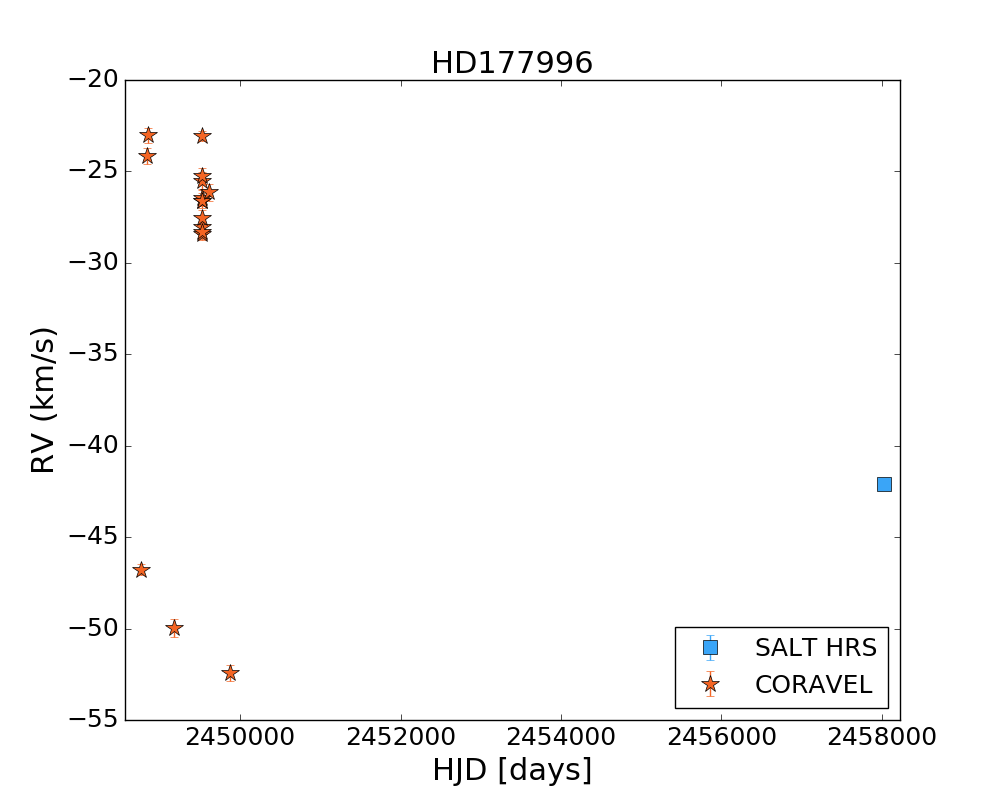}
\includegraphics[width=0.33\textwidth,clip]{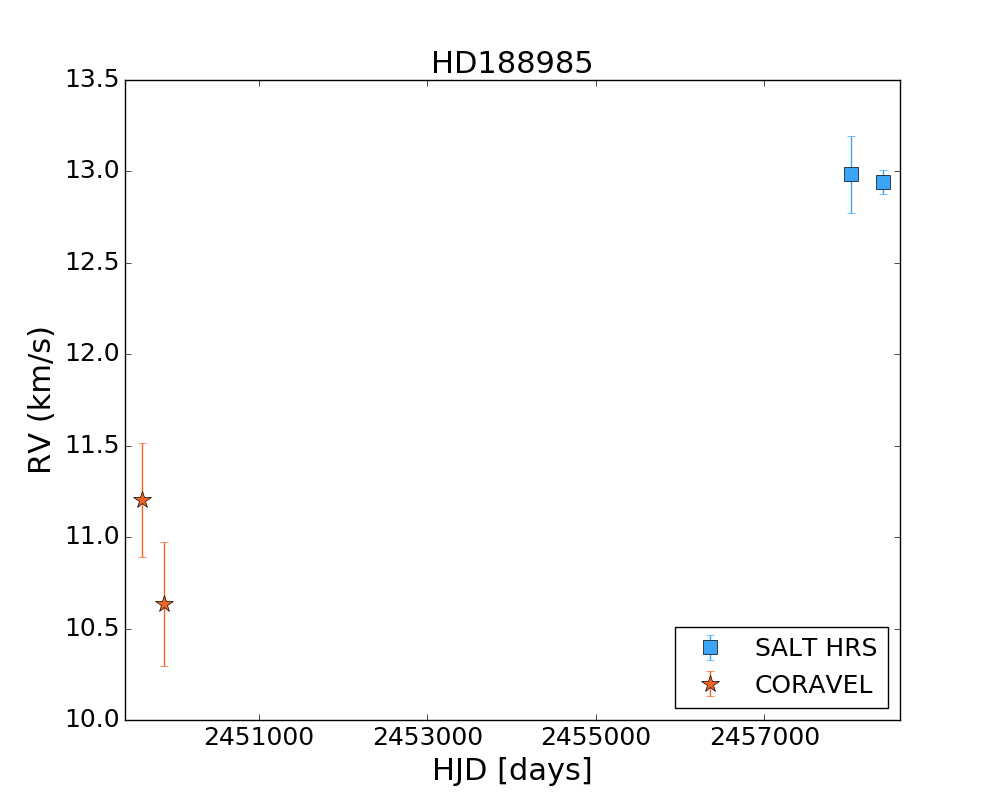}
\includegraphics[width=0.33\textwidth,clip]{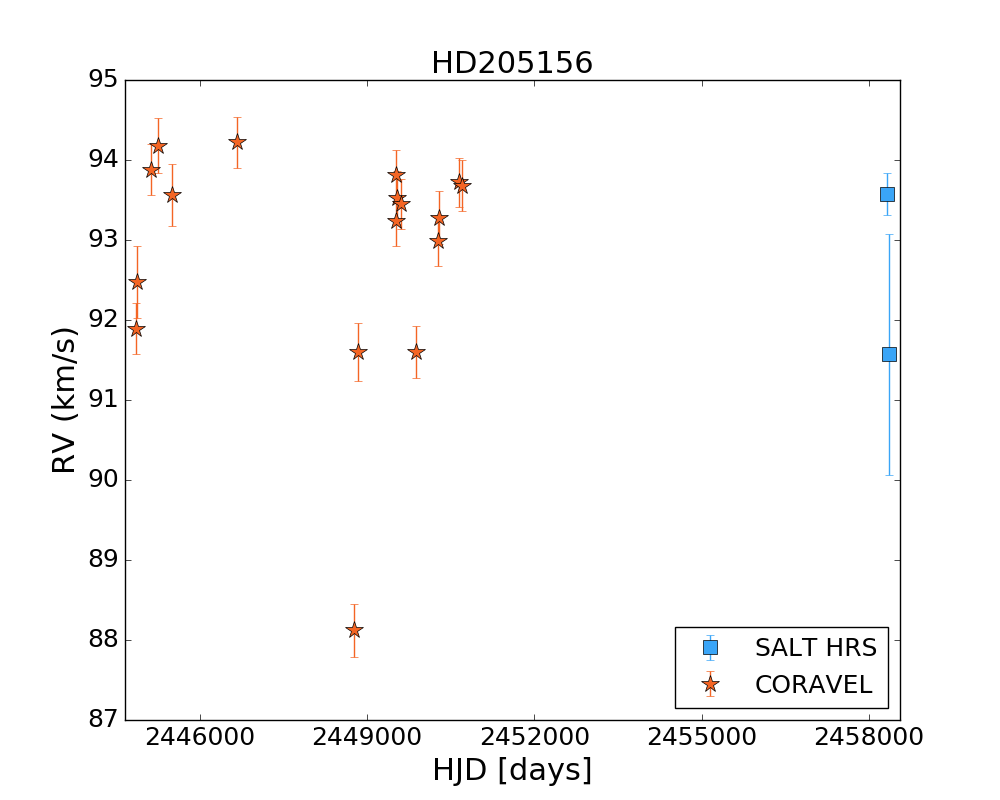}
\includegraphics[width=0.33\textwidth,clip]{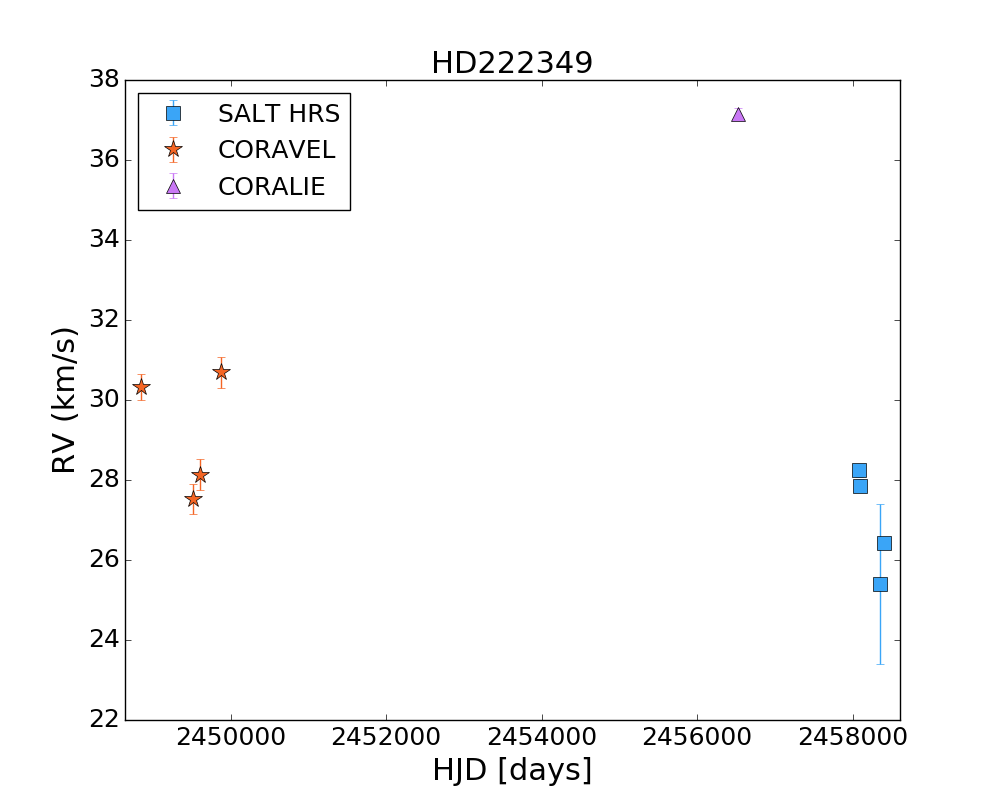}
\includegraphics[width=0.33\textwidth,clip]{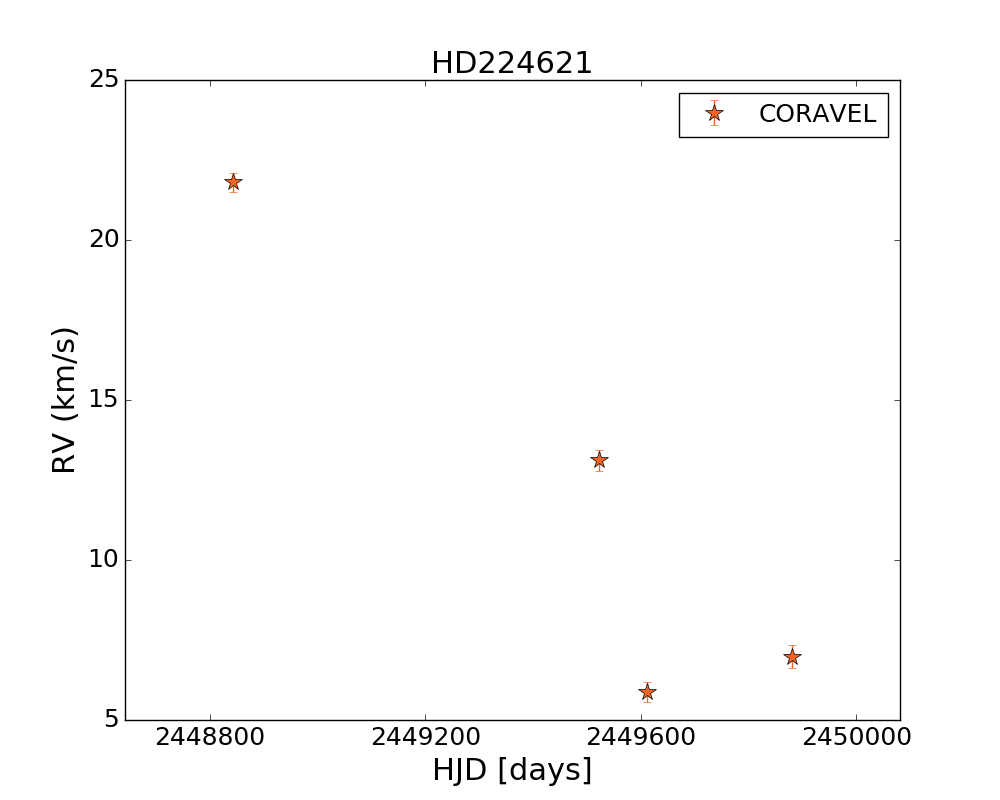}
\caption{\label{Fig:dBauncovered} HERMES, CORAVEL, CORALIE and SALT radial velocity data of SB with no orbit yet available}
\end{figure*}

Figure \ref{Fig:dBauncovered} shows the available data for the objects that, according to the criterion explained and applied in Sect. \ref{ssec:others}, are in binary systems even though we can not constrain their orbital element. Finally, Fig. \ref{Fig:dBaothers} collects the RV data of all the remaining objects. For some of them, we simply do not have enough data to be conclusive, while others are probably not binaries.

\clearpage

\clearpage

\begin{figure*}[!ht]
\includegraphics[width=0.33\textwidth,clip]{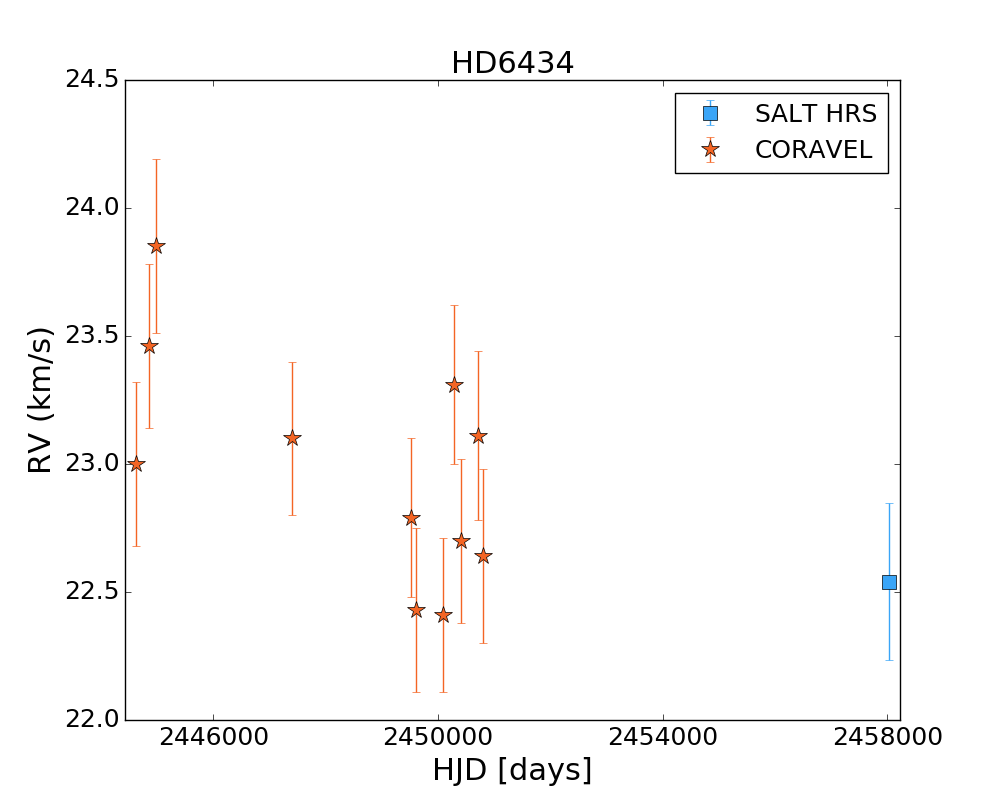}
\includegraphics[width=0.33\textwidth,clip]{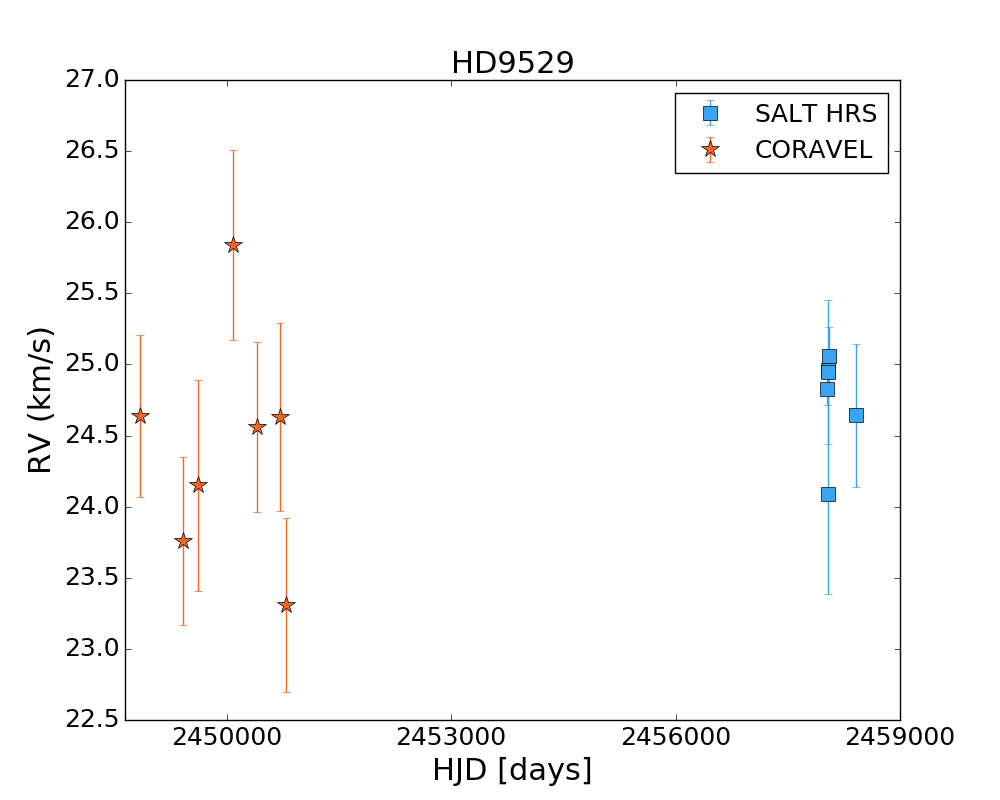}
\includegraphics[width=0.33\textwidth,clip]{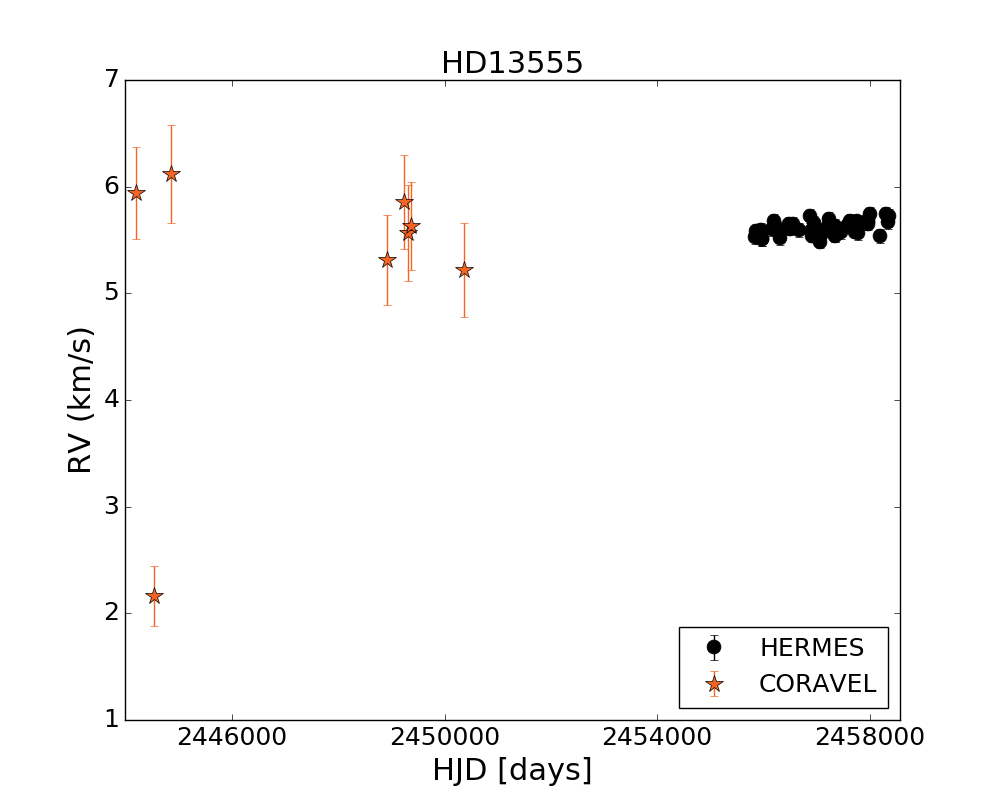}
\includegraphics[width=0.33\textwidth,clip]{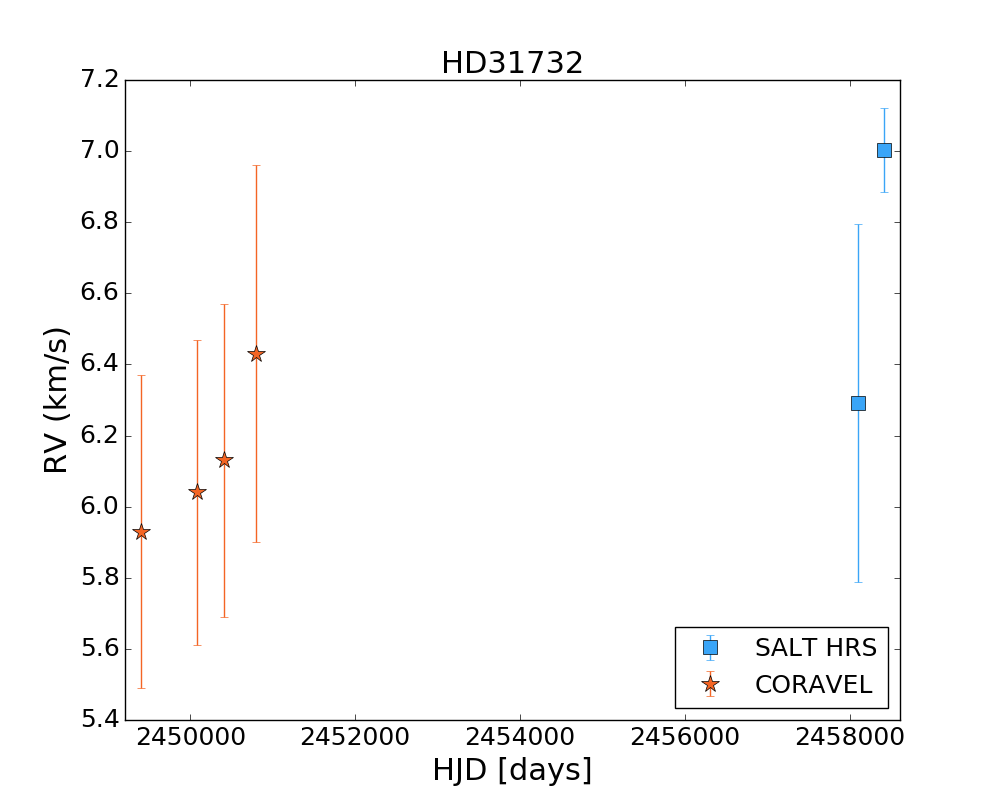}
\includegraphics[width=0.33\textwidth,clip]{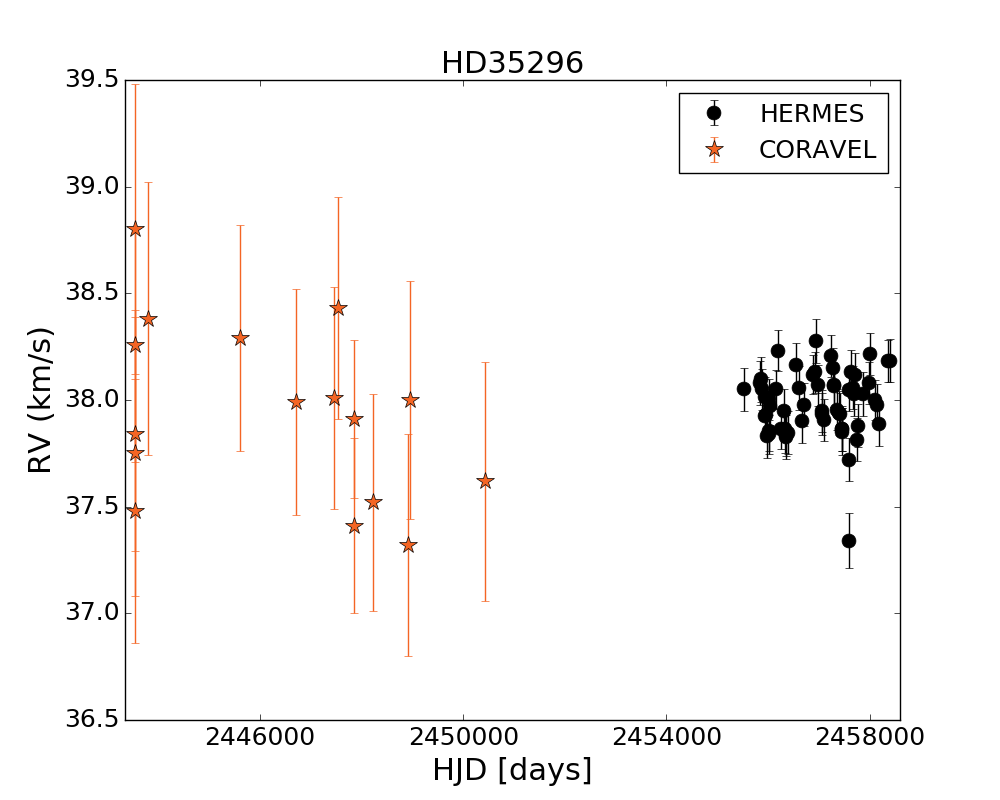}
\includegraphics[width=0.33\textwidth,clip]{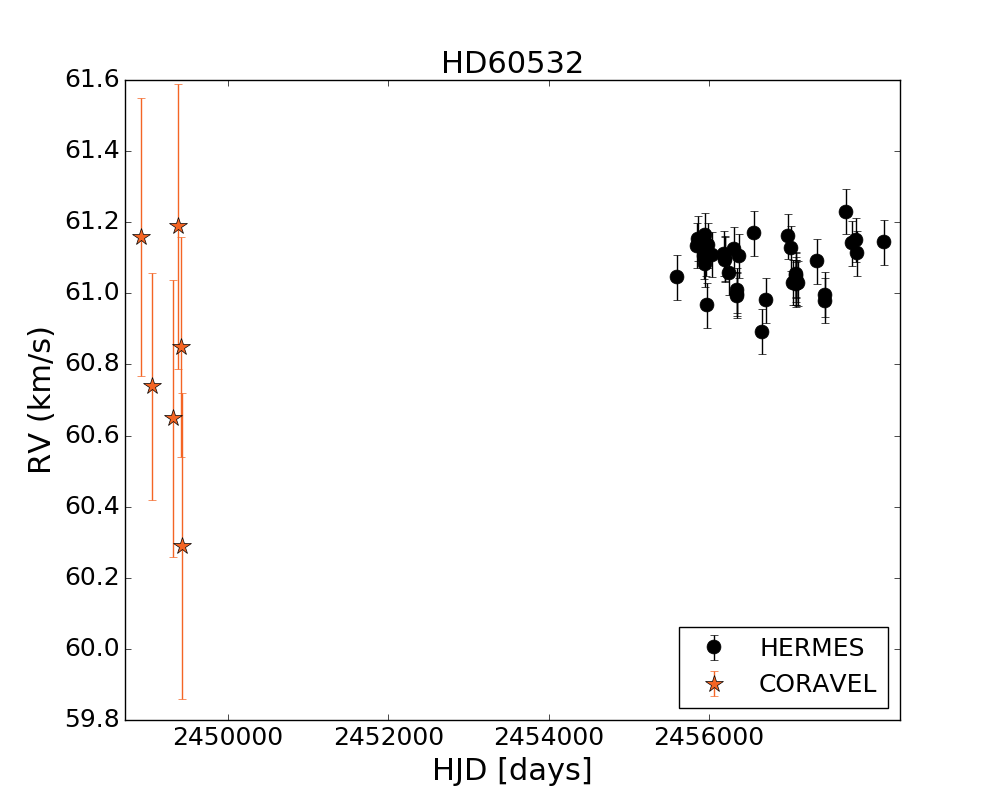}
\includegraphics[width=0.33\textwidth,clip]{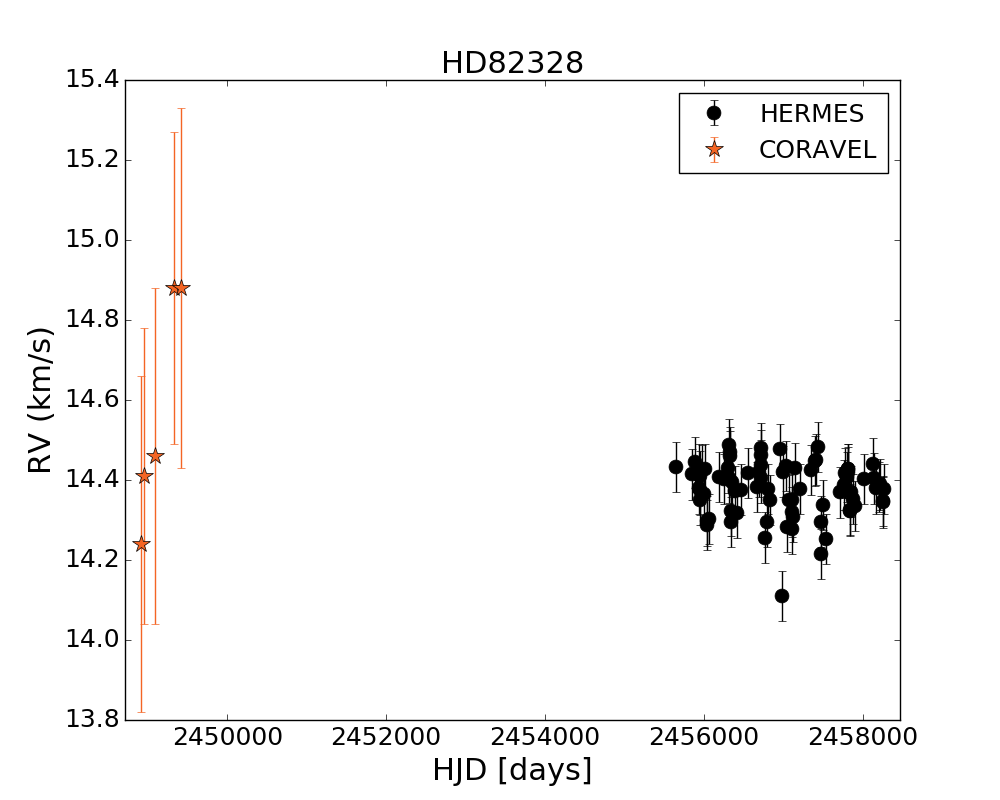}
\includegraphics[width=0.33\textwidth,clip]{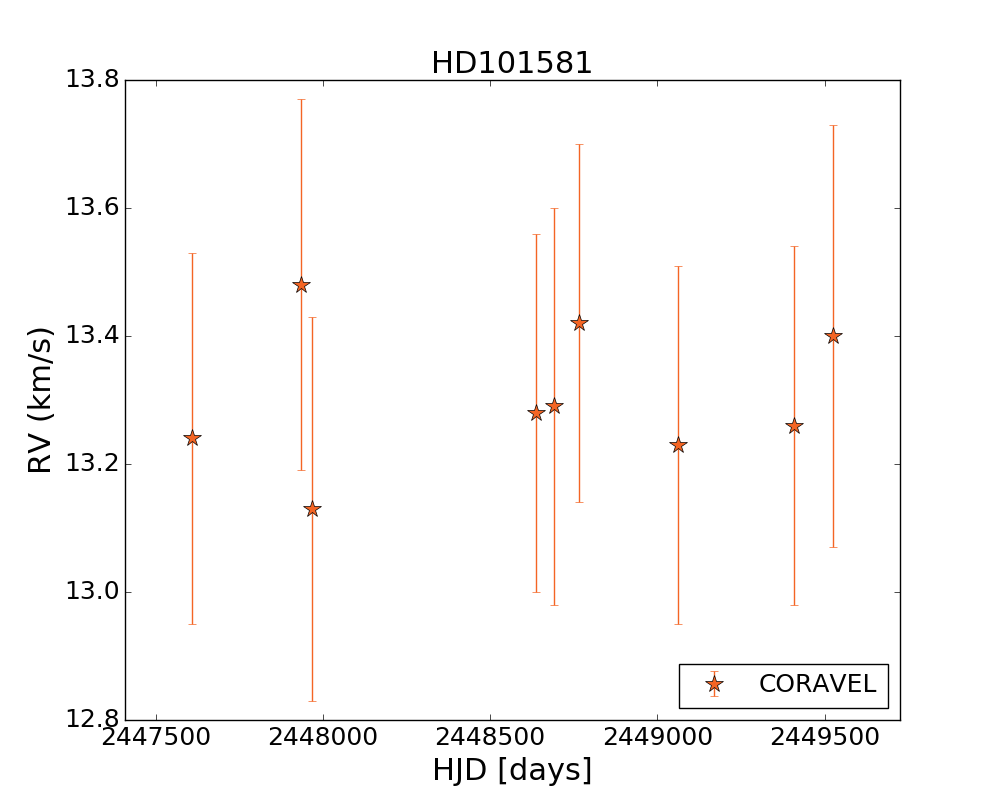}
\includegraphics[width=0.33\textwidth,clip]{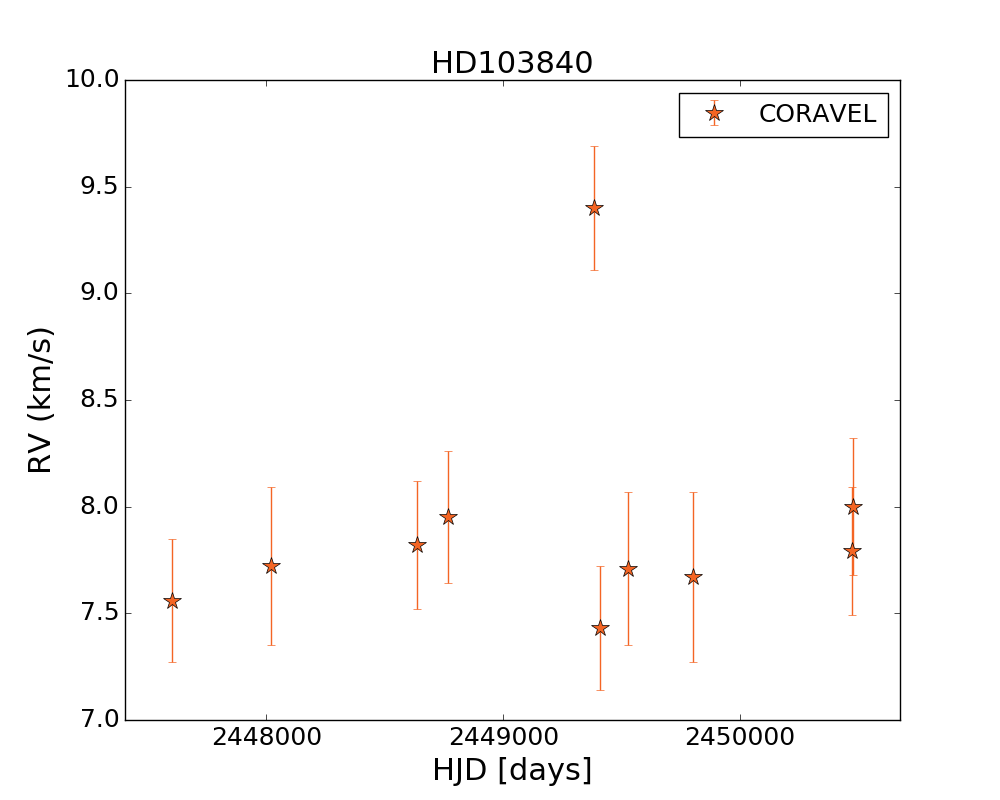}
\includegraphics[width=0.33\textwidth,clip]{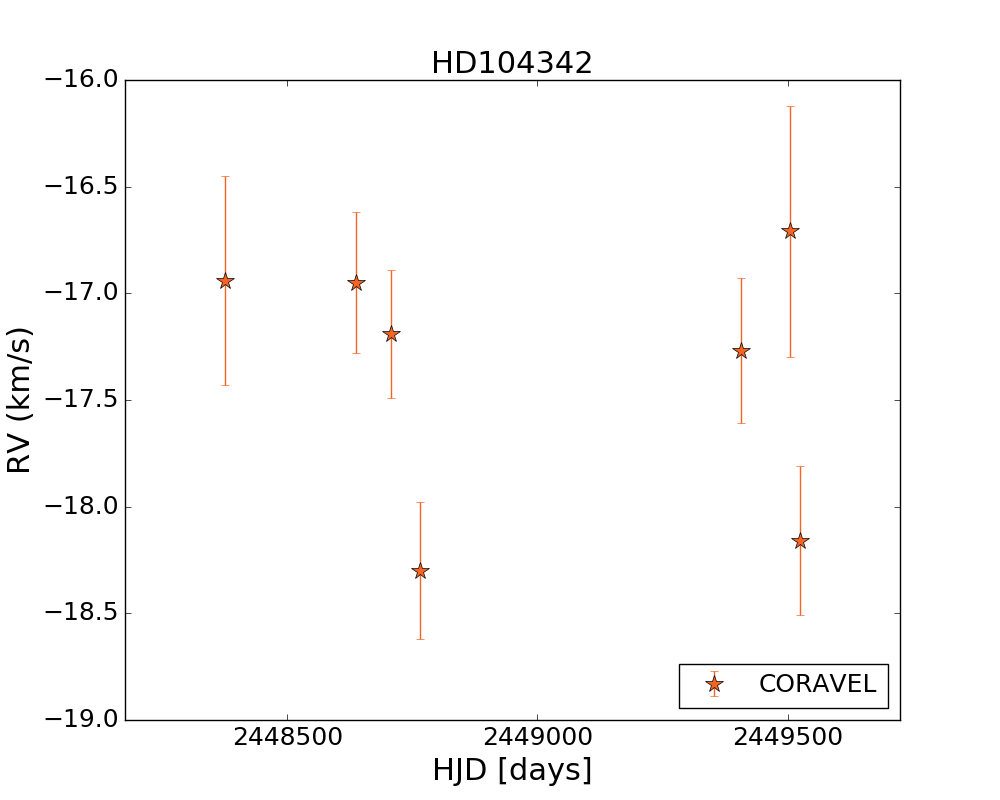}
\includegraphics[width=0.33\textwidth,clip]{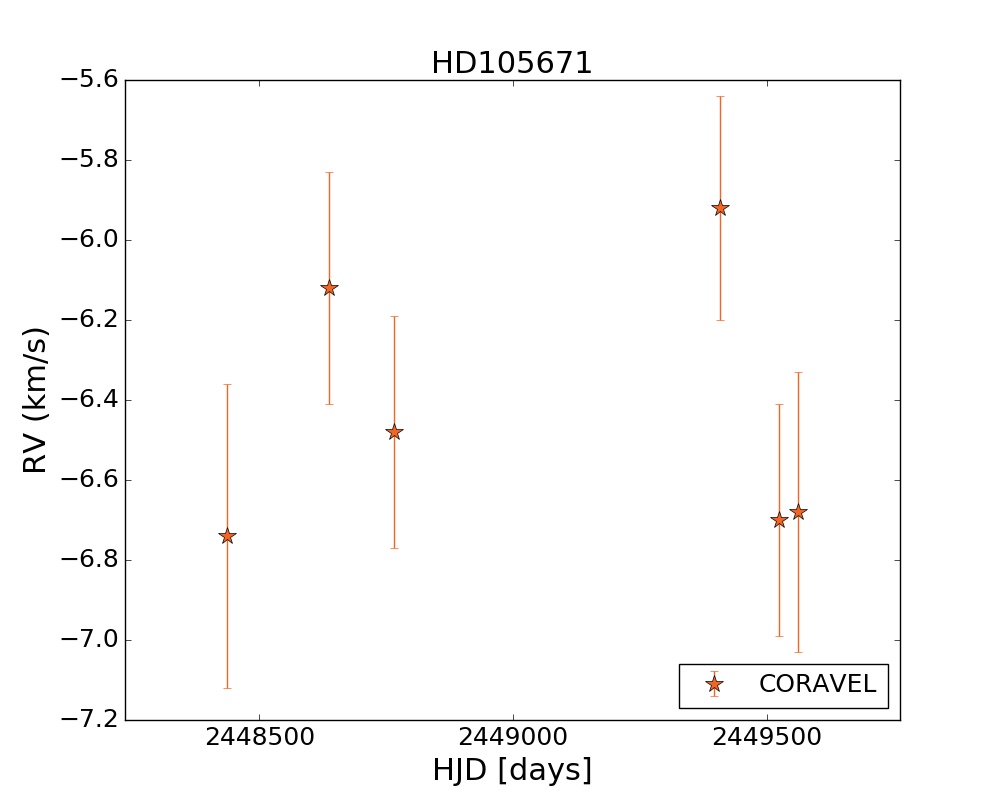}
\includegraphics[width=0.33\textwidth,clip]{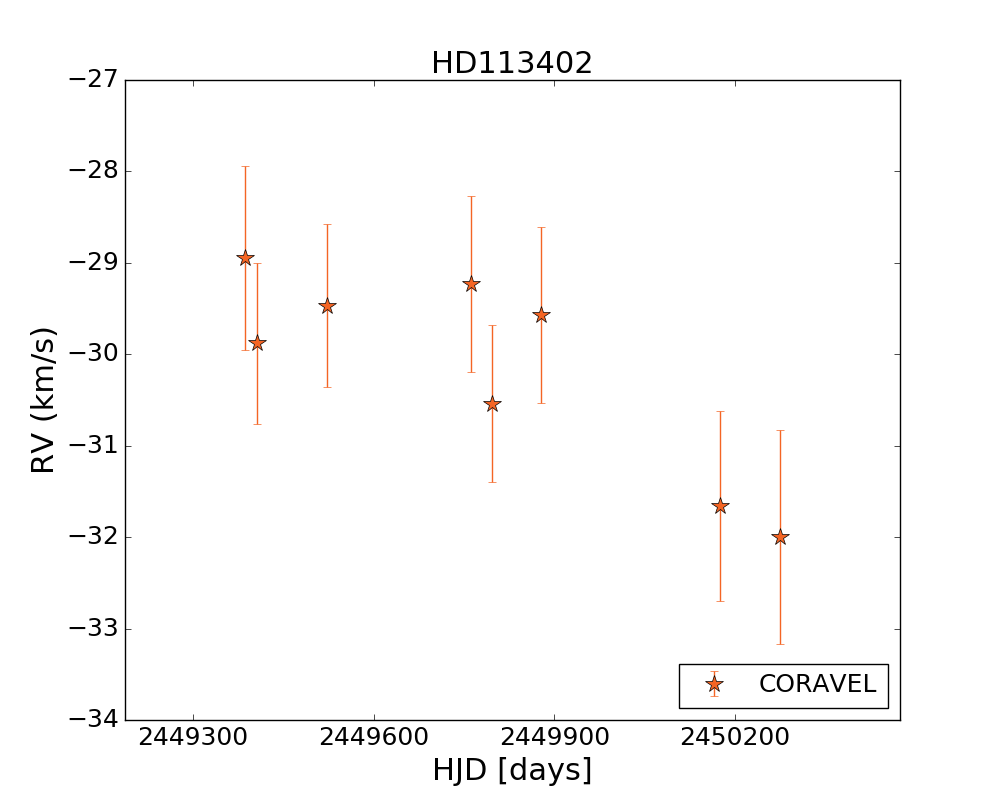}
\includegraphics[width=0.33\textwidth,clip]{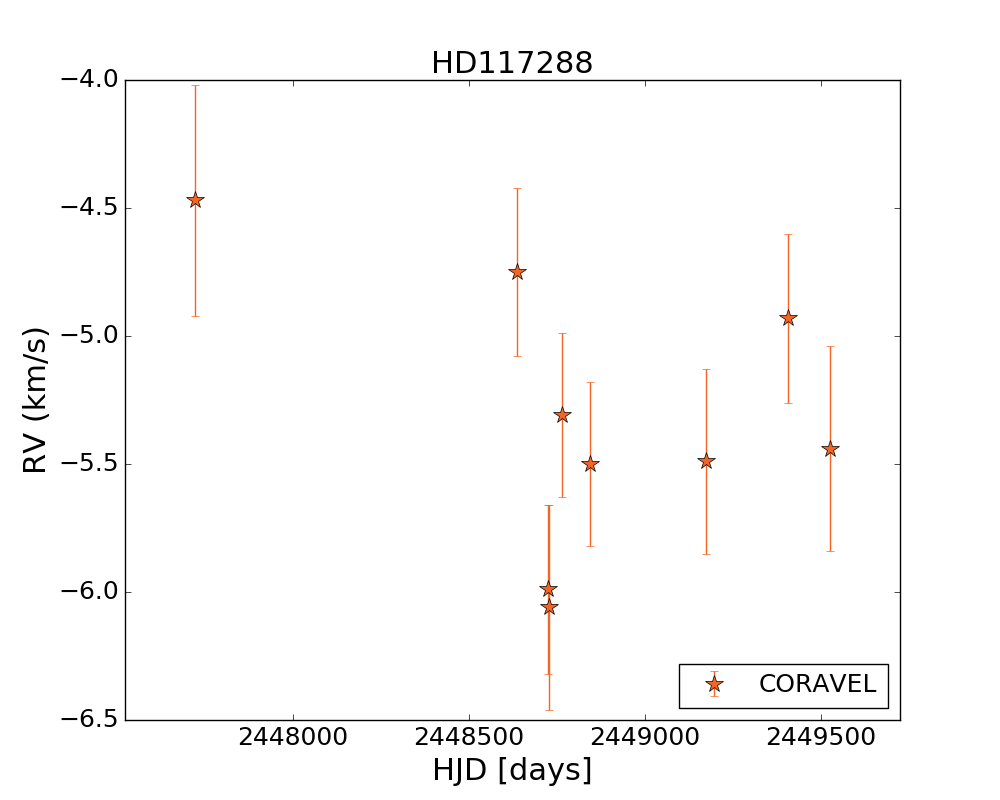}
\includegraphics[width=0.33\textwidth,clip]{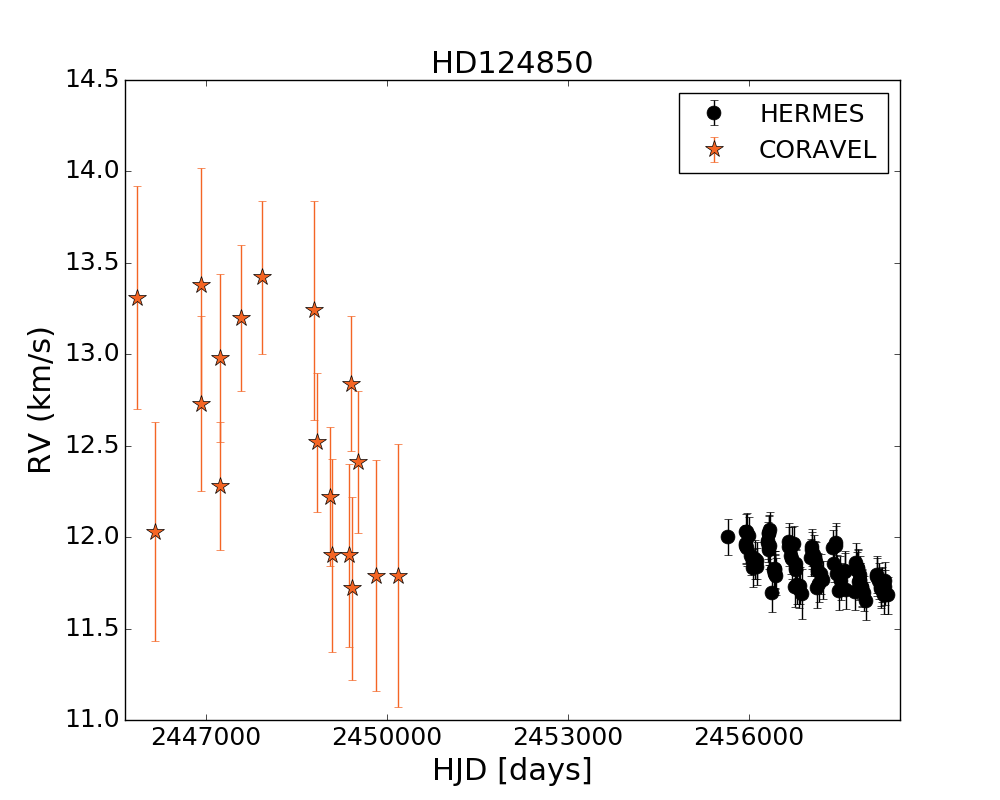}
\includegraphics[width=0.33\textwidth,clip]{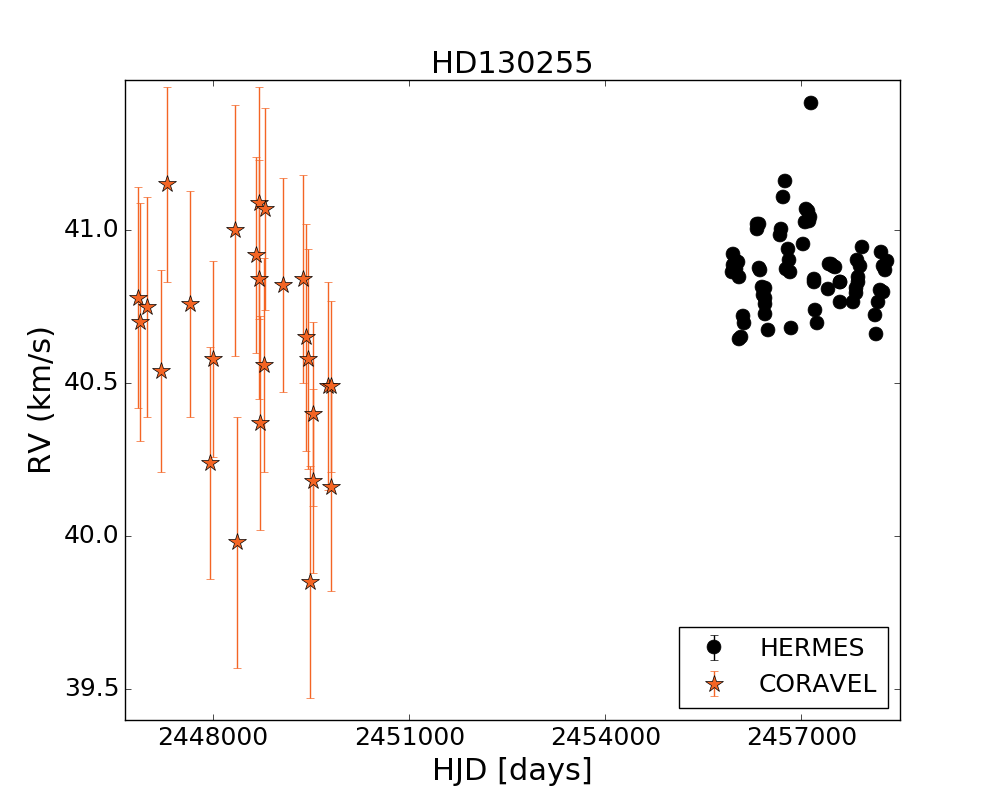}
\end{figure*}

\clearpage

\begin{figure*}[h!]
\includegraphics[width=0.33\textwidth,clip]{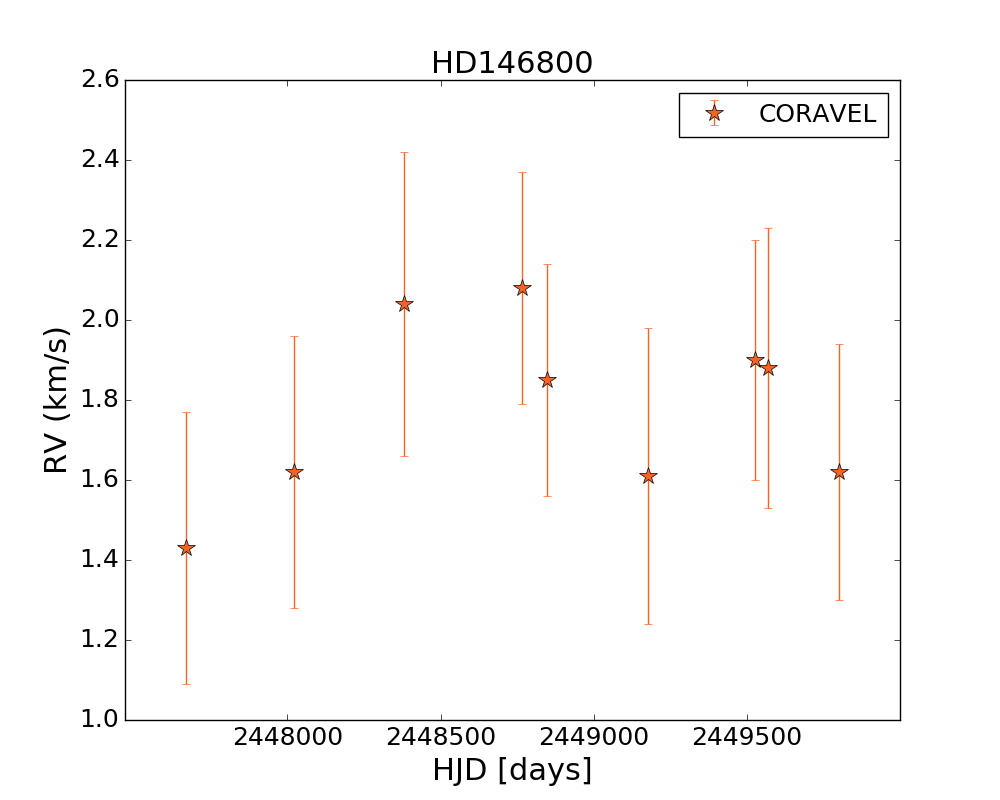}
\includegraphics[width=0.33\textwidth,clip]{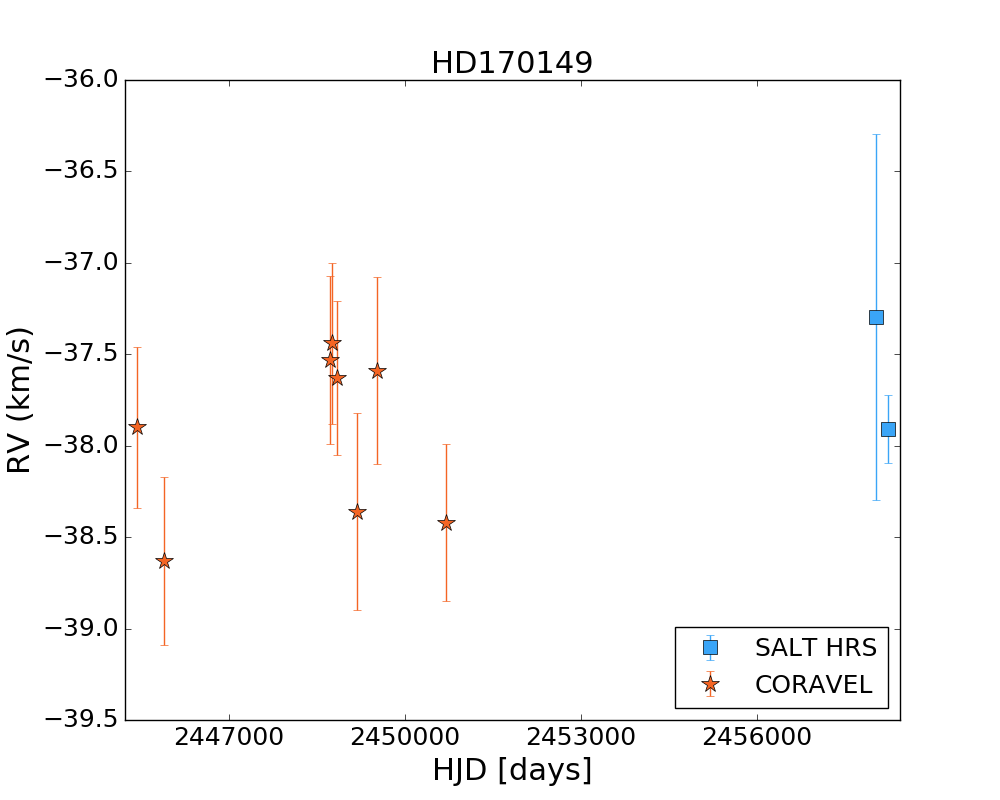}
\includegraphics[width=0.33\textwidth,clip]{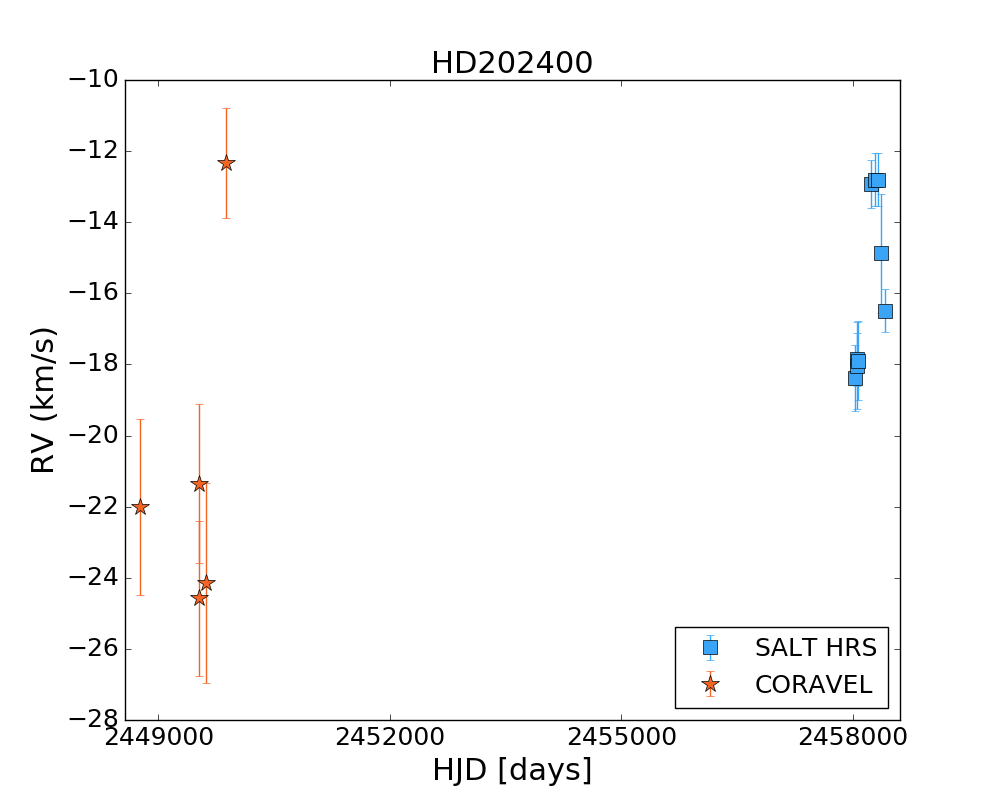}
\includegraphics[width=0.33\textwidth,clip]{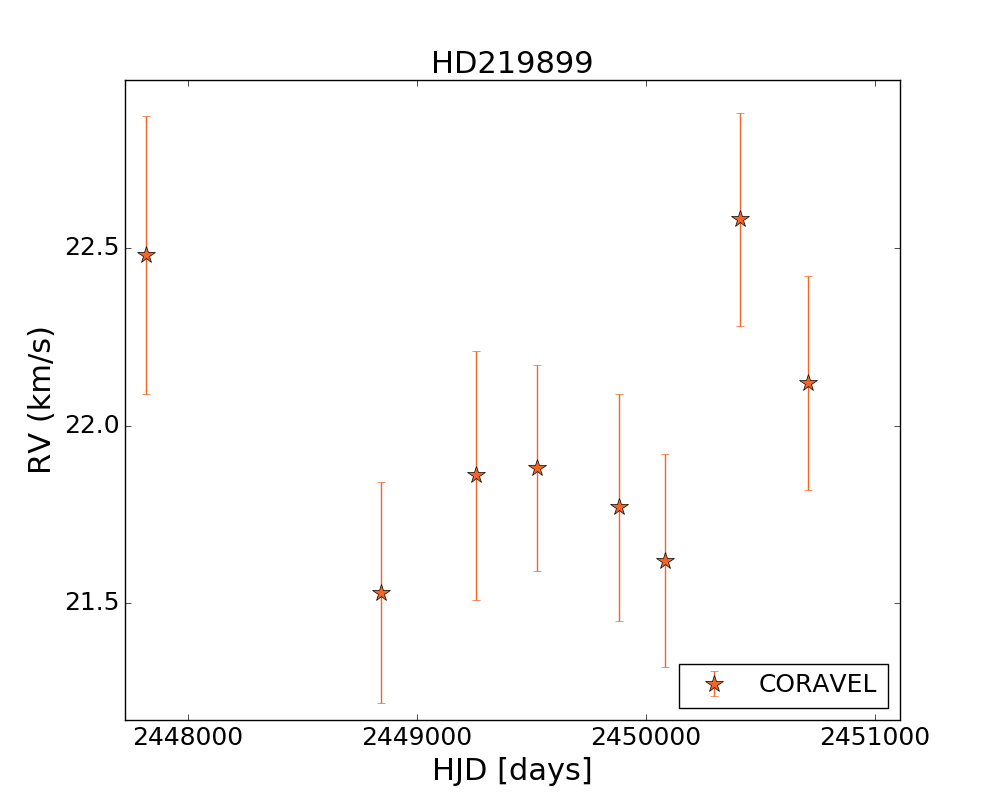}
\includegraphics[width=0.33\textwidth,clip]{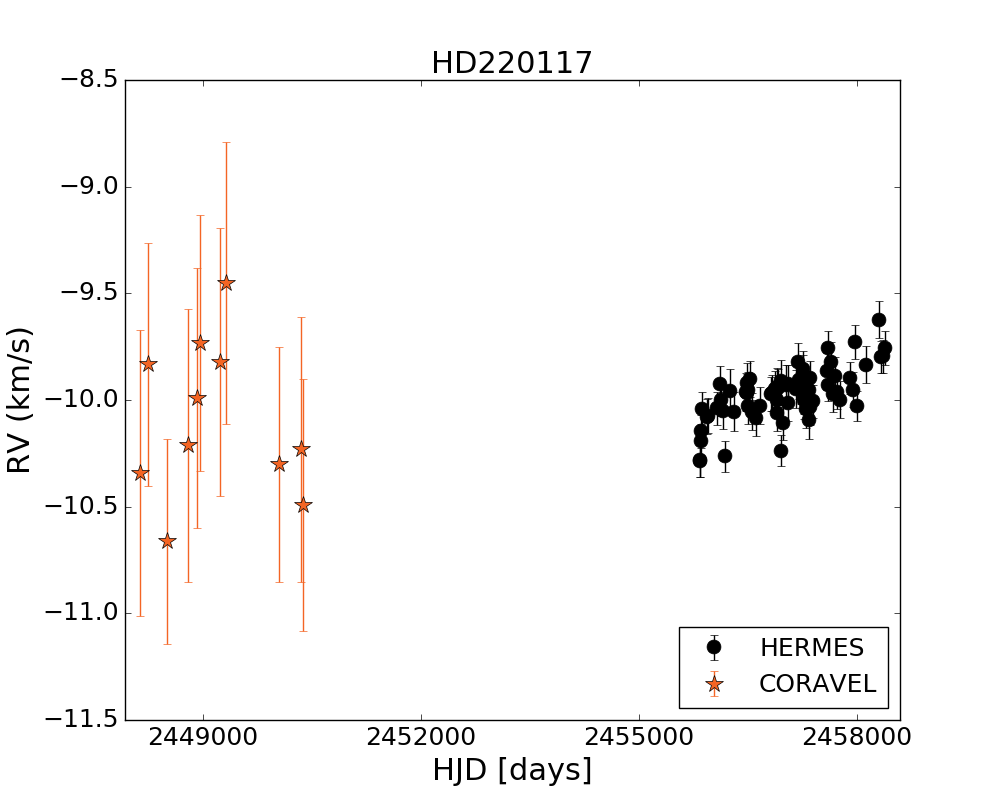}
\caption{\label{Fig:dBaothers} HERMES, CORAVEL, CORALIE and SALT radial velocity data of the targets with non-variable radial velocity data according to our criterion.}
\end{figure*}

\end{document}